# New Understanding of the Bethe Approximation and the Replica Method

by

Ryuhei Mori

A Dissertation Presented in Partial Fulfillment of the Requirements for the Degree
Doctor of Philosophy

Supervisor: Toshiyuki Tanaka

March 2013

KYOTO UNIVERSITY

# Abstract

In this thesis, new generalizations of the Bethe approximation and new understanding of the replica method are proposed. The Bethe approximation is an efficient approximation for graphical models, which gives an asymptotically accurate estimate of the partition function for many graphical models. The Bethe approximation explains the well-known message passing algorithm, belief propagation, which is exact for tree graphical models. It is also known that the cluster variational method gives the generalized Bethe approximation, called the Kikuchi approximation, yielding the generalized belief propagation. In the thesis, a new series of generalization of the Bethe approximation is proposed, which is named the asymptotic Bethe approximation. The asymptotic Bethe approximation is derived from the characterization of the Bethe free energy using graph covers, which was recently obtained by Vontobel. The asymptotic Bethe approximation can be expressed in terms of the edge zeta function by using Watanabe and Fukumizu's result about the Hessian of the Bethe entropy. The asymptotic Bethe approximation is confirmed to be better than the conventional Bethe approximation on some conditions. For this purpose, Chertkov and Chernyak's loop calculus formula is employed, which shows that the error of the Bethe approximation can be expressed as a sum of weights corresponding to generalized loops, and generalized for non-binary finite alphabets by using concepts of information geometry.

The replica method is a method invented in statistical physics for analyzing typical behaviors of random statistical models. Although the replica method is non-rigorous, it gives empirically correct results for various problems. However, many involved techniques employed in the replica method prevent study and understanding of it. The contribution of the second part of the thesis is regarding clarification of the replica method. The main tool for the purpose is the method of types, which is a well-known elementary tool in information theory. From the method of types, clear derivation and interpretation of the replica method are obtained. As a consequence, it is revealed that the replica method gives the same results as the cavity method, and that the replica method on the replica symmetry assumption implies the validity of Bethe approximation.

# Acknowledgments

I wish to thank all members of our laboratory. I would like to thank Toshiyuki Tanaka for his supervision. I am grateful to Kohichi Sakaniwa, Kenta Kasai and Takayuki Nozaki for their encouragement from 2007, the start of my research. I am thankful to all people who discussed with me in conferences. I would like to acknowledge Ruëdiger L. Urbanke, Nicolas Macris and Seyed Hamed Hassani for many discussions and for their hospitality on my visit in October 2010. I would like to acknowledge Andrea Montanari for insightful suggestions and for his hospitality on my visit in February 2012. I am grateful to Pascal O. Vontobel for many discussions, many insightful comments on my papers and hospitality on my visit in March 2012.

# Contents

# 1 Introduction

## 1.1 Graphical model and factor graph

The graphical model is a model for representing a probability measure with intuitive understanding. There are several ways for representing a probability measure by a graph. In this thesis, we deal with a *factor graph* as a graphical model for representing a probability measure. Let $\mathcal{X}$ be an alphabet for which probability measure is defined. A factor graph is a bipartite graph $G = (V, F, E, (f_a)_{a\in F})$ where $V$ is a set of variable nodes, $F$ is a set of factor nodes, $E \subseteq V \times F$ is a set of edges, and $f_a\colon \mathcal{X}^{d_a} \to \mathbb{R}_{\ge 0}$ is a non-negative function associated with a factor node $a \in F$. The neighborhoods of a factor node $a \in F$ and a variable node $i \in V$ are denoted by $\partial a \subseteq V$ and $\partial i \subseteq F$, respectively. Let $|\mathcal{A}|$ be the size of a set $\mathcal{A}$ and $N := |V|$. Here, $d_i$ and $d_a$ are the degrees of a variable node $i \in V$ and a factor node $a \in F$, respectively, i.e., $d_i = |\partial i|$ and $d_a = |\partial a|$. We also use notations $V(G)$, $F(G)$ and $E(G)$ for the set of variable nodes, the set of factor nodes and the set of edges in a factor graph $G$. For a discrete alphabet $\mathcal{X}$, the probability mass function $p(\boldsymbol{x}; G)$ defined by the factor graph $G$ is

$$p(\boldsymbol{x}; G) = \frac{1}{Z(G)} \prod_{a\in F} f_a(\boldsymbol{x}_{\partial a}), \qquad Z(G) = \sum_{\boldsymbol{x}\in\mathcal{X}^N} \prod_{a\in F} f_a(\boldsymbol{x}_{\partial a}). \tag{1.1}$$

Here, the constant $Z(G)$ for the normalization is called a *partition function*. For a continuous alphabet $\mathcal{X}$, the probability density function $p(\boldsymbol{x}; G)$ defined by the factor graph $G$ is the same as (1.1) except that the sum $\sum_{\boldsymbol{x}\in\mathcal{X}^N}$ is replaced by $\int_{\mathcal{X}^N} \mathrm{d}\boldsymbol{x}$. In this thesis, the alphabet $\mathcal{X}$ is assumed finite unless otherwise stated. One of the most classic and simplest examples of graphical models is the Ising model represented by a (non-factor) graph $G' = (V, E', ((J_{i,j} \in \mathbb{R})_{(i,j)\in E'}, (h_i \in \mathbb{R})_{i\in V}))$. The probability mass function $p_{\mathrm{Ising}}(\boldsymbol{x}; G', \beta)$ on $\{+1, -1\}^N$ is defined by a graph $G'$ and a parameter $\beta \in \mathbb{R}_{\ge 0}$ as

$$\begin{aligned} p_{\mathrm{Ising}}(\boldsymbol{x}; G', \beta) &= \frac{1}{Z_{\mathrm{Ising}}(G', \beta)} \exp\left\{ \beta\left( \sum_{(i,j)\in E'} J_{i,j} x_i x_j + \sum_{i\in V} h_i x_i \right) \right\} \\ Z_{\mathrm{Ising}}(G', \beta) &= \sum_{\boldsymbol{x}\in\{+1,-1\}^N} \exp\left\{ \beta\left( \sum_{(i,j)\in E'} J_{i,j} x_i x_j + \sum_{i\in V} h_i x_i \right) \right\}. \end{aligned} \tag{1.2}$$

A graph $G' = (V, E', (J_{i,j})_{(i,j)\in E'}, (h_i)_{i\in V})$ can be translated to a factor graph

$$G = \Big(V, F = E' \cup V, E = \{(i,(j,k)) \subseteq V \times E' \mid i = j \text{ or } i = k\},$$
$$\big((f_{(i,j)}(x_i, x_j) = \exp\{\beta J_{i,j} x_i x_j\})_{(i,j)\in E'}, (f_i(x_i) = \exp\{\beta h_i x_i\})_{i\in V}\big)\Big).$$

When $J_{i,j} \geq 0$ ($J_{i,j} \leq 0$) for all $(i,j) \in E$, the Ising model is said to be *ferromagnetic* (*antiferromagnetic*). The parameters $(h_i)_{i\in V}$ are called *(external) magnetic fields*. The Ising model is a simple model but often exhibits non-trivial behaviors in large-size limit, i.e., $N \to \infty$ as shown in Section 1.3. The parameter $\beta$ is called *inverse temperature*. A family of probability measures with a parameter $\beta \in \mathbb{R}_{\geq 0}$ defining a probability mass function $p(\boldsymbol{x}; G, \beta)$

$$p(\boldsymbol{x}; G, \beta) = \frac{1}{Z(G,\beta)} \prod_{a\in F} f_a(\boldsymbol{x}_{\partial a})^\beta, \qquad Z(G,\beta) = \sum_{x\in\mathcal{X}^N} \prod_{a\in F} f_a(\boldsymbol{x}_{\partial a})^\beta \tag{1.3}$$

is called the *Boltzmann-Gibbs distribution*. For the Boltzmann-Gibbs distribution, the partition function $Z(G,\beta)$ of a fixed factor graph $G$ can be regarded as a function of $\beta \in \mathbb{R}_{\geq 0}$ (or more broadly $\beta \in \mathbb{C}$). This is the reason why $Z(G)$ in (1.1) is generally called a partition "function."

Another important class of factor graphs is a *constraint satisfaction problem* (CSP). When $f_a(\boldsymbol{x}_{\partial a})$ takes values in $\{0,1\}$ for all $a \in F$, the probability measure defined by the factor graph is given as $p(\boldsymbol{x}; G) = 1/Z(G)$ if $f_a(\boldsymbol{x}_{\partial a}) = 1$ for all $a \in F$ and $p(\boldsymbol{x}; G) = 0$ otherwise. In this case, $Z(G)$ is the number of solutions of CSP, which is of great interest in theoretical computer science.

## 1.2 Partition function

The partition function plays a fundamental role in statistical physics. For the Boltzmann-Gibbs distribution (1.3), it holds

$$\frac{\partial \log Z(G,\beta)}{\partial \beta} = \Big\langle \sum_{a\in F} \log f_a(\boldsymbol{X}_{\partial a}) \Big\rangle =: -\mathcal{U}(G,\beta)$$

where $\boldsymbol{X}_{\partial a}$ denotes a random variable corresponding to $\boldsymbol{x}_{\partial a}$ and where $\langle\cdot\rangle$ denotes the expectation with respect to $p(\cdot; G, \beta)$[1]. The quantity $\mathcal{U}(G,\beta)$ is called the *internal energy* in thermodynamics and statistical physics. Similarly, it holds

$$\frac{\partial \frac{1}{\beta} \log Z(G,\beta)}{\partial \beta} = \frac{1}{\beta^2} \Big\langle \log p(\boldsymbol{X}; G, \beta) \Big\rangle =: -\frac{1}{\beta^2} \mathcal{H}(p(\,\cdot\,; G, \beta))$$

[1] In this paper, $0 \log 0$ is regarded as 0.

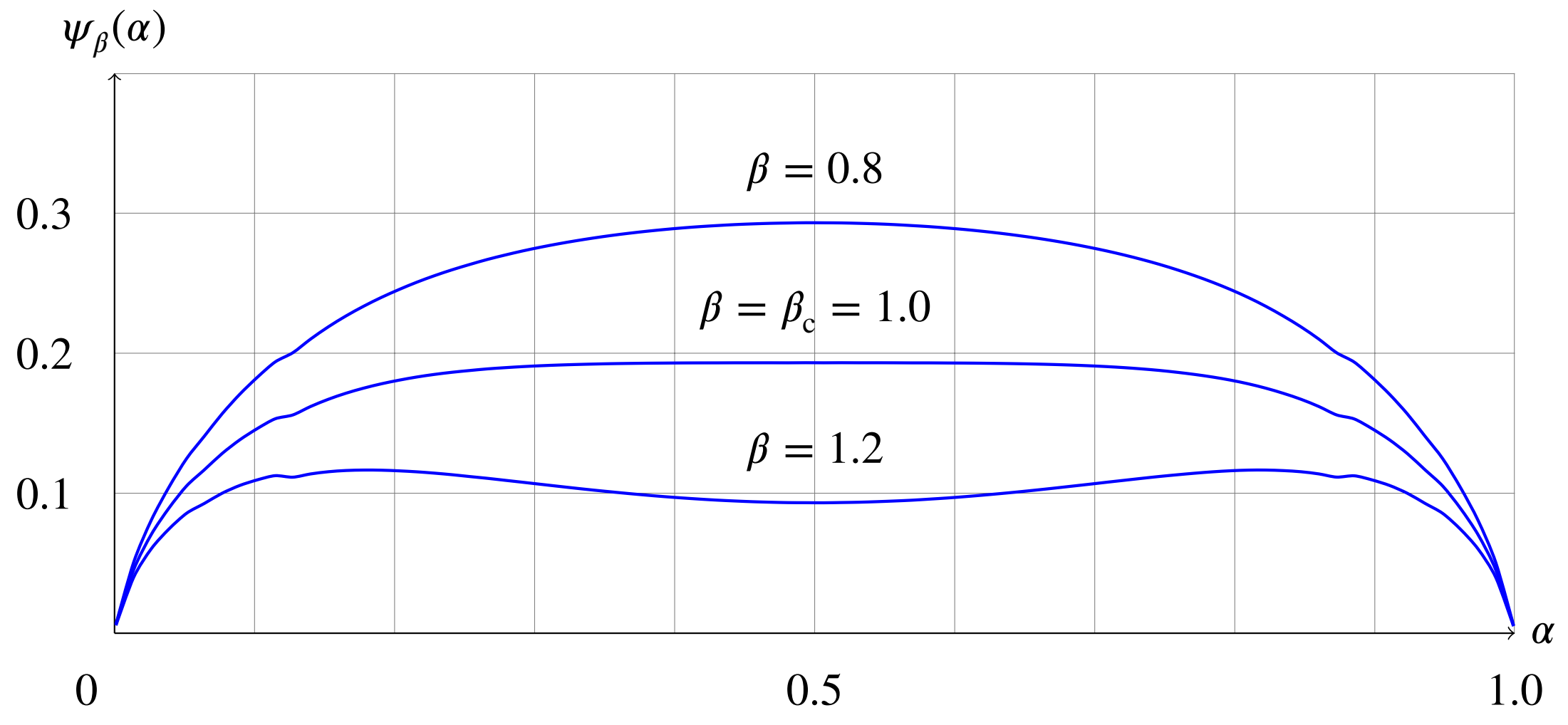

Figure 1.1: The Curie-Weiss model.

where $\boldsymbol{X}$ denotes a random variable corresponding to $\boldsymbol{x} \in \mathcal{X}^N$. The quantity $\mathcal{F}(G, \beta) := -(1/\beta) \log Z(G, \beta)$ is called the *Helmholtz free energy* in thermodynamics and statistical physics. The value $\mathcal{H}(\cdot)$ is called the *Shannon entropy* in information theory and the (canonical) *entropy* in thermodynamics and statistical physics. In physics, the entropy is often denoted by $\mathcal{S}(\cdot)$ rather than $\mathcal{H}(\cdot)$. The Helmholtz free energy, the internal energy and the entropy satisfy the equation

$$\mathcal{F}(G, \beta) = \mathcal{U}(G, \beta) - \frac{1}{\beta} \mathcal{H}(p(\,\cdot\,;, G, \beta)). \tag{1.4}$$

Furthermore, the second derivative of the logarithm of the partition function gives the variance of energy

$$\frac{\partial^2 \log Z(G, \beta)}{(\partial \beta)^2} = \left\langle \left( -\sum_{a \in F} \log f_a(\boldsymbol{X}_{\partial a}) - \mathcal{U}(G, \beta) \right)^2 \right\rangle.$$

Since the internal energy and the Shannon entropy are of great interest in both statistical physics and information theory, the Helmholtz free energy and equivalently the partition function are also meaningful. Due to the reason, analysis of the Helmholtz free energy is one of the central problems in information theory and statistical physics.

## 1.3 Phase transition

*Phase transition* is an important phenomenon well considered in statistical physics. Let $(G_N)_{N=1,2,\ldots,}$ be a deterministic or probabilistic sequence of factor graphs where the number of variable nodes in $G_N$ is $N$. For a finite $N$, $Z(G_N, \beta)$ is analytic in the whole complex plane $\beta \in \mathbb{C}$. However, in the limit $N \to \infty$, the analyticity can be lost at some point $\beta = \beta_c \in \mathbb{R}$. Such $\beta_c$ is called *critical temperature* and this phenomenon of lost analyticity is called phase transition. In the following, we review one of the most classic and simplest example including phase transition in statistical physics called the Curie-Weiss model. The Curie-Weiss model is the Ising model (1.2) where the graph is the complete graph, i.e., $V = \{1, \dots, N\}$, $E = \{(i, j) \in V \times V \mid i < j\}$, and where $J_{i,j} = 1/N$ for all $(i, j) \in E$. The partition function of the Curie-Weiss model of size $N$, denoted by $Z_{\mathrm{CW}}(N, \beta)$, is

$$Z_{\mathrm{CW}}(N,\beta) = \sum_{k=0}^{N} \binom{N}{k} \exp\left\{ \frac{\beta}{2N}\Big(k(k-1) + (N-k)(N-k-1) - 2k(N-k)\Big)\right\}$$
$$\doteq \max_{k=0,\dots,N}\left[\exp\left\{ N\left(H\left(\frac{k}{N}\right) + \frac{\beta}{2}\left(\frac{k}{N}\frac{k-1}{N} + \frac{N-k}{N}\frac{N-k-1}{N} - 2\frac{k}{N}\frac{N-k}{N}\right)\right)\right\}\right]$$

where $H(\alpha) := -\alpha\log\alpha - (1-\alpha)\log(1-\alpha)$ is the binary entropy function for $\alpha \in [0, 1]$. Here, $A(N) \doteq B(N) \overset{\text{def}}{\iff} \lim_{N\to\infty}(1/N)\log A(N) = \lim_{N\to\infty}(1/N)\log B(N)$. In the above equation, $k \in \{0, 1, \dots, N\}$ corresponds to the number of 1s in an assignment $\boldsymbol{x} \in \{+1, -1\}^N$. The derivation of the above equation will be explained in Section 4.1 in a more general setting. Then, the exponent of the partition function is obtained as

$$\phi(\beta) := \lim_{N\to\infty}\frac{1}{N}\log Z_{\mathrm{CW}}(N,\beta) = \max_{\alpha\in[0,1]}\left\{ H(\alpha) + \frac{\beta}{2}(\alpha^2 + (1-\alpha)^2 - 2\alpha(1-\alpha))\right\}$$
$$= \max_{\alpha\in[0,1]}\left\{ H(\alpha) + \frac{\beta}{2}(1-2\alpha)^2\right\} =: \frac{\beta}{2} + \max_{\alpha\in[0,1]}\left\{\psi_\beta(\alpha)\right\}.$$

The function $\psi_\beta(\alpha)$ is depicted in Figure 1.1. Let $\alpha(\beta) := \arg\max_\alpha \psi_\beta(\alpha)$. Then, $\alpha(\beta)$ must satisfy

$$\log\frac{1-\alpha(\beta)}{\alpha(\beta)} = 2\beta(1-2\alpha(\beta))$$

for the condition $\mathrm{d}\psi_\beta(\alpha)/\mathrm{d}\alpha = 0$. The left-hand side and the right-hand side of the above equation are depicted in Figure 1.2 as functions of $\alpha$. For $\beta \le 1$, $\alpha(\beta) = 1/2$ is the unique solution. For $\beta > 1$, there is a pair of non-trivial solutions, which must be chosen as shown in Figure 1.1. Then, the exponent of the partition function is

$$\phi(\beta) = \begin{cases} \log 2, & \beta \le 1 \\ H(\alpha(\beta)) + \frac{\beta}{2}(1-2\alpha(\beta))^2, & \beta > 1 \end{cases}$$

which is shown in Figure 1.3. Hence, the critical temperature of the Curier-Weiss model

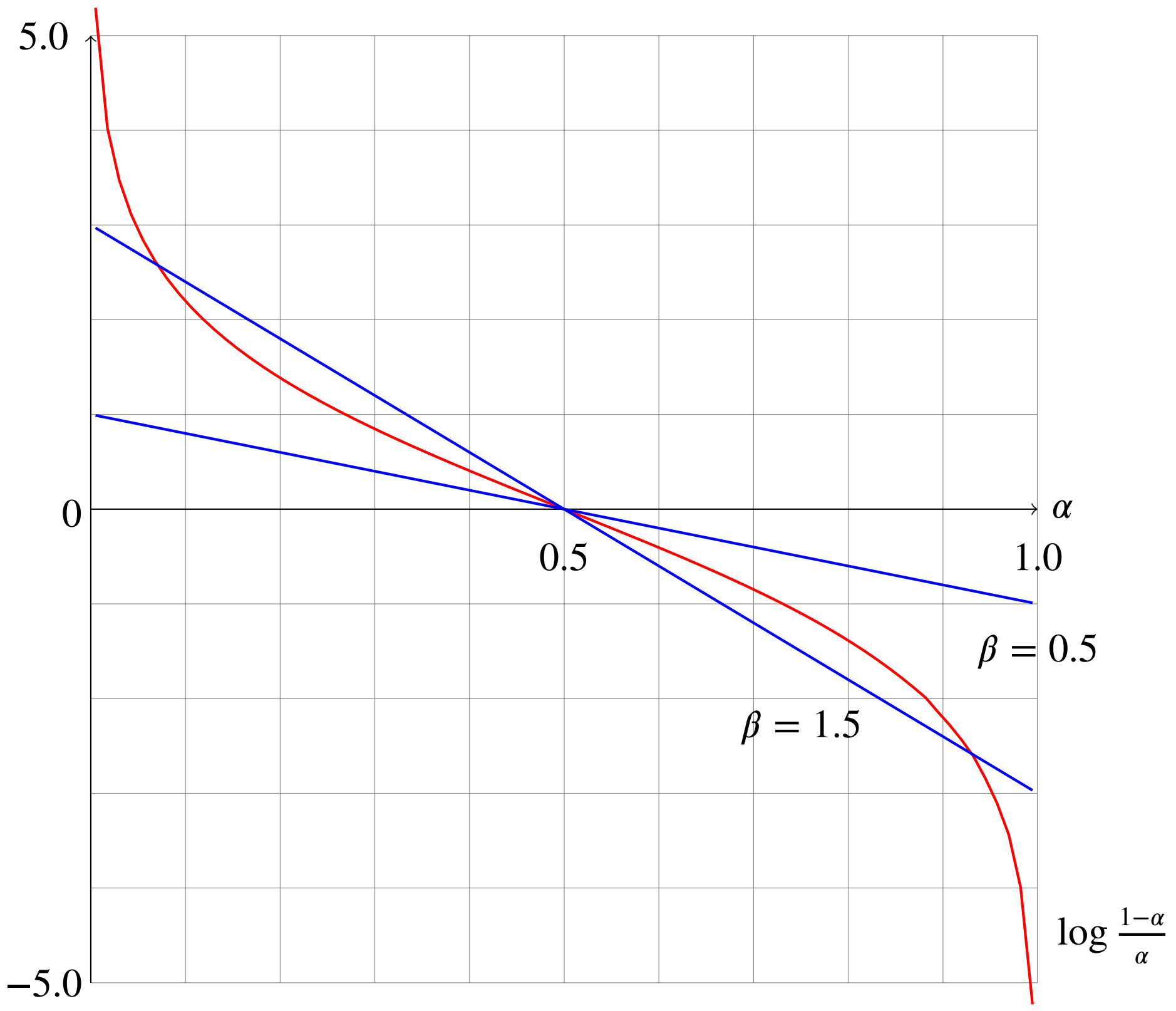

Figure 1.2: Red curve: $\log(1-\alpha)/\alpha$. Blue lines: $2\beta(1-2\alpha)$.

is $\beta_c = 1$. While the first derivative of $\phi(\beta)$

$$\frac{\mathrm{d}\phi(\beta)}{\mathrm{d}\beta} = \frac{1}{2}(1-2\alpha(\beta))^2$$

is a continuous function, the second derivative of $\phi(\beta)$ is discontinuous at $\beta = \beta_\mathrm{c}$ since

$$\alpha^{-1}(x) = \frac{1}{2(1-2x)} \log \frac{1-x}{x}$$

has zero slope at $x = 1/2$. When $\mathrm{d}\phi(\beta)/\mathrm{d}\beta$ is discontinuous at $\beta_\mathrm{c}$, a model is said to have a *first-order phase transition*. When $\mathrm{d}\phi(\beta)/\mathrm{d}\beta$ is continuous but $\mathrm{d}^2\phi(\beta)/\mathrm{d}\beta^2$ is discontinuous at $\beta_\mathrm{c}$, a model is said to have a *second-order phase transition*.

## 1.4 Complexities of computation and approximation of the partition function

Valiant defined the complexity class #P and showed that the computation of the permanents of (0,1)-matrices is #P-complete [Valiant, 1979]. The permanent of an $N \times$

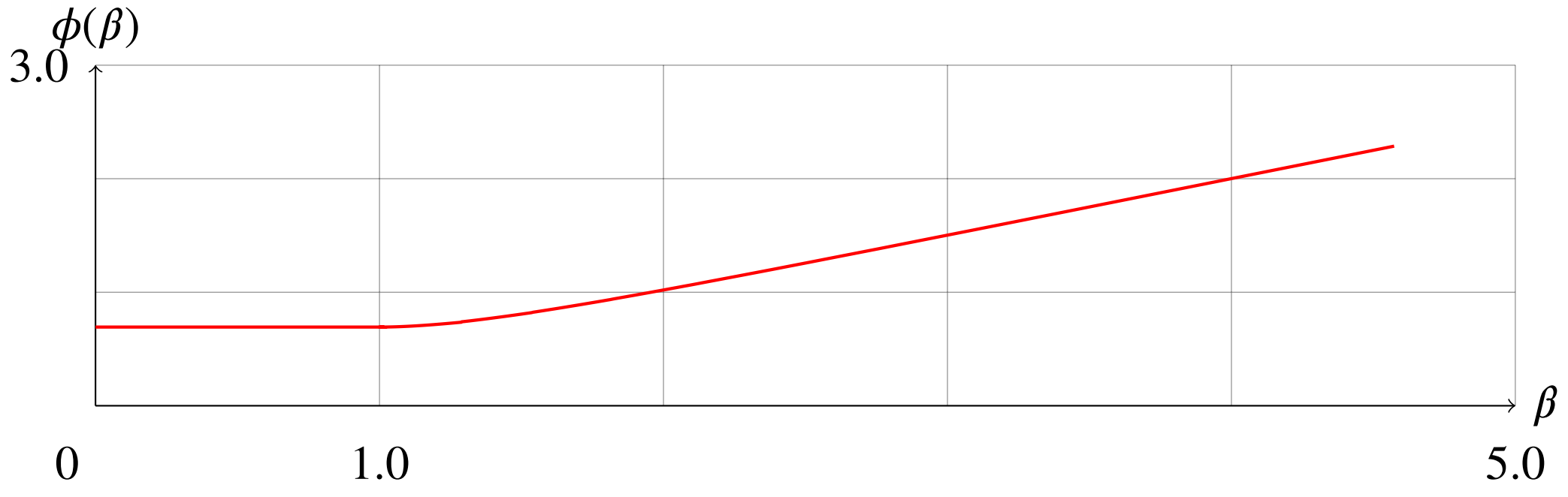

Figure 1.3: Free energy of the Curie-Weiss model.

$N$ square matrix $A$, which is similar to the determinant, is

$$\mathrm{perm}(A) = \sum_{\sigma} \prod_{i=1}^{n} A_{i,\sigma(i)}$$

where $\sigma$ runs over all permutations on $\{1, 2, \ldots, n\}$. The permanent of a non-negative matrix can be regarded as a partition function as follows

$$Z(G) = \sum_{x \in \{0,1\}^{n^2}} \prod_{i=1}^{n} \mathbb{1}\left\{ \sum_{j=1}^{n} x_{i,j} = 1 \right\} \prod_{j=1}^{n} \mathbb{1}\left\{ \sum_{i=1}^{n} x_{i,j} = 1 \right\} \prod_{i=1}^{n} \prod_{j=1}^{n} A_{i,j}^{x_{i,j}}. \tag{1.5}$$

Hence, the computation of partition function includes #P problems, which are considered to be harder than NP-complete problems due to Toda's theorem [Toda, 1991]. Even if the degrees of variable and factor nodes are bounded by a finite constant, counting the number of solutions for CSP is still #P-complete [Vadhan, 2001]. Hence, an accurate approximation is considered to be a realistic goal. For accurate and efficient approximations, the notion of *fully polynomial-time randomized approximation scheme* (FPRAS) is useful, which is a randomized algorithm computing an approximation $\bar{Z}$ for a partition function $Z$ in polynomial time with respect to the size of the problem and with respect also to $1/\epsilon$ where $\epsilon$ satisfies $Z(1+\epsilon)^{-1} \leq \bar{Z} \leq Z(1+\epsilon)$. Although some #P-complete problems do not have FPRAS unless NP =RP, fortunately, some #P-complete problems have FPRAS [Dyer et al., 2004]. Most of FPRAS for #P-complete problems are based on the Markov chain Monte Carlo (MCMC) approach. The MCMC algorithm gives FPRAS for the computation of permanent of non-negative matrix [Jerrum and Sinclair, 1989; Jerrum et al., 2004] and the computation of partition function of the ferromagnetic Ising model [Jerrum and Sinclair, 1993]. Even if the degrees are bounded, there are #P-complete problems which do not admit an FPRAS unless NP =RP, e.g., counting number of independent sets on $\Delta$-regular graph for $\Delta \geq 6$ [Dyer et al., 2002; Sly and Sun, 2012]. Other properties of graph, e.g., girth, expander, etc., may be useful for admitting FPRAS [Chandrasekaran et al., 2011]. *Fully polynomial-time approximation scheme* (FPTAS), which is the deterministic version of FPRAS, is also found for the independent set problem [Weitz, 2006] and the Ising model [Sinclair et al., 2012].

# 1.5 Background and contributions of the thesis

## 1.5.1 Background

**Bethe approximation:** The *Bethe approximation* is a popular approximation invented in statistical physics for the partition function of many graphical models, e.g., low-density parity-check codes [Richardson and Urbanke, 2008], code division multiple access channel [Kabashima, 2003], compressed sensing [Donoho et al., 2010], etc. The Bethe approximation yields the well-known efficient message passing algorithm *belief propagation*. Although theoretical aspects of the Bethe approximation have not been well understood, recently it is proved that for some models, the Bethe approximation gives exact asymptotic behaviors of the partition function in the large-size limit [Dembo and Montanari, 2010], [Dembo et al., 2011]. Since the belief propagation is a more efficient algorithm than MCMC algorithms, studies of the Bethe approximation are considered to be useful also in a practical point of view although no FPTAS using the Bethe approximation has been known. Recently, the Bethe approximation is analyzed for some problems in computer science [Vontobel, 2011b], [Chandrasekaran et al., 2011]. In this thesis, theoretical aspects of the Bethe approximation are discussed.

**Replica method:** The replica method is a non-rigorous method invented in statistical physics for evaluation of $\lim_{N\to\infty}(1/N)\mathbb{E}[\log Z(G)]$ where $\mathbb{E}[\cdot]$ denotes the expectation with respect to a random factor graph $G$ [Mézard et al., 1987]. In the replica method, first $\lim_{N\to\infty}(1/N)\log\mathbb{E}[Z(G)^n]$ is evaluated for $n \in \mathbb{N}$. Then, $\lim_{N\to\infty}(1/N)\mathbb{E}[\log Z(G)]$ is obtained as $\lim_{n\to 0}(1/n)\lim_{N\to\infty}(1/N)\log\mathbb{E}[Z(G)^n]$ on the basis of several ansatz. Although the replica method is not rigorous, empirically it always gives correct results [Mézard et al., 1987], [Monasson and Zecchina, 1997], [Nishimori, 2001], [Tanaka, 2002]. It is empirically known that the replica method on the replica symmetry assumption implies the asymptotic exactness of the Bethe approximation [Nishimori, 2001], [Mézard and Montanari, 2009]. However, the relationship between the replica method and the Bethe approximation has not been clearly understood.

### 1.5.2 Contributions of the thesis

**Generalization of the loop calculus formula:** Recently, an exact equality called the *loop calculus formula* between the true partition function and its Bethe approximation for binary models is found in [Chertkov and Chernyak, 2006a]. This result is useful for improvement of the Bethe approximation [Chertkov and Chernyak, 2006c], [Gómez et al., 2007] and analysis of errors of the Bethe approximation [Chandrasekaran et al., 2011]. Generalization of the formula for non-binary finite alphabets is considered in [Chernyak and Chertkov, 2007]. However, this result does not give an explicit representation of the formula. In this thesis, explicit representations of the loop calculus formula for non-binary finite alphabets are given by using concepts of information geometry. This result is useful for many purposes similarly to the binary case. This result is presented in Chapter 3.

**Proposal of new generalizations of the Bethe approximation:** New series of generalizations of the Bethe approximation is proposed. The idea of the generalizations is based on the method of graph covers [Vontobel, 2010b]. The generalized Bethe approximations are represented by using the edge zeta function. For some problems, the new generalized Bethe approximation is provably better than the original Bethe approximation. In order to explain the relationship between the new generalized Bethe approximation and the true partition function, the loop calculus formula for non-binary finite alphabets is used, which is obtained in Chapter 3. This result is presented in Chapter 5.

**New derivations of expected log-partition function by the replica method:** New derivations of expected log-partition function by the replica method are proposed. In the derivation, the method of types for a factor graph is used, which is introduced in [Vontobel, 2010b]. From this derivation, one can understand that the replica symmetry assumption implies the asymptotic exactness of the Bethe approximation. This understanding is considered to be useful for study of the replica method for non-physicists. This result is presented in Chapter 6.

## 1.6 Organization of the thesis

**Chapter 2:** The Bethe approximation is introduced by the commonly accepted cluster variation method. The relationship between the Bethe approximation and the belief propagation is also shown.

**Chapter 3:** The characterization of the Bethe approximation using loop calculus is introduced, which is recently obtained by [Chertkov and Chernyak, 2006a]. Equalities between the partition function and the Bethe approximated partition function are also obtained. Here, the loop calculus formula is generalized for non-binary finite alphabets by using tools of information geometry.

**Chapter 4:** The method of graph covers, used in [Vontobel, 2010b], is reviewed, which gives a novel characterization of the Bethe entropy and the Bethe free energy. This idea is used for generalization of the Bethe approximation in Chapter 5.

**Chapter 5:** New generalizations of the Bethe approximation are introduced by using the method of graph covers. In the new generalization, the prefactor for improving the Bethe approximation is represented by the edge zeta function. The representation by the edge zeta function is obtained from the formula between the edge zeta function and the determinant of the Hessian of the Bethe free energy shown in [Watanabe, 2010]. It is shown that the new approximation is better than the Bethe approximation on some problem settings both theoretically and empirically.

**Chapter 6:** New calculations in the replica method is proposed in this chapter which clarify the relationship between the replica symmetric free energy and the Bethe free energy.

# Part I

# The Bethe Approximation

# 2 Definition of the Bethe Free Energy by the Cluster Variation Method

**In this chapter, the Bethe approximation is introduced for an efficient approximation of the partition function. The cluster variation method is used for the definition of the Bethe free energy, which is the most traditional and commonly accepted definition. Furthermore, the belief propagation is introduced as an algorithm that tries to find the minimum of the Bethe free energy.**

## 2.1 Exactly solvable factor graphs

### 2.1.1 Tree factor graph

In this section, the computation of the partition function of a tree factor graph is considered. A *leaf variable node* is defined as a degree-one variable node. When factor node $a \in F$ has one and only one neighboring variable node $i \in \partial a$ which is not a leaf variable node, $a$ is said to be a *leaf factor node*. A tree factor graph $G$ includes a leaf factor node unless $G$ only includes the unique factor node. The following tree decimation algorithm outputs the partition function $Z(G)$ of a tree factor graph $G$.

Step 0: If the factor graph includes the unique factor node $a$, output $Z(G) = \sum_{\boldsymbol{x}_{\partial a} \in \mathcal{X}^{d_a}} f_a(\boldsymbol{x}_{\partial a})$.

Step 1: Choose a leaf factor node $a \in F$.

Step 2: If the degree of $a$ is not one, remove all leaf variable nodes connected to $a$ and replace $f_a(\boldsymbol{x}_{\partial a})$ by $\sum_{\boldsymbol{x}_{\partial a \setminus \{i\}}} f_a(\boldsymbol{x}_{\partial a})$ where $i \in \partial a$ is the non-leaf variable node connected to $a$. Now, $a$ is a degree-one factor node.

Step 3: Choose a factor node $b \in \partial i \setminus \{a\}$ where $\{i\} = \partial a$. Remove a factor node $a$ and replace $f_b(\boldsymbol{x}_{\partial b})$ by $f_b(\boldsymbol{x}_{\partial b}) f_a(x_i)$.

Step 4: Go to Step 0.

The complexity of this algorithm is $O(|E|)$.

### 2.1.2 Single-cycle factor graph

By using the tree decimation algorithm in the previous section, without loss of generality, a single-cycle factor graph $G$ can be assumed to be a cycle graph, i.e., all variable nodes and all factor nodes in a factor graph are degree-two. For a chain factor graph, i.e., all variable nodes and all factor nodes in a factor graph are degree-two except for two degree-one variable nodes at the ends of the chain, the partition function can be easily calculated by the method of transfer matrix.

**Lemma 2.1** (Partition function of a chain factor graph)**.**

$$\sum_{\boldsymbol{x}\in\mathcal{X}^N, x_1=x, x_N=x'} \prod_{i=1}^{N-1} f_i(x_i, x_{i+1}) = (F^{(1)}F^{(2)} \cdots F^{(N-1)})_{x,x'}$$

*where $F^{(i)}$ is a $|\mathcal{X}| \times |\mathcal{X}|$ matrix with $F^{(i)}_{x,x'} = f_i(x, x')$*

*Proof.* Let $Z^{(K)}$ be a $|\mathcal{X}| \times |\mathcal{X}|$ matrix whose $(x_1, x_K)$-element is

$$Z^{(K)}_{x_1,x_K} := \sum_{(x_2,x_3,\ldots,x_{K-1})\in\mathcal{X}^{K-2}} \prod_{i=1}^{K-1} f_i(x_i, x_{i+1}).$$

Then, it holds

$$Z^{(2)} = F^{(1)}, \qquad Z^{(K)} = Z^{(K-1)}F^{(K-1)}. \qquad \square$$

From this lemma, the partition function of a cycle factor graph can be easily calculated.

**Lemma 2.2** (Partition function of a cycle factor graph)**.**

$$\sum_{\boldsymbol{x}\in\mathcal{X}^N} f_N(x_N, x_1) \prod_{i=1}^{N-1} f_i(x_i, x_{i+1}) = \mathrm{tr}(F^{(1)}F^{(2)} \cdots F^{(N)})$$

*Proof.* The lemma is obtained from

$$\begin{aligned}\sum_{\boldsymbol{x}\in\mathcal{X}^N} f_N(x_N, x_1) \prod_{i=1}^{N-1} f_i(x_i, x_{i+1}) &= \sum_{x\in\mathcal{X}} \sum_{\boldsymbol{x}\in\mathcal{X}^{N+1}, x_1=x_{N+1}=x} \prod_{i=1}^{N} f_i(x_i, x_{i+1}) \\ &= \sum_{x\in\mathcal{X}} (F^{(1)} \cdots F^{(N)})_{x,x} = \mathrm{tr}(F^{(1)}F^{(2)} \cdots F^{(N)}). \qquad \square\end{aligned}$$

### 2.1.3 Planar factor graph without magnetic field

A partition function of two-dimensional Ising model without magnetic field is for the first time shown by [Onsager, 1944]. Generally a partition function of the Ising model with a planar factor graph can be calculated in polynomial time if it does not have magnetic field. A combinatorial method for the problems is shown by [Kac and Ward, 1952]. Another method is invented and conjectured by Feynman and proved by [Sherman, 1960]. The method of dimer statistics and Pfaffian is shown by [Kasteleyn, 1961] and [Fisher, 1961]. Here, the result of [Kac and Ward, 1952] is shown without a proof.

**Lemma 2.3** ([Kac and Ward, 1952])**.** *The partition function of the Ising model without magnetic field represented by a planar graph is*

$$Z_{\text{Ising}}(G', \beta) = \left[ 2^N \prod_{(i,j)\in E} \cosh(\beta J_{i,j}) \right] \sqrt{\det\left(I_{2|E|} - M\right)}$$

*where $M$ is a square matrix whose rows and columns are indexed by directed edges and is defined by*

$$M_{i\to j, k\to l} = \begin{cases} \exp\left\{ \sqrt{-1}\gamma_{i\to j,k\to l}/2 \right\} \tanh(\beta J_{i,j}), & \text{if } j = k, i \neq l \\ 0, & \text{otherwise.} \end{cases}$$

*Here, $\gamma_{i\to j,k\to l}$ is the angle of the edge connection between $i \to j$ and $k \to l$.*

## 2.2 Exponential family and Legendre transform

### 2.2.1 Legendre transform

**Definition 2.4** (Legendre transform)**.** For a continuous function $f : U \to \mathbb{R}$ where $U$ is an open subset of $\mathbb{R}^d$, the Legendre transform $f^\star$ of $f$ is defined as

$$f^\star(\boldsymbol{\eta}) := \sup_{\theta} \left\{ \sum_{k=1}^{d} \eta_k \theta_k - f(\boldsymbol{\theta}) \right\}. \tag{2.1}$$

From the definition, $f^\star$ is always convex.

**Lemma 2.5** (Duality of Legendre transform)**.**

$$f^{\star\star} = \text{conv}(f)$$

*where* conv$(f)$ *denotes the convex hull of $f$.*

*Proof.* Since $f^{\star\star}$ is convex, it is sufficient to prove

$$\mathrm{conv}(f)(\boldsymbol{\theta}) \leq f^{\star\star}(\boldsymbol{\theta}) \leq f(\boldsymbol{\theta}).$$

Since

$$\begin{aligned} f^{\star\star}(\boldsymbol{\theta}) &= \sup_{\boldsymbol{\eta}} \left\{ \sum_{k=1}^{d} \theta_k \eta_k - \sup_{\boldsymbol{\theta}'} \left\{ \sum_{k=1}^{d} \eta_k \theta'_k - f(\boldsymbol{\theta}') \right\} \right\} \\ &= \sup_{\boldsymbol{\eta}} \left\{ \inf_{\boldsymbol{\theta}'} \left\{ \sum_{k=1}^{d} (\theta_k - \theta'_k)\eta_k + f(\boldsymbol{\theta}') \right\} \right\} \end{aligned}$$

the upper bound $f^{\star\star} \leq f$ is obtained by fixing $\boldsymbol{\theta}' = \boldsymbol{\theta}$ and the lower bound $\mathrm{conv}(f) \leq f^{\star\star}$ is obtained by restricting $\eta_k$ to the left or the right partial derivatives of $\mathrm{conv}(f)$ at $\boldsymbol{\theta}$ with respect to $\theta_k$. □

When $f(\boldsymbol{\theta})$ is strictly convex, the supremum in (2.1) is achieved at the unique point, say $\boldsymbol{\theta}(\boldsymbol{\eta})$. In this case, if $f(\boldsymbol{\theta})$ is differentiable, $\boldsymbol{\theta}(\boldsymbol{\eta})$ must satisfy

$$\left.\frac{\mathrm{d}f(\boldsymbol{\theta})}{\mathrm{d}\boldsymbol{\theta}}\right|_{\boldsymbol{\theta}=\boldsymbol{\theta}(\boldsymbol{\eta})} = \boldsymbol{\eta}$$

where $\frac{\mathrm{d}f(\boldsymbol{\theta})}{\mathrm{d}\boldsymbol{\theta}}$ is the gradient vector, i.e., its $k$-th element is $\frac{\partial f(\boldsymbol{\theta})}{\partial \theta_k}$. In the same way, $\boldsymbol{\eta}(\boldsymbol{\theta})$ is defined by

$$\left.\frac{\mathrm{d}f^{\star}(\boldsymbol{\eta})}{\mathrm{d}\boldsymbol{\eta}}\right|_{\boldsymbol{\eta}=\boldsymbol{\eta}(\boldsymbol{\theta})} = \boldsymbol{\theta}$$

when $f^{\star}(\boldsymbol{\eta})$ is strictly convex and differentiable. Let $\frac{\partial^2 f(\boldsymbol{\theta})}{\partial \boldsymbol{\theta}^2}$ be the Hessian matrix of $f(\boldsymbol{\theta})$, i.e., its $(k,l)$-element is $\frac{\partial^2 f(\boldsymbol{\theta})}{\partial \theta_k \partial \theta_l}$, and $\frac{\partial \boldsymbol{\theta}(\boldsymbol{\eta})}{\partial \boldsymbol{\eta}}$ be the Jacobian matrix, i.e., its $(k,l)$-element is $\frac{\partial \theta_k(\boldsymbol{\eta})}{\partial \eta_l}$ where $\theta_k(\boldsymbol{\eta}) := (\boldsymbol{\theta}(\boldsymbol{\eta}))_k$. If $f$ is a $C^2$ function and the Hessian matrix of $f$ is positive-definite, $\boldsymbol{\theta}(\boldsymbol{\eta})$ is differentiable since $\boldsymbol{\theta}(\boldsymbol{\eta}) = f'^{-1}(\boldsymbol{\eta})$ where $f'(\boldsymbol{\theta}) := \frac{\mathrm{d}f(\boldsymbol{\theta})}{\mathrm{d}\boldsymbol{\theta}}$.

**Lemma 2.6.** *If $f$ is a $C^2$ function with positive-definite Hessian matrix,*

$$\frac{\mathrm{d}f(\boldsymbol{\theta})}{\mathrm{d}\boldsymbol{\theta}} = \boldsymbol{\eta}(\boldsymbol{\theta}), \qquad \frac{\mathrm{d}f^{\star}(\boldsymbol{\eta})}{\mathrm{d}\boldsymbol{\eta}} = \boldsymbol{\theta}(\boldsymbol{\eta}).$$

*Hence, $\boldsymbol{\theta}(\boldsymbol{\eta})$ is the inverse function of $\boldsymbol{\eta}(\boldsymbol{\theta})$. Furthermore, it holds*

$$\frac{\partial f(\boldsymbol{\theta})}{\partial \boldsymbol{\theta}^2} = \frac{\partial \boldsymbol{\eta}(\boldsymbol{\theta})}{\partial \boldsymbol{\theta}} = \left(\frac{\partial \boldsymbol{\theta}(\boldsymbol{\eta})}{\partial \boldsymbol{\eta}}\right)^{-1} = \left(\frac{\partial f^{\star}(\boldsymbol{\eta})}{\partial \boldsymbol{\eta}^2}\right)^{-1}.$$

*Proof.* It is sufficient to prove the first equation. The first equation is obtained by

$$\begin{aligned} \frac{\mathrm{d}f(\boldsymbol{\theta})}{\mathrm{d}\boldsymbol{\theta}} &= \frac{\mathrm{d}\left[\sum_{k=1}^{d} \theta_k \eta_k(\boldsymbol{\theta}) - f^{\star}(\boldsymbol{\eta}(\boldsymbol{\theta}))\right]}{\mathrm{d}\boldsymbol{\theta}} \\ &= \boldsymbol{\eta}(\boldsymbol{\theta}) + \sum_{k=1}^{d} \theta_k \frac{\partial \eta_k(\boldsymbol{\theta})}{\partial \boldsymbol{\theta}} - \left.\frac{\partial f^{\star}(\boldsymbol{\eta})}{\partial \boldsymbol{\eta}}\right|_{\boldsymbol{\eta}=\boldsymbol{\eta}(\boldsymbol{\theta})} \frac{\partial \eta(\boldsymbol{\theta})}{\partial \boldsymbol{\theta}} = \boldsymbol{\eta}(\boldsymbol{\theta}). \end{aligned}$$

□

### 2.2.2 Exponential family

In this section, the exponential family is introduced, which is a class of parametric families of probability measures. The domain of the probability measures is assumed to be $\mathcal{X}$ rather than $\mathcal{X}^N$ since any graphical structure is not assumed in this section. The parametric family of probability measures is a family of probability measures having a parameter $\boldsymbol{\theta} \in \Lambda \subseteq \mathbb{R}^d$ where $d \in \mathbb{N}$ is the dimension of the parameters and where $\Lambda$ is a space of the parameters which is an open subset of $\mathbb{R}^d$. Usually, the existence of the first and the second derivatives of probability mass (density) functions with respect to the parameter is assumed.

**Definition 2.7** (Fisher information matrix)**.** The Fisher information matrix $\mathcal{J}(\boldsymbol{\theta})$ is a $d \times d$ matrix whose $(k, l)$ element is

$$\mathcal{J}_{k,l}(\boldsymbol{\theta}) := \left\langle \frac{\partial \log p(X \mid \boldsymbol{\theta})}{\partial \theta_k} \frac{\partial \log p(X \mid \boldsymbol{\theta})}{\partial \theta_l} \right\rangle$$

where $X \sim p(x \mid \boldsymbol{\theta})$.

**Definition 2.8** (Exponential family)**.** The exponential family is a parametric family of probability measures whose probability mass (density) functions can be expressed in the form

$$p_{\mathrm{E}}(x; \boldsymbol{\theta}) = \frac{1}{Z_{\mathrm{E}}(\boldsymbol{\theta})} \exp\left\{ C(x) + \sum_{k=1}^{d} \theta_k t_k(x) \right\}, \quad Z_{\mathrm{E}}(\boldsymbol{\theta}) = \sum_{x \in \mathcal{X}} \exp\left\{ C(x) + \sum_{k=1}^{d} \theta_k t_k(x) \right\} \tag{2.2}$$

using a set of functions $(t_k : \mathcal{X} \to \mathbb{R})_{k=1,\dots,d}$ called a *sufficient statistic* and a function $C(x) : \mathcal{X} \to \mathbb{R}$.

The Ising model can be regarded as the exponential family with the single parameter $\beta$. For an exponential family, it holds

$$\frac{\partial \log Z_{\mathrm{E}}(\boldsymbol{\theta})}{\partial \theta_k} = \langle t_k(\boldsymbol{X}) \rangle_{\theta}$$

$$\frac{\partial^2 \log Z_{\mathrm{E}}(\boldsymbol{\theta})}{\partial \theta_k \partial \theta_l} = \Big\langle \big(t_k(X) - \langle t_k(X) \rangle\big) \big(t_l(X) - \langle t_l(X) \rangle\big) \Big\rangle_{\theta} = \mathcal{J}(\boldsymbol{\theta})_{k,l}.$$

In the following, we assume that $C(x) = 0$ and consider the Legendre transform of $f(\boldsymbol{\theta}) := \log Z_{\mathrm{E}}(\boldsymbol{\theta})$. Since $\mathcal{J}(\boldsymbol{\theta})$ is positive-semidefinite, $f(\boldsymbol{\theta})$ is convex. The supremum of $\sum_{k=1}^{d} \eta_k \theta_k - f(\boldsymbol{\theta})$ is taken at the stationary points $\{\boldsymbol{\theta} \in \Theta \mid \langle \boldsymbol{t}(X) \rangle_{\theta} = \boldsymbol{\eta}\}$. Hence, the Legendre transform is obtained as

$$f^{\star}(\boldsymbol{\eta}) = \sum_{k=1}^{d} \eta_k \theta_k^* - \log Z_{\mathrm{E}}(\boldsymbol{\theta}^*) = \langle \log p_{\mathrm{E}}(\boldsymbol{X}; \boldsymbol{\theta}^*) \rangle_{\theta^*}$$

which is the minus Shannon entropy of $p_{\mathrm{E}}(x;\boldsymbol{\theta}^*)$ where $\boldsymbol{\theta}^* \in \{\boldsymbol{\theta} \in \Theta \mid \langle \boldsymbol{t}(X)\rangle_{\boldsymbol{\theta}} = \boldsymbol{\eta}\}$.

When the Fisher information matrix $\mathcal{J}(\boldsymbol{\theta})$ is positive-definite, $f(\boldsymbol{\theta})$ is strictly convex. In this case, $\boldsymbol{\eta}(\boldsymbol{\theta})$ is called an *expectation parameter*. From Lemma 2.6, one obtains the following lemma.

**Lemma 2.9.** *If $\mathcal{J}(\boldsymbol{\theta})$ is positive-definite, it holds*

$$\frac{\mathrm{d}\log Z_{\mathrm{E}}(\boldsymbol{\theta})}{\mathrm{d}\boldsymbol{\eta}} = \boldsymbol{\eta}(\boldsymbol{\theta}), \qquad \frac{\mathrm{d}\langle \log p(\boldsymbol{X};\boldsymbol{\theta}(\boldsymbol{\eta}))\rangle_{\boldsymbol{\theta}(\boldsymbol{\eta})}}{\mathrm{d}\boldsymbol{\eta}} = \boldsymbol{\theta}(\boldsymbol{\eta})$$

$$\frac{\partial^2 \log Z_{\mathrm{E}}(\boldsymbol{\theta})}{\partial \boldsymbol{\eta}^2} = \frac{\partial \boldsymbol{\eta}(\boldsymbol{\theta})}{\partial \boldsymbol{\theta}} = \mathcal{J}(\boldsymbol{\theta}), \qquad \frac{\partial^2 \langle \log p(\boldsymbol{X};\boldsymbol{\theta}(\boldsymbol{\eta}))\rangle_{\boldsymbol{\theta}(\boldsymbol{\eta})}}{\partial \boldsymbol{\eta}^2} = \frac{\partial \boldsymbol{\theta}(\boldsymbol{\eta})}{\partial \boldsymbol{\eta}} = \mathcal{J}(\boldsymbol{\eta})$$

*and hence, $\mathcal{J}(\boldsymbol{\eta}) = \mathcal{J}(\boldsymbol{\theta})^{-1}$.*

**Example 2.10** (Distribution on a finite alphabet)**.** The family of distributions on a finite set $\mathcal{X} = \{0, 1, \ldots, |\mathcal{X}| - 1\}$ can be regarded as the exponential family with a sufficient statistic $(t_z(x) = \mathbb{I}\{x = z\})_{z \in \mathcal{X}\setminus\{0\}}$. In this case, $\eta_z = p(z \mid \boldsymbol{\theta})$ for $z \in \mathcal{X} \setminus \{0\}$.

# 2.3 Gibbs free energy, mean-field approximation and variational bounds

## 2.3.1 Gibbs free energy

As shown in the previous section, for the exponential family, $\log Z_{\mathrm{E}}(\boldsymbol{\theta})$ is the Legendre transform of the negative Shannon entropy $\langle \log p(\boldsymbol{X};\boldsymbol{\theta}^*)\rangle$. Assume that alphabet $\mathcal{X}$ is finite. Then, from Lemma 2.5, it holds

$$\log Z(G) = \log Z_{\mathrm{E}}(\boldsymbol{\theta}) = \sup_{\boldsymbol{\eta}} \left\{ \sum_{k=1}^{d} \eta_k \theta_k - \langle \log p_{\mathrm{E}}(\boldsymbol{X};\boldsymbol{\theta}^*)\rangle_{\boldsymbol{\theta}^*} \right\}$$

where $d = |\mathcal{X}|^N$, $C(\boldsymbol{x}) = 0$, $t_{\boldsymbol{z}}(\boldsymbol{x}) = \mathbb{I}\{\boldsymbol{x} = \boldsymbol{z}\}$ and $\theta_{\boldsymbol{z}} = \log \prod_{a \in F} f_a(\boldsymbol{z}_{\partial a})$ for $\boldsymbol{z} \in \mathcal{X}^N$ in the exponential family. Here, $\boldsymbol{\theta}^*$ is an arbitrary value satisfying $\langle t_{\boldsymbol{z}}(\boldsymbol{X})\rangle_{\boldsymbol{\theta}^*} = p_{\mathrm{E}}(\boldsymbol{z};\boldsymbol{\theta}^*) = \eta_{\boldsymbol{z}}$ for $\boldsymbol{z} \in \mathcal{X}^N$. The above equation can be rewritten as

$$\log Z(G) = - \min_{q \in \mathcal{P}(\mathcal{X}^N)} \mathcal{F}_{\mathrm{Gibbs}}(q) \tag{2.3}$$

for $q \in \mathcal{P}(\mathcal{X}^N)$ where $\mathcal{F}_{\mathrm{Gibbs}}(q) := \mathcal{U}_{\mathrm{Gibbs}}(q) - \mathcal{H}_{\mathrm{Gibbs}}(q)$, and where

$$\mathcal{U}_{\mathrm{Gibbs}}(q) := - \sum_{\boldsymbol{x} \in \mathcal{X}^N} q(\boldsymbol{x}) \log \prod_{a \in F} f_a(\boldsymbol{x}_{\partial a}), \qquad \mathcal{H}_{\mathrm{Gibbs}}(q) := - \sum_{\boldsymbol{x} \in \mathcal{X}^N} q(\boldsymbol{x}) \log q(\boldsymbol{x}).$$

The quantities $\mathcal{U}_{\text{Gibbs}}(q)$, $\mathcal{H}_{\text{Gibbs}}(q)$ and $\mathcal{F}_{\text{Gibbs}}(q)$ are called the *Gibbs average energy*, the *Gibbs entropy* and the *Gibbs free energy*, respectively. The Gibbs free energy $\mathcal{F}_{\text{Gibbs}}(q)$ takes the minimum $-\log Z(G)$ at $q = p$. For $q = p$, the equation $\mathcal{F}_{\text{Gibbs}}(p) = \mathcal{U}_{\text{Gibbs}}(p) - \mathcal{H}_{\text{Gibbs}}(p)$ is equivalent to (1.4) for $\beta = 1$. In the variational method, we deal with $\min_{q \in \mathcal{P}(\mathcal{X}^N)} \mathcal{F}_{\text{Gibbs}}(q)$ instead of $-\log Z(G)$. In [Yedidia et al., 2005], this representation is explained as the minimization of the Kullback-Leibler divergence.

By restricting the domain of the minimization problem (2.3), an upper bound of $-\log Z(G)$ can be obtained. The most popular bound is called the *mean-field approximation*, which is

$$\min_{(q_i \in \mathcal{P}(\mathcal{X}))_{i \in V}} \mathcal{F}_{\text{Gibbs}}\left(\prod_{i \in V} q_i\right) \geq \min_{q \in \mathcal{P}(\mathcal{X}^N)} \mathcal{F}_{\text{Gibbs}}(q) = -\log Z(G).$$

### 2.3.2 Variational lower bound

In this section, the variational lower bound considered in [Wainwright et al., 2005] is introduced. Let $\Gamma$ be a sample space and $\rho$ be a probability measure on $\Gamma$. Let $(e(\boldsymbol{x};\theta) : \mathcal{X}^N \to \mathbb{R})_{\theta \in \Gamma}$ be a list of functions satisfying

$$\langle e(\boldsymbol{x};\Theta)\rangle_\rho = \log \prod_{a \in F} f_a(\boldsymbol{x}_{\partial a}).$$

Then, one obtains

$$\begin{aligned}
\mathcal{F}_{\text{Gibbs}}(q) &= -\sum_{\boldsymbol{x} \in \mathcal{X}^N} q(\boldsymbol{x}) \langle e(\boldsymbol{x};\Theta)\rangle_\rho + \sum_{\boldsymbol{x} \in \mathcal{X}^N} q(\boldsymbol{x}) \log q(\boldsymbol{x}) \\
&= \left\langle -\sum_{\boldsymbol{x} \in \mathcal{X}^N} q(\boldsymbol{x}) e(\boldsymbol{x};\Theta) + \sum_{\boldsymbol{x} \in \mathcal{X}^N} q(\boldsymbol{x}) \log q(\boldsymbol{x}) \right\rangle_\rho =: \langle \mathcal{F}_{\text{Gibbs}}(q;\theta)\rangle_\rho.
\end{aligned}$$

From this representation of the Gibbs free energy, a lower bound of the Helmholtz free energy $-\log Z(G)$ is obtained by exchanging the expectation and the minimization as

$$\min_{q \in \mathcal{P}(\mathcal{X}^N)} \mathcal{F}_{\text{Gibbs}}(q) = \min_{q \in \mathcal{P}(\mathcal{X}^N)} \langle \mathcal{F}_{\text{Gibbs}}(q;\Theta)\rangle_\rho \geq \left\langle \min_{q \in \mathcal{P}(\mathcal{X}^N)} \mathcal{F}_{\text{Gibbs}}(q;\Theta) \right\rangle_\rho. \tag{2.4}$$

If $\min_{q \in \mathcal{P}(\mathcal{X}^N)} \mathcal{F}_{\text{Gibbs}}(q;\theta)$ can be evaluated efficiently for each $\theta \in \Gamma$, the lower bound (2.4) can be evaluated practically. Now, we assume that $e(\boldsymbol{x};\theta) = \sum_{a \in F} e_a(\boldsymbol{x}_{\partial a};\theta)$. In [Wainwright et al., 2005], decomposition to spanning trees is suggested. For a given distribution over spanning trees, by optimizing the upper bound with respect to the energy $e(\boldsymbol{x};\boldsymbol{\theta})$ for each spanning tree, it is shown that one does not have to solve the maximization problems for each spanning tree and that the optimized upper bound is expressed by a solution of a single convex optimization problem. The obtained free energy is also generalized as the parameterized Bethe free energy (without theoretical motivation) [Wiegerinck and Heskes, 2003]. In [Globerson and Jaakkola, 2007], decomposition to planar factor graphs is suggested.

## 2.4 Cluster variation method

In this section, the cluster variation method (CVM) is introduced, which is closely related to the Bethe approximation. In contrast to the variational lower bound in the previous section, the entropic term is approximated in the CVM. The CVM is for the first time suggested by [Kikuchi, 1951], reformulated by [Morita, 1957] and further simplified by [An, 1988] by using the Möbius inversion formula, whose relationship with the CVM was mentioned by [Schlijper, 1983]. In CVM, the Shannon entropy is approximated by using the Möbius inversion formula.

**Definition 2.11** (Poset)**.** A poset is a set $P$ with a binary relation $\leq$ which is reflective, antisymmetric and transitive, i.e.,

- (reflective): $a \leq a$.
- (antisymmetric): if $a \leq b$ and $b \leq a$, then $a = b$.
- (transitive): if $b \leq a$ and $c \leq b$, then $c \leq a$.

A poset $P$ is said to be *locally finite* if an interval $[x, y] := \{z \in P \mid x \leq z, z \leq y\}$ is finite for any $x,\ y \in P$.

**Lemma 2.12** (Möbius inversion formula)**.** *Let $P$ be a locally finite poset. Let $\omega : (P \to \mathbb{R}) \to (P \to \mathbb{R})$ be defined as*

$$(\omega f)(x) = \sum_{y \leq x} f(y).$$

*Then, the inverse function $\omega^{-1}$ of $\omega$ is*

$$(\omega^{-1} g)(y) = \sum_{x \leq y} \mu_{x,y} g(x)$$

*where $(\mu_{x,y})_{x \leq y}$ is determined from*

$$\sum_{x \leq z \leq y} \mu_{x,z} = \sum_{x \leq z \leq y} \mu_{z,y} = \delta(x, y), \qquad \text{for } x \leq y$$

*where $\delta(x, y)$ denotes the function which takes 1 if $x = y$ and takes 0 otherwise for $x, y \in \mathcal{X}$.*

Let $S_{\mathcal{A}}(q) := -q_{\mathcal{A}}(\boldsymbol{x}_{\mathcal{A}}) \log q_{\mathcal{A}}(\boldsymbol{x}_{\mathcal{A}})$ for $\mathcal{A} \in 2^V \setminus \{\emptyset\}$ and $S_{\emptyset}(q) := 0$ where $q_{\mathcal{A}}$ be the marginal probability mass function for $\boldsymbol{x}_{\mathcal{A}}$, i.e., $q_{\mathcal{A}}(\boldsymbol{x}_{\mathcal{A}}) := \sum_{\boldsymbol{x}_{V\setminus\mathcal{A}}} q(\boldsymbol{x})$. Let $\mathcal{R} \subseteq 2^V \setminus \{V\}$ be a set of clusters. Then, $(\mathcal{R} \cup \{V\}, \subseteq)$ is a finite poset. Then, $S_V(q) = \mathcal{H}_{\text{Gibbs}}(q)$ and hence, $\mathcal{H}_{\text{Gibbs}}(q) = \sum_{\mathcal{B}\in\mathcal{R}\cup\{V\}} \tilde{S}_{\mathcal{B}}(q)$ from Lemma 2.12 where $\tilde{S}_{\mathcal{B}}(q) := \sum_{\mathcal{C}\subseteq\mathcal{B}} \mu_{\mathcal{C},\mathcal{B}} S_{\mathcal{C}}(q)$. The partial sum $\sum_{\mathcal{B}\in\mathcal{R}} \tilde{S}_{\mathcal{B}}(q)$ is regarded as an efficient approximation for the Gibbs entropy in the CVM. The approximation for the Gibbs entropy is

$$\sum_{\mathcal{B}\in\mathcal{R}} \tilde{S}_{\mathcal{B}}(q) = \sum_{\mathcal{B}\in\mathcal{R}} \sum_{\mathcal{C}\subseteq\mathcal{B}} \mu_{\mathcal{C},\mathcal{B}} S_{\mathcal{C}}(q) = \sum_{\mathcal{C}\in\mathcal{R}} S_{\mathcal{C}}(q) \sum_{\mathcal{C}\subseteq\mathcal{B}\in\mathcal{R}} \mu_{\mathcal{C},\mathcal{B}}.$$

Let $C_{\mathcal{C}} := \sum_{\mathcal{C}\subseteq\mathcal{B}\in\mathcal{R}} \mu_{\mathcal{C},\mathcal{B}}$. Then, for any $\mathcal{A} \in \mathcal{R}$, it holds

$$\sum_{\mathcal{A}\subseteq\mathcal{C}\in\mathcal{R}} C_{\mathcal{C}} = \sum_{\mathcal{A}\subseteq\mathcal{B}\in\mathcal{R}} \sum_{\mathcal{A}\subseteq\mathcal{C}\subseteq\mathcal{B}} \mu_{\mathcal{C},\mathcal{B}} = \sum_{\mathcal{A}\subseteq\mathcal{B}\in\mathcal{R}} \delta(\mathcal{A},\mathcal{B}) = 1$$

from Lemma 2.12.

**Definition 2.13** (The CVM / Kikuchi free energy)**.** The CVM (or Kikuchi) free energy is defined for a set of clusters $\mathcal{R}$ satisfying $\{\partial a \mid a \in F\} \subseteq \mathcal{R}$ and

$$\bigcap_{\mathcal{A}\in\mathcal{S}} \mathcal{A} \in \mathcal{R} \qquad \text{for any } \mathcal{S} \subseteq \mathcal{R} \tag{2.5}$$

as

$$\mathcal{F}_{\text{CVM}}((q_{\mathcal{A}})_{\mathcal{A}\in\mathcal{R}}) := \mathcal{U}_{\text{CVM}}((q_{\mathcal{A}})_{\mathcal{A}\in\mathcal{R}}) - \mathcal{H}_{\text{CVM}}((q_{\mathcal{A}})_{\mathcal{A}\in\mathcal{R}})$$

for $\left(q_{\mathcal{A}} \in \mathcal{P}(\mathcal{X}^{|\mathcal{A}|})\right)_{\mathcal{A}\in\mathcal{R}}$ satisfying

$$q_{\mathcal{C}}(\boldsymbol{x}_{\mathcal{C}}) = \sum_{\boldsymbol{x}_{\mathcal{B}\setminus\mathcal{C}}} q_{\mathcal{B}}(\boldsymbol{x}_{\mathcal{B}}) \qquad \text{for } \mathcal{C} \subseteq \mathcal{B}$$

where $\mathcal{U}_{\text{CVM}}((q_{\mathcal{A}})_{\mathcal{A}\in\mathcal{R}})$ and $\mathcal{H}_{\text{CVM}}((q_{\mathcal{A}})_{\mathcal{A}\in\mathcal{R}})$ are the *CVM average energy* and the *CVM entropy*, respectively, defined by

$$\mathcal{U}_{\text{CVM}}((q_{\mathcal{A}})_{\mathcal{A}\in\mathcal{R}}) := -\sum_{a\in F} \sum_{\boldsymbol{x}_{\partial a}\in\mathcal{X}^{d_a}} q_{\partial a}(\boldsymbol{x}_{\partial a}) \log f_a(\boldsymbol{x}_{\partial a})$$

$$\mathcal{H}_{\text{CVM}}((q_{\mathcal{A}})_{\mathcal{A}\in\mathcal{R}}) := \sum_{\mathcal{A}\in\mathcal{R}} C_{\mathcal{A}} S_{\mathcal{A}}(q_{\mathcal{A}})$$

and where $(C_{\mathcal{A}})_{\mathcal{A}\in\mathcal{R}}$ is defined by $\sum_{\mathcal{A}\subseteq\mathcal{C}\in\mathcal{R}} C_{\mathcal{C}} = 1$ for any $\mathcal{A} \in \mathcal{R}$.

Here, $\left(q_{\mathcal{A}} \in \mathcal{P}(\mathcal{X}^{|\mathcal{A}|})\right)_{\mathcal{A}\in\mathcal{R}}$ are called *pseudo-marginals*. Generally, the CVM entropy $\mathcal{H}_{\text{CVM}}((q_{\mathcal{A}})_{\mathcal{A}\in\mathcal{R}})$ is not concave, and hence the minimization of the CVM free energy is not easy. In [Pakzad and Anantharam, 2002], a sufficient condition for the concavity of the CVM entropy is shown without using the condition (2.5) as

$$\sum_{\substack{\mathcal{A}\in\mathcal{R} \\ \exists\mathcal{B}\in\mathcal{S},\mathcal{B}\subseteq\mathcal{A}}} C_{\mathcal{A}} \geq 0 \tag{2.6}$$

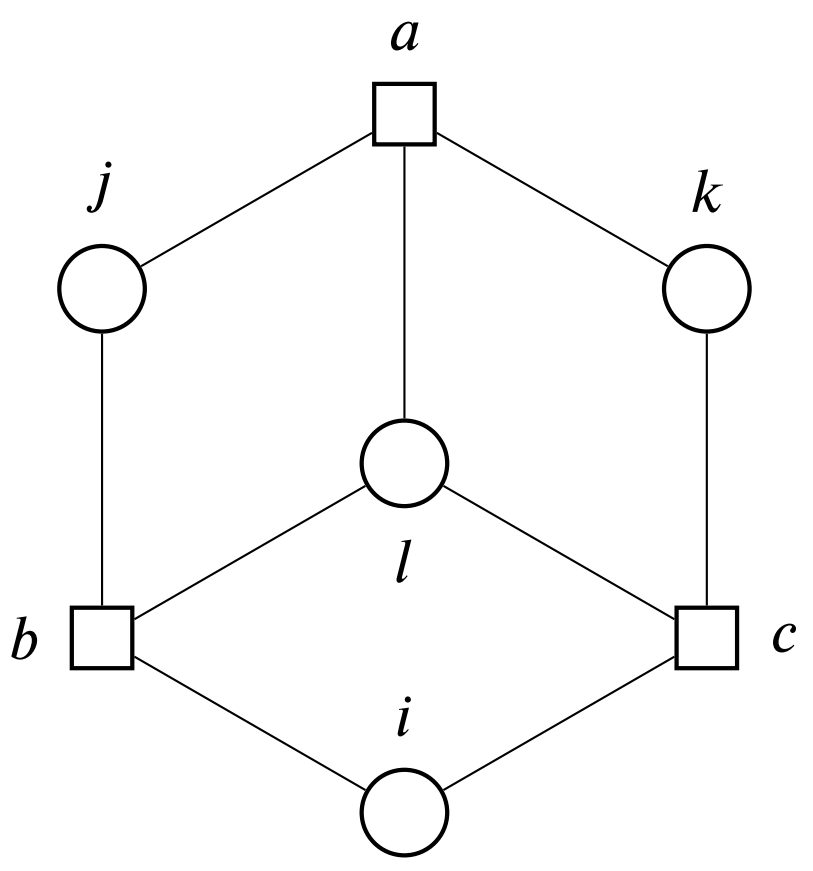

Figure 2.1: Example of factor graph

for any $\mathcal{S} \subseteq \mathcal{R}$. In [Pelizzola, 2005], examples for which the CVM is exact are shown. If one chooses $\mathcal{R}$ as the minimum set of clusters including $\{\partial a \mid a \in F\}$ and satisfying (2.5), the CVM is closely related with the Bethe approximation. For the factor graph in Figure 2.1, the maximal sets are $\{j,k,l\}$, $\{i,j,l\}$ and $\{i,k,l\}$. In this case,

$$\mathcal{R} = \{\{j,k,l\},\{i,j,l\},\{i,k,l\},\{j,l\},\{i,l\},\{k,l\},\{l\}\}$$

and the approximation of entropy is

$$\sum_{\mathcal{B}\in\mathcal{R}} \tilde{S}_{\mathcal{B}}(q) = S_{\{a,j,k,l\}}(q)+S_{\{b,i,j,l\}}(q)+S_{\{c,i,k,l\}}(q)-S_{\{j,l\}}(q)-S_{\{i,l\}}(q)-S_{\{k,l\}}(q)+S_{\{l\}}(q).$$

In the next section, the Bethe free energy is introduced on the basis of the CVM with the above choice of the maximal clusters.

## 2.5 Bethe approximation

In the Bethe approximation, the condition (2.5) of CVM is violated. The set of clusters is $\mathcal{R} = \{\partial a \mid a \in F\} \cup \{\{i\} \mid i \in V\}$ in the Bethe approximation. For the factor graph shown in Figure 2.1, the approximated entropy is

$$\sum_{\mathcal{B}\in\mathcal{R}} \tilde{S}_{\mathcal{B}}(q) = S_{\{a,j,k,l\}}(q)+S_{\{b,i,j,l\}}(q)+S_{\{c,i,k,l\}}(q)-S_{\{i\}}(q)-S_{\{j\}}(q)-S_{\{k\}}(q)-2S_{\{l\}}(q).$$

If any two factor nodes $a, b \in F$ do not connect to more than one common variable node, i.e., $|\partial a \cap \partial b| \leq 1$, the Bethe approximation is equivalent to the CVM in the last of the previous section. The general form of the Bethe free energy is defined in the following.

**Definition 2.14** (The Bethe free energy)**.** The Bethe free energy is defined as

$$\mathcal{F}_{\text{Bethe}}((b_i)_{i\in V},(b_a)_{a\in F}) = \mathcal{U}_{\text{Bethe}}((b_i)_{i\in V},(b_a)_{a\in F}) - \mathcal{H}_{\text{Bethe}}((b_i)_{i\in V},(b_a)_{a\in F})$$

for $\big((b_i \in \mathcal{P}(\mathcal{X}))_{i\in V},(b_a \in \mathcal{P}(\mathcal{X}^{d_a}))_{a\in F}\big)$ satisfying

$$b_i(z_i) = \sum_{\boldsymbol{x}_{\partial a}\in\mathcal{X}^{d_a},x_i=z_i} b_a(\boldsymbol{x}_{\partial a}), \qquad \text{for } (i,a)\in E. \tag{2.7}$$

where

$$\mathcal{U}_{\text{Bethe}}((b_i)_{i\in V},(b_a)_{a\in F}) := -\sum_{a\in F}\sum_{\boldsymbol{x}_{\partial a}\in\mathcal{X}^{d_a}} b_a(\boldsymbol{x}_{\partial a})\log f_a(\boldsymbol{x}_{\partial a})$$

$$\mathcal{H}_{\text{Bethe}}((b_i)_{i\in V},(b_a)_{a\in F}) := -\sum_{a\in F}\sum_{\boldsymbol{x}_{\partial a}\in\mathcal{X}^{d_a}} b_a(\boldsymbol{x}_{\partial a})\log b_a(\boldsymbol{x}_{\partial a}) + \sum_{i\in V}(d_i-1)\sum_{x_i\in\mathcal{X}} b_i(x_i)\log b_i(x_i).$$

From the sufficient condition (2.6) for concavity of the approximated entropy, the Bethe entropy is concave if the factor graph $G$ has at most one cycle. In fact, this is also a necessary condition if $f(\boldsymbol{x}_{\partial a}) > 0$ for all $\boldsymbol{x}_{\partial a} \in \mathcal{X}^{d_a}$ [Watanabe, 2010] where the Watanabe-Fukumizu formula, introduced in Appendix A, is used for the proof.

## 2.6 Belief propagation

It is shown that the fixed point of the message passing algorithm, belief propagation (BP), is a stationary point of the Bethe free energy in [Yedidia et al., 2005]. In the followings of the thesis, the factor graph model (1.1) is modified for the BP as follows

$$p(\boldsymbol{x};G) = \frac{1}{Z(G)}\prod_{a\in F} f_a(\boldsymbol{x}_{\partial a})\prod_{i\in V} f_i(x_i), \quad Z(G) = \sum_{\boldsymbol{x}\in\mathcal{X}^N}\prod_{a\in F} f_a(\boldsymbol{x}_{\partial a})\prod_{i\in V} f_i(x_i). \tag{2.8}$$

Here, a degree of any factor node indexed by $a \in F$ is greater than one, i.e., $|\partial a| \geq 2$. The set $E$ of edges only includes edges between $i \in V$ and $a \in F$. Then, the Lagrangian of the Bethe free energy is

$$\begin{aligned}
\mathcal{L}_{\text{Bethe}}((b_m)_{m\in F};(\lambda_m)_{m\in F},(\rho_{i,a})_{(i,a)\in E}) &= -\sum_{a\in F}\sum_{\boldsymbol{x}_{\partial a}\in\mathcal{X}^{d_a}} b_a(\boldsymbol{x}_{\partial a})\log f_a(\boldsymbol{x}_{\partial a}) \\
&- \sum_{i\in V}\sum_{x_i\in\mathcal{X}} b_i(x_i)\log f_i(x_i) + \sum_{a\in F}\sum_{\boldsymbol{x}_{\partial a}\in\mathcal{X}^{d_a}} b_a(\boldsymbol{x}_{\partial a})\log b_a(\boldsymbol{x}_{\partial a}) \\
&- \sum_{i\in V}(d_i-1)\sum_{x_i\in\mathcal{X}} b_i(x_i)\log b_i(x_i) + \sum_{i\in V}\lambda_i\left(\sum_{x\in\mathcal{X}} b_i(x) - 1\right) \\
&+ \sum_{a\in F}\lambda_a\left(\sum_{\boldsymbol{x}_{\partial a}\in\mathcal{X}^{d_a}} b_a(\boldsymbol{x}_{\partial a}) - 1\right) + \sum_{(i,a)\in E}\sum_{x_i\in\mathcal{X}}\rho_{i,a}(x_i)\left(\sum_{\boldsymbol{x}_{\partial a\setminus\{i\}}\in\mathcal{X}^{d_a-1}} b_a(\boldsymbol{x}_{\partial a}) - b_i(x_i)\right).
\end{aligned}$$

The partial derivatives of the Lagrangian are

$$\frac{\partial \mathcal{L}_{\mathrm{Bethe}}}{\partial b_i(x_i)} = -\log f_i(x_i) - (d_i - 1)(1 + \log b_i(x_i)) + \lambda_i - \sum_{a \in \partial i} \rho_{i,a}(x_i)$$

$$\frac{\partial \mathcal{L}_{\mathrm{Bethe}}}{\partial b_a(\boldsymbol{x}_{\partial a})} = -\log f_a(\boldsymbol{x}_{\partial a}) + (1 + \log b_a(\boldsymbol{x}_{\partial a})) + \lambda_a + \sum_{i \in \partial a} \rho_{i,a}(x_i)$$

for $i \in V$, $a \in F$. The conditions of stationary point are

$$\frac{\partial \mathcal{L}_{\mathrm{Bethe}}}{\partial b_i(x_i)} = -\log f_i(x_i) - (d_i - 1)(1 + \log b_i(x_i)) + \lambda_i - \sum_{a \in \partial i} \rho_{i,a}(x_i) = 0$$

$$\iff b_i(x_i) = \exp\left\{ -\frac{1}{d_i - 1}\left( \sum_{a \in \partial i} \rho_{i,a}(x_i) + \log f_i(x_i) - \lambda_i \right) - 1 \right\}$$

$$\frac{\partial \mathcal{L}_{\mathrm{Bethe}}}{\partial b_a(\boldsymbol{x}_{\partial a})} = -\log f_a(\boldsymbol{x}_{\partial a}) + (1 + \log b_a(\boldsymbol{x}_{\partial a})) + \lambda_a + \sum_{i \in \partial a} \rho_{i,a}(x_i) = 0$$

$$\iff b_a(\boldsymbol{x}_{\partial a}) = f_a(\boldsymbol{x}_{\partial a}) \exp\left\{ -\lambda_a - \sum_{i \in \partial a} \rho_{i,a}(x_i) - 1 \right\}.$$

Let $m_{i \to a}(x_i) := C_{i,a} f_i(x_i) \exp\{-\rho_{i,a}(x_i)\}$ satisfying the condition $\sum_{x_i \in \mathcal{X}} m_{i \to a}(x_i) = 1$ for all $(i,a) \in E$ and $m_{i \to a}(x_i) =: \frac{1}{Z_{i \to a}} f_i(x_i) \prod_{b \in \partial i \setminus \{a\}} m_{b \to i}(x_i)$ satisfying the condition $\sum_{x_i \in \mathcal{X}} m_{a \to i}(x_i) = 1$ for all $(i,a) \in E$. Then, it holds

$$b_i(x_i) = \exp\left\{ \frac{\lambda_i - \sum_{a \in \partial i} \log C_{i,a} Z_{i \to a}}{d_i - 1} - 1 \right\} \prod_{a \in \partial i} f_i(x_i) m_{a \to i}(x_i)$$

$$b_a(x_a) = \exp\{-\lambda_a - 1\} f_a(\boldsymbol{x}_{\partial a}) \prod_{i \in \partial a} \frac{m_{i \to a}(x_i)}{C_{i,a}}.$$

For satisfying the conditions $\sum_{x_i \in \mathcal{X}} b_i(x_i) = 1$ and $\sum_{\boldsymbol{x}_{\partial a} \in \mathcal{X}^{d_a}} b_a(\boldsymbol{x}_{\partial a}) = 1$, $(\lambda_i)_{i \in V}$ and $(\lambda_a)_{a \in F}$ are determined by $(m_{i \to a})_{(i,a) \in E}$ and $(m_{a \to i})_{(i,a) \in E}$ and hence

$$\begin{aligned} b_i(x_i) &= \frac{1}{Z_i((m_{a \to i})_{a \in \partial i})} f_i(x_i) \prod_{a \in \partial i} m_{a \to i}(x_i) \\ &= \frac{1}{Z_{i,a}(m_{a \to i}, m_{i \to a})} m_{a \to i}(x_i) m_{i \to a}(x_i), \qquad \text{for any } a \in \partial i \qquad (2.9) \\ b_a(x_a) &= \frac{1}{Z_a((m_{i \to a})_{i \in \partial a})} f_a(\boldsymbol{x}_{\partial a}) \prod_{i \in \partial a} m_{i \to a}(x_i). \end{aligned}$$

where

$$Z_a((m_{i \to a})_{i \in \partial a}) := \sum_{\boldsymbol{x} \in \mathcal{X}^{d_a}} f_a(\boldsymbol{x}) \prod_{i \in \partial a} m_{i \to a}(x_i), \quad Z_i((m_{a \to i})_{a \in \partial i}) := \sum_{x \in \mathcal{X}} f_i(x) \prod_{a \in \partial i} m_{a \to i}(x)$$

$$Z_{i,a}(m_{a \to i}, m_{i \to a}) := \sum_{x \in \mathcal{X}} m_{a \to i}(x) m_{i \to a}(x).$$

Here, $(m_{i\to a})_{(i,a)\in E}$ and $(m_{a\to i})_{(i,a)\in E}$ must satisfy

$$\begin{aligned} m_{i\to a}(x) &\propto f_i(x) \prod_{b\in\partial i\setminus\{a\}} m_{b\to i}(x) \\ m_{a\to i}(x) &\propto \sum_{\boldsymbol{x}\in\mathcal{X}^{d_a}, x_i = x} f_a(\boldsymbol{x}) \prod_{j\in\partial a\setminus\{i\}} m_{j\to a}(x_j) \end{aligned} \tag{2.10}$$

for their definition and the condition $\sum_{\boldsymbol{x}_{\partial a\setminus\{i\}}\in\mathcal{X}^{d_a-1}} b_a(\boldsymbol{x}_{\partial a}) = b_i(x_i)$. The equation (2.10) is the equation for fixed points of a well-known message passing algorithm, the belief propagation.

**Definition 2.15** (Belief propagation)**.** Choose initial messages $(m^{(0)}_{i\to a} \in \mathcal{P}(\mathcal{X}))_{(i,a)\in E}$ and $(m^{(0)}_{a\to i} \in \mathcal{P}(\mathcal{X}))_{(i,a)\in E}$ arbitrarily. For $t = 1, 2, \dots$, the messages are updated by the following rule

$$\begin{aligned} m^{(t)}_{i\to a}(x) &\propto f_i(x) \prod_{b\in\partial i\setminus\{a\}} m^{(t-1)}_{b\to i}(x) \\ m^{(t)}_{a\to i}(x) &\propto \sum_{\boldsymbol{x}\in\mathcal{X}^{d_a}, x_i = x} f_a(\boldsymbol{x}) \prod_{j\in\partial a\setminus\{i\}} m^{(t-1)}_{j\to a}(x_j). \end{aligned}$$

Belief propagation does not necessarily converge while sufficient conditions of convergence have been well investigated [Tatikonda and Jordan, 2002], [Mooij, 2008]. In [Heskes, 2002], it is shown that a locally stable fixed point of BP is a local minimum of the Bethe free energy. This relationship between local stability of BP and local convexity of the Bethe free energy is clarified in [Watanabe, 2010].

Let

$$\begin{aligned} \mathrm{Int}(\mathcal{F}_{\mathrm{Bethe}}) := \Big\{ \big((b_i)_{i\in V}, (b_a)_{a\in F}\big) \in \mathrm{Stat}(\mathcal{F}_{\mathrm{Bethe}}) \mid b_i(x_i) > 0, b_a(\boldsymbol{x}_{\partial a}) > 0, \\ \forall x_i \in \mathcal{X}, \forall \boldsymbol{x}_{\partial a} \in \mathrm{Supp}(f_a), \forall i \in V, \forall a \in F \Big\} \end{aligned}$$

where $\mathrm{Stat}(\mathcal{F}_{\mathrm{Bethe}})$ denotes the set of stationary points of the Bethe free energy and $\mathrm{Supp}(f_a) \subseteq \mathcal{X}^{d_a}$ denotes the support of $f_a$. From (2.9) and (2.10), it holds

$$\mathrm{Int}(\mathcal{F}_{\mathrm{Bethe}}) = \left\{ \big((b_i)_{i\in V}, (b_a)_{a\in F}\big) \in \mathrm{Stat}(\mathcal{F}_{\mathrm{Bethe}}) \mid b_i(x_i) > 0, \forall x_i \in \mathcal{X}, \forall i \in V \right\}.$$

Note that if $\mathrm{Supp}(f_a) = \mathcal{X}^{d_a}$ for all $a \in F$, it obviously holds $\mathrm{Stat}(\mathcal{F}_{\mathrm{Bethe}}) = \mathrm{Int}(\mathcal{F}_{\mathrm{Bethe}})$. By substituting (2.9) to Definition 2.14, one obtains the following alternative definition of the Bethe free energy for $((b_i)_{i\in V}, (b_a)_{a\in F}) \in \mathrm{Int}(\mathcal{F}_{\mathrm{Bethe}})$.

**Definition 2.16** (Alternative definition of the Bethe free energy)**.** For $((b_i)_{i\in V}, (b_a)_{a\in F}) \in \mathrm{Int}(\mathcal{F}_{\mathrm{Bethe}})$,

$$\mathcal{F}_{\mathrm{Bethe}}((m_{i\to a}, m_{a\to i})_{(i,a)\in E}) = -\sum_{i\in V} \log Z_i((m_{a\to i})_{a\in\partial i}) - \sum_{a\in F} \log Z_a((m_{i\to a})_{i\in\partial a}) + \sum_{(i,a)\in E} \log Z_{i,a}(m_{i\to a}, m_{a\to i}).$$

Note that (2.10) is also the stationary condition of $\mathcal{F}_{\mathrm{Bethe}}((m_{i\to a}, m_{a\to i})_{(i,a)\in E})$. Since the minimum of the Bethe free energy is an approximation of $-\log Z(G)$, the *Bethe partition function* $Z_{\mathrm{Bethe}}((m_{i\to a}, m_{a\to i})_{(i,a)\in E})$ is defined as

$$Z_{\mathrm{Bethe}}((m_{i\to a}, m_{a\to i})_{(i,a)\in E}) := \prod_{i\in V} Z_i((m_{a\to i})_{a\in\partial i}) \prod_{a\in F} Z_a((m_{i\to a})_{i\in\partial a}) \cdot \prod_{(i,a)\in E} \frac{1}{Z_{i,a}(m_{i\to a}, m_{a\to i})}. \tag{2.11}$$

We also use the notations $Z_{\mathrm{Bethe}}((b_i)_{i\in V}, (b_a)_{a\in F})$ for the Bethe partition function, and $Z_{\mathrm{Bethe}}$ and $Z_{\mathrm{Bethe}}(G)$ for the maximum of $Z_{\mathrm{Bethe}}((b_i)_{i\in V}, (b_a)_{a\in F})$ among all stationary points $((b_i)_{i\in V}, (b_a)_{a\in F})$ of the Bethe free energy. Then, one obtains the following lemma.

**Lemma 2.17** ([Wainwright et al., 2003])**.** *For* $((b_i)_{i\in V}, (b_a)_{a\in F}) \in \mathrm{Int}(\mathcal{F}_{\mathrm{Bethe}})$*, it holds*

$$\prod_{i\in V} f_i(x_i) \prod_{a\in F} f_a(\boldsymbol{x}_{\partial a}) = Z_{\mathrm{Bethe}}((b_i)_{i\in V}, (b_a)_{a\in F}) \prod_{i\in V} b_i(x_i) \prod_{a\in F} \frac{b_a(\boldsymbol{x}_{\partial a})}{\prod_{i\in\partial a} b_i(x_i)}$$

*and hence*

$$Z(G) = Z_{\mathrm{Bethe}}((b_i)_{i\in V}, (b_a)_{a\in F}) \sum_{\boldsymbol{x}\in\mathcal{X}^N} \prod_{i\in V} b_i(x_i) \prod_{a\in F} \frac{b_a(\boldsymbol{x}_{\partial a})}{\prod_{i\in\partial a} b_i(x_i)} \tag{2.12}$$

$$Z(G)p_{\mathcal{A}}(\boldsymbol{z}) = Z_{\mathrm{Bethe}}((b_i)_{i\in V}, (b_a)_{a\in F}) \sum_{\boldsymbol{x}\in\mathcal{X}^N, \boldsymbol{x}_{\mathcal{A}}=\boldsymbol{z}} \prod_{i\in V} b_i(x_i) \prod_{a\in F} \frac{b_a(\boldsymbol{x}_{\partial a})}{\prod_{i\in\partial a} b_i(x_i)} \tag{2.13}$$

*for any* $\mathcal{A} \subseteq V$*,* $\boldsymbol{z} \in \mathcal{X}^{|\mathcal{A}|}$*.*

*Proof.* It is sufficient to prove the first equality. The first equality is obtained by

$$\prod_{i\in V} b_i(x_i) \prod_{a\in F} \frac{b_a(\boldsymbol{x}_{\partial a})}{\prod_{i\in\partial a} b_i(x_i)} = \prod_{i\in V} \frac{f_i(x_i)\prod_{a\in\partial i} m_{a\to i}(x_i)}{Z_i} \prod_{a\in F} \frac{\frac{1}{Z_a} f_a(\boldsymbol{x}_{\partial a}) \prod_{i\in\partial a} m_{i\to a}(x_i)}{\prod_{i\in\partial a} \frac{m_{i\to a}(x_i) m_{a\to i}(x_i)}{Z_{i,a}}}$$

$$= \prod_{i\in V} \frac{f_i(x_i)\prod_{a\in\partial i} m_{a\to i}(x_i)}{Z_i} \prod_{a\in F} \frac{\frac{1}{Z_a} f_a(\boldsymbol{x}_{\partial a})}{\prod_{i\in\partial a} \frac{m_{a\to i}(x_i)}{Z_{i,a}}}$$

$$= \prod_{i\in V} \frac{f_i(x_i)}{Z_i} \prod_{a\in F} \frac{f_a(\boldsymbol{x}_{\partial a}) \prod_{i\in\partial a} Z_{i,a}}{Z_a} = \frac{\prod_{i\in V} f_i(x_i) \prod_{a\in F} f_a(\boldsymbol{x}_{\partial a})}{Z_{\mathrm{Bethe}}}.$$

□

**Lemma 2.18** (The Bethe approximation is exact for a tree factor graph)**.** *For a tree factor graph, if* $\left|\mathrm{Int}(\mathcal{F}_{\mathrm{Bethe}})\right| \geq 1$*, it holds* $\left|\mathrm{Int}(\mathcal{F}_{\mathrm{Bethe}})\right| = 1$*. On the unique stationary point in* $\mathrm{Int}(\mathcal{F}_{\mathrm{Bethe}})$*, the Bethe approximation is exact, i.e.,* $Z(G) = Z_{\mathrm{Bethe}}(G)$*.*

*Proof.* First, it is proved that the Bethe approximation is exact for an arbitrary stationary point of the Bethe free energy. For any pseudo-marginals $((b_i)_{i\in V}, (b_a)_{a\in F})$, it holds

$$\begin{aligned}\sum_{\boldsymbol{x}\in\mathcal{X}^N}\prod_{i\in V} b_i(x_i)\prod_{a\in F}\frac{b_a(\boldsymbol{x}_{\partial a})}{\prod_{i\in\partial a} b_i(x_i)} &= \sum_{\boldsymbol{x}\in\mathcal{X}^N}\prod_{i\in V} b_i(x_i)\prod_{a\in F}\left[1+\frac{b_a(\boldsymbol{x}_{\partial a})-\prod_{i\in\partial a} b_i(x_i)}{\prod_{i\in\partial a} b_i(x_i)}\right]\\ &= \sum_{\boldsymbol{x}\in\mathcal{X}^N}\prod_{i\in V} b_i(x_i)\sum_{F'\subseteq F}\prod_{a\in F'}\frac{b_a(\boldsymbol{x}_{\partial a})-\prod_{i\in\partial a} b_i(x_i)}{\prod_{i\in\partial a} b_i(x_i)}\\ &= \sum_{F'\subseteq F}\sum_{\boldsymbol{x}\in\mathcal{X}^N}\prod_{i\in V} b_i(x_i)\prod_{a\in F'}\frac{b_a(\boldsymbol{x}_{\partial a})-\prod_{i\in\partial a} b_i(x_i)}{\prod_{i\in\partial a} b_i(x_i)} =: \sum_{F'\subseteq F}\mathcal{Z}(F'). \end{aligned} \tag{2.14}$$

Let $d_i(F') := |\partial i \cap F'|$ for $F' \subseteq F$. For any $F' \subseteq F$, if there exists $a \in F'$ such that all but at most one $i \in \partial a$ satisfy $d_i(F') = 1$, then $\mathcal{Z}(F') = 0$ when the pseudo-marginals satisfy (2.7). Hence, for a tree factor graph, $\mathcal{Z}(F') = 0$ unless $F' \neq \emptyset$. Then, one obtains

$$\sum_{\boldsymbol{x}\in\mathcal{X}^N}\prod_{i\in V} b_i(x_i)\prod_{a\in F}\frac{b_a(\boldsymbol{x}_{\partial a})}{\prod_{i\in\partial a} b_i(x_i)} = \mathcal{Z}(\emptyset) = 1.$$

From (2.12), it holds $Z(G) = Z_{\mathrm{Bethe}}((b_i)_{i\in V}, (b_a)_{a\in F})$. Similarly, from (2.13), all pseudo-marginals $((b_i)_{i\in V}, (b_a)_{a\in F})$ must be the true marginal. Hence, $\left|\mathrm{Int}(\mathcal{F}_{\mathrm{Bethe}})\right| = 1$. □

**Example 2.19** (Ising model)**.** By letting

$$h_{a\to i} := \frac{1}{2}\log\frac{m_{a\to i}(+1)}{m_{a\to i}(-1)}, \qquad h_{i\to a} := \frac{1}{2}\log\frac{m_{i\to a}(+1)}{m_{i\to a}(-1)}$$

for $(i,a) \in E$, the Bethe free energy in Definition 2.16 can be written as

$$\begin{aligned}\mathcal{F}_{\mathrm{Bethe}}((h_{i\to a}, h_{a\to i})_{(i,a)\in E}) = &-\sum_{i\in V}\log\left(\mathrm{e}^{h_i}\prod_{a\in\partial i}\frac{1+\tanh(h_{a\to i})}{2}+\mathrm{e}^{-h_i}\prod_{a\in\partial i}\frac{1-\tanh(h_{a\to i})}{2}\right)\\ &-\sum_{a\in F}\log\left(\cosh(J_{i,j})+\sinh(J_{i,j})\prod_{i\in a}\tanh(h_{i\to a})\right)\\ &+\sum_{(i,a)\in E}\log\left(\frac{1+\tanh(h_{a\to i})\tanh(h_{i\to a})}{2}\right).\end{aligned}$$

The stationary condition (2.10) is

$$h_{i\to(i,j)} = h_i + \sum_{k\neq j} h_{(i,k)\to i}, \qquad h_{(i,j)\to i} = \tanh^{-1}\left(\tanh\left(J_{i,j}\right)\tanh(h_{j\to a})\right).$$

## 2.7 Thouless-Anderson-Palmer approximation

If a factor graph is dense, belief propagation can be further approximated by using asymptotic analysis. The Sherrington-Kirkpatrick (SK) model is known as a good example, which is the fully connected pairwise Ising model defined by

$$p_{\mathrm{SK}}(\boldsymbol{x}) \propto \exp\left\{ \frac{1}{\sqrt{N}} \sum_{i<j} J_{i,j} x_i x_j + \sum_i h_i x_i \right\}$$

where $(J_{i,j})_{i<j}$ are i.i.d. random variables obeying the normal distribution with the mean zero and the variance $J_0$. Let

$$m_i = b_i(+1) - b_i(-1) = \tanh\left( h_i + \sum_{j \neq i} h_{(i,j)\to i} \right) = \tanh\left( h_{i\to(i,j)} + h_{(i,j)\to i} \right).$$

From

$$\begin{aligned} h_{(i,j)\to i} &= \tanh^{-1}\left( \tanh\left( \frac{J_{i,j}}{\sqrt{N}} \right) \tanh(h_{j\to(i,j)}) \right) \\ &= \frac{J_{i,j}}{\sqrt{N}} \tanh(h_{j\to(i,j)}) + O\left( \frac{1}{N^{\frac{3}{2}}} \right) \end{aligned}$$

it holds $h_{(i,j)\to i} = O\big(1/\sqrt{N}\big)$. Furthermore, one obtains

$$\begin{aligned} h_{(i,j)\to i} &= \frac{J_{i,j}}{\sqrt{N}} \tanh\left( (h_{j\to(i,j)} + h_{(i,j)\to j}) - h_{(i,j)\to j} \right) + O\left( \frac{1}{N^{\frac{3}{2}}} \right) \\ &= \frac{J_{i,j}}{\sqrt{N}} \left( m_j - \tanh'\left( h_{j\to(i,j)} + h_{(i,j)\to j} \right) h_{(i,j)\to j} + O\left( \frac{1}{N} \right) \right) + O\left( \frac{1}{N^{\frac{3}{2}}} \right) \\ &= \frac{J_{i,j}}{\sqrt{N}} \left( m_j - (1 - m_j^2) h_{(i,j)\to j} \right) + O\left( \frac{1}{N^{\frac{3}{2}}} \right). \end{aligned}$$

By using the same asymptotic equality again, one obtains

$$h_{(i,j)\to i} = \frac{J_{i,j}}{\sqrt{N}} \left( m_j - (1 - m_j^2) \frac{J_{i,j}}{\sqrt{N}} m_i \right) + O\left( \frac{1}{N^{\frac{3}{2}}} \right).$$

Hence,

$$\begin{aligned} m_i &= \tanh\left( h_i + \sum_{j\neq i} h_{(i,j)\to i} \right) \\ &= \tanh\left( h_i + \sum_{j\neq i} \frac{J_{i,j}}{\sqrt{N}} m_j - \sum_{j\neq i} \left( \frac{J_{i,j}}{\sqrt{N}} \right)^2 (1 - m_j^2) m_i + O\left( \frac{1}{N^{\frac{1}{2}}} \right) \right) \\ &= \tanh\left( h_i + \sum_{j\neq i} \frac{J_{i,j}}{\sqrt{N}} m_j - J_0 \frac{1}{N} \sum_{j\neq i} (1 - m_j^2) m_i + O\left( \frac{1}{N^{\frac{1}{2}}} \right) \right). \end{aligned}$$

In the last equality, the law of large numbers is used. This equation is called the Thouless-Anderson-Palmer (TAP) equation. The TAP equation can also be obtained by the cavity method [Opper and Winther, 2001] and Plefka expansion [Plefka, 1982].

## 2.8 Applications

### 2.8.1 Low-density parity-check codes

The most popular and practical application of the Bethe approximation is low-density parity-check (LDPC) codes [Richardson and Urbanke, 2008]. A binary LDPC code of rate $1-(M/N)$ is a binary linear code defined by an $M \times N$ binary parity-check matrix $H$. The set of codewords of an LDPC code is defined as the kernel of $H$, i.e., $\left\{\boldsymbol{x} \in \{0,1\}^N \mid H\boldsymbol{x} = 0\right\}$ where the operations are taken on the binary field. Each of the $M$ linear constraints corresponds to a factor node. Let the set of factor nodes be $F = \{a_1, a_2, \dots, a_M\}$. A factor node $a_k$ is connected to variable nodes corresponding to non-zero entries of the $k$-th row of $H$ for $k \in \{1, 2, \dots, M\}$. For binary LDPC codes, the a posteriori distribution given $y \in \mathcal{Y}^N$ over memoryless symmetric channel $P_{Y|X}$ is

$$\begin{aligned} p_{\mathrm{LDPC}}(\boldsymbol{x}; \boldsymbol{y}) &:\propto \mathbb{I}\{H\boldsymbol{x} = 0\} \prod_{i=1}^{N} P_{Y|X}(y_i \mid x_i) \\ &\propto \prod_{a\in F} \mathbb{I}\left\{\sum_{i\in\partial a} x_i = 0\right\} \prod_{i=1}^{N} \mathrm{e}^{h_i(1-2x_i)} \end{aligned}$$

for $\boldsymbol{x} \in \{0,1\}^N$ where $h_i = \frac{1}{2}\log\big(P_{Y|X}(y \mid 0)/P_{Y|X}(y \mid 1)\big)$ [Murayama et al., 2000], [Vicente et al., 2003]. Since the alphabet is binary, each message of BP can be expressed by a single parameter. Let

$$h_{a\to i} := \frac{1}{2}\log\frac{m_{a\to i}(0)}{m_{a\to i}(1)}, \qquad h_{i\to a} := \frac{1}{2}\log\frac{m_{i\to a}(0)}{m_{i\to a}(1)}$$

for $(i,a) \in E$. The Bethe free energy is expressed as

$$\begin{aligned} \mathcal{F}_{\mathrm{Bethe}}((h_{i\to a}, h_{a\to i})_{(i,a)\in E}) = &-\sum_{i\in V} \log\left(\mathrm{e}^{h_i} \prod_{a\in\partial i} \frac{1+\tanh(h_{a\to i})}{2} + \mathrm{e}^{-h_i} \prod_{a\in\partial i} \frac{1-\tanh(h_{a\to i})}{2}\right) \\ &- \sum_{a\in F} \log\left(\frac{1+\prod_{i\in a}\tanh(h_{i\to a})}{2}\right) + \sum_{(i,a)\in E} \log\left(\frac{1+\tanh(h_{a\to i})\tanh(h_{i\to a})}{2}\right). \end{aligned}$$

Then, the stationary condition (2.10) can be written as

$$h_{i\to a} = h_i + \sum_{b\in\partial i\setminus\{a\}} h_{b\to i}, \qquad h_{a\to i} = \tanh^{-1}\left(\prod_{j\in\partial a\setminus\{i\}} \tanh(h_{j\to a})\right).$$

For LDPC codes on memoryless symmetric channel, it is believed and partially proved that the Bethe approximation is asymptotically exact [Montanari, 2001], [Méasson et al., 2009], [Montanari, 2005], [Macris, 2007], [Kudekar, 2009].

### 2.8.2 Code division multiple access channel and compressed sensing

In [Kabashima, 2003], the TAP equation is derived for code division multiple access (CDMA) channel on binary phase-shift-keying (BPSK) modulation, whose probabilistic model is

$$p_{\mathrm{CDMA}}(\boldsymbol{x} \mid \boldsymbol{y}) := \frac{1}{Z_{\mathrm{CDMA}}} \int \delta\{\boldsymbol{z} - S\boldsymbol{x}\} \prod_{a=1}^{K} P_{Y|Z}(y_a \mid z_a)\, \mathrm{d}\boldsymbol{z}$$

where $\boldsymbol{x} \in \{+1, -1\}^N$ and $P_{Y|Z}(y_a \mid z_a)$ denotes the transition probability of a channel. Here, $S$ is a $K \times N$ matrix on $\mathbb{R}$ corresponding to spreading codes. Similarly, the probabilistic model of compressed sensing [Donoho et al., 2010] is

$$p_{\mathrm{CS}}(\boldsymbol{x} \mid \boldsymbol{y}; \beta) := \frac{1}{Z_{\mathrm{CS}}(\beta)} \int \delta\{\boldsymbol{z} - S\boldsymbol{x}\} \prod_{i=1}^{N} \mathrm{e}^{-\beta |x_i|} \prod_{a=1}^{K} P_{Y|Z}(y_a \mid z_a)\, \mathrm{d}\boldsymbol{z}$$

where $S$ is a $K \times N$ matrix on $\mathbb{R}$ corresponding to a sensing matrix. Here, one has to take the limit $\beta \to \infty$. The difference between the two models lies in the prior distributions of the original signal $\boldsymbol{x}$. In the probabilistic model of CDMA with BPSK modulation, the prior distribution is the uniform distribution on $\{+1, -1\}^N$. In the probabilistic model of compressed sensing, the prior distribution is the Laplace distribution $\mathrm{e}^{-\beta \|\boldsymbol{x}\|_1}$. In both the models, the channel is usually assumed to be the additive white Gaussian noise channel.

## 2.9 Additional historical remarks on the Bethe approximation

The Bethe free energy is suggested in [Bethe, 1935] and [Peierls, 1936]. The message passing algorithm, belief propagation, is proposed in the area of artificial intelligence [Pearl, 1988]. In [Yedidia et al., 2005], it is revealed that the belief propagation algorithm can be regarded as a simple algorithm which tries to find the minimum of the Bethe free energy. Other algorithms have been considered for minimization of the Bethe free energy, e.g., concave-convex procedure (CCCP) [Yuille, 2002], unified propagation and scaling (UPS) [Teh and Welling, 2002], etc. Remarks about convexity of the Bethe free energy and local stability of belief propagation are mentioned in Appendix A.

# 3 Characterization of the Bethe Free Energy by Loop Calculus

**In this chapter, the characterization of the Bethe free energy using loop calculus is introduced, which is obtained by Chertkov and Chernyak. The expression of loop calculus formula for non-binary finite alphabets is shown by using tangent vectors of information manifolds for exponential family.**

## 3.1 Linear transform

The basic and general idea of this chapter is the use of linear transform. Let $\mathcal{V}$ be a linear space and $\boldsymbol{x} \in \mathcal{V}$ be a vector. By using an invertible linear transform $A : \mathcal{V} \to \mathcal{W}$ for some linear space $\mathcal{W}$, one obtains the trivial equation

$$\boldsymbol{x} = A^{-1} A \boldsymbol{x} \tag{3.1}$$

where $A^{-1}$ is the inverse transform of $A$. This idea can yield a non-trivial equality. The most popular one is the method of generating function, i.e.,

$$(Ap)(z) = \sum_{n=1}^{\infty} p(n) z^n$$
$$(A^{-1}g)(n) = \frac{1}{2\pi i} \oint \frac{g(z)}{z^{n+1}} \mathrm{d}z.$$

The method of generating function is useful for many problems in number theory and combinatorial theory [Flajolet and Sedgewick, 2009]. The Riemann prime number formula is also obtained by (3.1), which shows that the number of prime numbers can be expressed as a sum with respect to zeros of the Riemann zeta function.

The equation (3.1) gives a non-trivial equality even when $A$ is a linear transform between finite-dimensional linear spaces. Now, we consider (3.1) for the partition function of factor graph defined in (2.8). Assume that $f_a$ for $a \in F$ has the following expression

$$f_a(\boldsymbol{x}_{\partial a}) = \sum_{\boldsymbol{y}_{\partial a} \in \mathcal{Y}^{d_a}} \hat{f}_a(\boldsymbol{y}_{\partial a}) \prod_{i \in \partial a} \phi_{i,a}(x_i, y_i). \tag{3.2}$$

When $\mathcal{Y} = \mathcal{X}$, this representation is obtained by letting

$$\hat{f}_a(\boldsymbol{y}_{\partial a}) = \sum_{\boldsymbol{x}_{\partial a} \in \mathcal{X}^{d_a}} f_a(\boldsymbol{x}_{\partial a}) \prod_{i \in \partial a} \hat{\phi}_{i,a}(y_i, x_i)$$

for $\{(\phi_{i,a}, \hat{\phi}_{i,a})\}_{i \in \partial a}$ satisfying

$$\sum_{y \in \mathcal{X}} \phi_{i,a}(x, y) \hat{\phi}_{i,a}(y, z) = \delta(x, z)$$

or equivalently

$$\sum_{x \in \mathcal{X}} \hat{\phi}_{i,a}(y, x) \phi_{i,a}(x, w) = \delta(y, w). \tag{3.3}$$

Then, the partition function can be rewritten by using the transform (3.2) as

$$\begin{aligned} Z(G) &= \sum_{\boldsymbol{x} \in \mathcal{X}^N} \prod_{a \in F} \left( \sum_{\boldsymbol{y}_{\partial a} \in \mathcal{Y}^{d_a}} \hat{f}_a(\boldsymbol{y}_{\partial a}) \prod_{i \in \partial a} \phi_{i,a}(x_i, y_i) \right) \prod_{i \in V} f_i(x_i) \\ &= \sum_{\boldsymbol{y} \in \mathcal{Y}^{|E|}} \prod_{a \in F} \hat{f}_a(\boldsymbol{y}_{\partial a, a}) \prod_{i \in V} \left( \sum_{x \in \mathcal{X}} f_i(x) \prod_{a \in \partial i} \phi_{i,a}(x, y_{i,a}) \right) \\ &=: \sum_{\boldsymbol{y} \in \mathcal{Y}^{|E|}} \prod_{a \in F} \hat{f}_a(\boldsymbol{y}_{\partial a, a}) \prod_{i \in V} \hat{f}_i(\boldsymbol{y}_{i, \partial i}). \end{aligned} \tag{3.4}$$

Although $\hat{f}_a(\boldsymbol{y}_{\partial a,a})$ and $\hat{f}_i(\boldsymbol{y}_{i,\partial i})$ are not necessarily non-negative, the representation (3.4) can be regarded as a partition function of another factor graph if one ignores the sign of the functions. In the new factor graph associated with the representation (3.4), the variables $\boldsymbol{y} \in \mathcal{Y}^{|E|}$ are associated with edges of the original factor graph. Both the variable nodes and factor nodes in the original graph can be regarded as factor nodes in the new representation (3.4). When $Z(G)$ is the number of solutions of CSP, (3.4) is called the *Holant theorem* and used for polynomial-time algorithm called the *holographic algorithm* in theoretical computer science [Valiant, 2008]. In [Al-Bashabsheh and Mao, 2011], (3.4) for normal factor graphs is called the *generalized Holant theorem*. In mathematics, this kind of formula is generally understood as the *trace formula*, which, roughly speaking, states

$$\mathrm{tr}(X) = \mathrm{tr}(AXA^{-1})$$

for linear operators $X : \mathcal{V} \to \mathcal{V}$ and $A : \mathcal{V} \to \mathcal{W}$ on Hilbert spaces $\mathcal{V}, \mathcal{W}$.

## 3.2 Linear constraints and the MacWilliams identity

The most popular application of the transform of the partition function (3.4) is the MacWilliams identity. Assume that an alphabet $\mathcal{X}$ is a prime ring $\mathbb{Z}/p\mathbb{Z}$ for a prime number $p$. The Fourier transform is suitable for linear function, i.e., $f_a(\boldsymbol{x}_{\partial a})$ depends on $\boldsymbol{x}_{\partial a}$ only through some linear combination of $x_i$ for $i \in \partial a$.

**Lemma 3.1** (Poisson summation formula). *Let $\omega_p \in \mathbb{C}$ be a p-th root of unity for a prime number p. Then, it holds*

$$\mathbb{I}\{x = 0\} = \frac{1}{p} \sum_{w \in \mathbb{Z}/p\mathbb{Z}} \omega_p^{w \cdot x}$$

*Proof.* When $x = 0$ the equation is trivial. When $x \neq 0$, the statement is obtained from $\{w \cdot x \mid w \in \mathbb{Z}/p\mathbb{Z}\} = \mathbb{Z}/p\mathbb{Z}$ and $\sum_{w \in \mathbb{Z}/p\mathbb{Z}} \omega_p^w = 0$. □

The following lemma is easy to confirm.

**Lemma 3.2** (Fourier transform and inverse Fourier transform). *For $f(x) : \mathcal{X} \to \mathbb{C}$, the Fourier transform $\bar{f}$ of $f$ is defined by*

$$\bar{f}(y) := \frac{1}{p} \sum_{x \in \mathcal{X}} \omega_p^{y \cdot x} f(x).$$

*Then, the original function $f$ can be obtained from $\bar{f}$ by the inverse Fourier transform*

$$f(x) := \sum_{y \in \mathcal{X}} \omega_p^{-x \cdot y} \bar{f}(y).$$

Then, MacWilliams identity can be obtained as follows.

**Lemma 3.3** (Generalized MacWilliams identity).

$$\begin{aligned}
&\sum_{\boldsymbol{v} \in (\mathbb{Z}/p\mathbb{Z})^{|F|}} \sum_{\boldsymbol{x} \in (\mathbb{Z}/p\mathbb{Z})^N} \mathbb{I}\{H\boldsymbol{x} = \boldsymbol{v}\} \prod_{i \in V} \lambda_i(x_i) \prod_{a \in F} \rho_a(v_a) \\
&= p^N \sum_{\boldsymbol{z} \in (\mathbb{Z}/p\mathbb{Z})^N} \sum_{\boldsymbol{y} \in (\mathbb{Z}/p\mathbb{Z})^{|F|}} \mathbb{I}\left\{H^t \boldsymbol{y} = \boldsymbol{z}\right\} \prod_{i \in V} \bar{\lambda}_i(-z_i) \prod_{a \in F} \bar{\rho}_a(y_a)
\end{aligned}$$

*where $\bar{\lambda}_i(z)$ and $\bar{\rho}_a(y)$ are the Fourier transforms of $\lambda_i(x)$ and $\rho_a(v)$, respectively, for $i \in V$ and $a \in F$.*

*Proof.* Since

$$\mathbb{I}\left\{\sum_{i \in \partial a} H_{a,i} \cdot x_i = v_a\right\} = \frac{1}{p} \sum_{w \in \mathcal{X}} \omega_p^{w \cdot (\sum_{i \in \partial a} H_{a,i} \cdot x_i - v_a)}$$

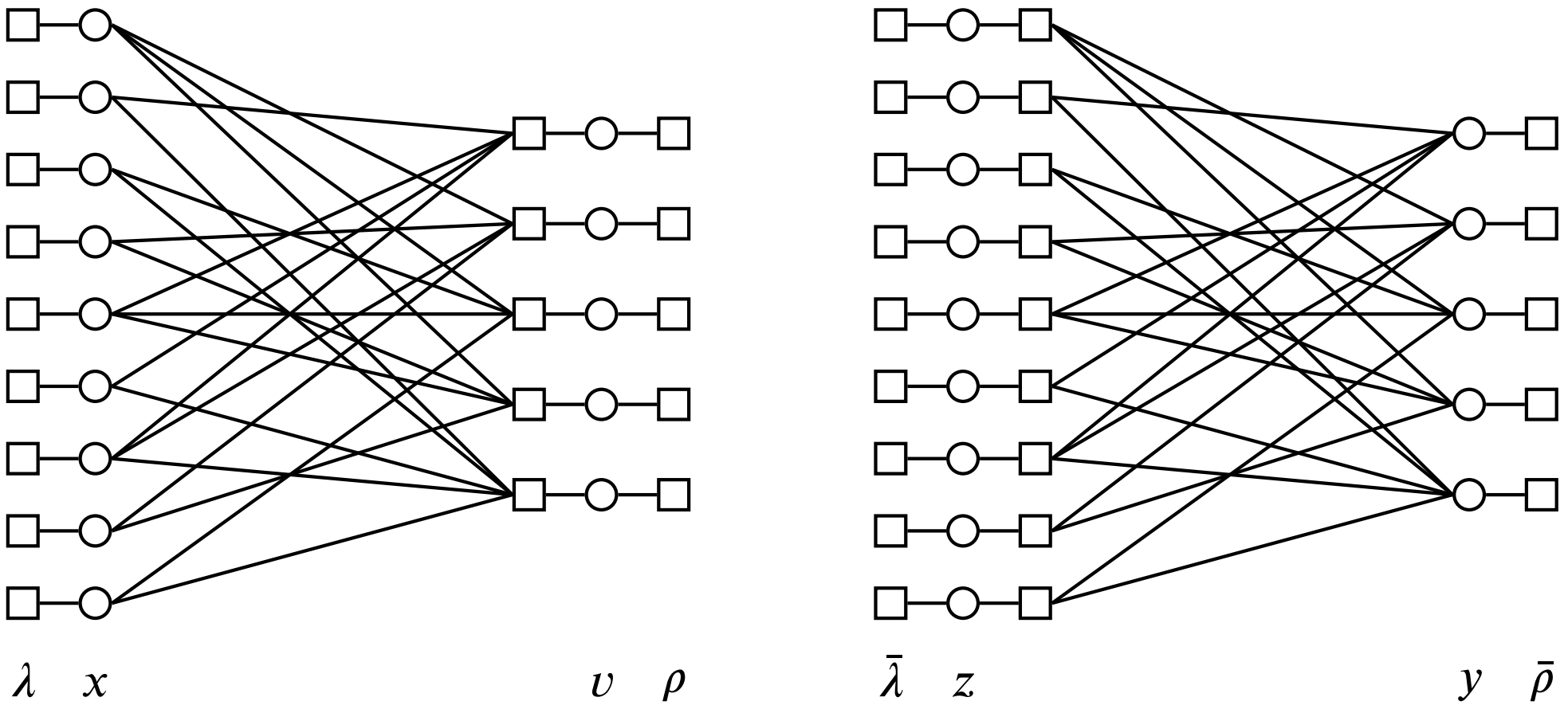

Figure 3.1: Left: Factor graph corresponding to the general linear model. Right: Its equivalent dual factor graph obtained by the MacWilliams identity.

the result is obtained from Lemmas 3.1 and 3.2 as

$$
\begin{aligned}
Z(G) = &\sum_{\boldsymbol{v}\in\mathcal{X}^{|F|}} \sum_{\boldsymbol{x}\in\mathcal{X}^N} \prod_{a\in F} \left( \frac{1}{p} \sum_{w\in\mathcal{X}} \omega_p^{w\cdot(\sum_{i\in\partial a} H_{a,i}\cdot x_i - v_a)} \right) \\
&\cdot \prod_{i\in V} \left( \sum_{z\in\mathcal{X}} \bar{\lambda}_i(z)\omega_p^{-x_i\cdot z} \right) \prod_{a\in F} \left( \sum_{y\in\mathcal{X}} \bar{\rho}_a(y)\omega_p^{-v_a\cdot y} \right) \\
= &\frac{1}{p^{|F|}} \sum_{\boldsymbol{z}\in\mathcal{X}^N} \sum_{\boldsymbol{y}\in\mathcal{X}^{|F|}} \sum_{\boldsymbol{w}\in\mathcal{X}^{|F|}} \prod_{i\in V} \bar{\lambda}_i(z_i) \prod_{a\in F} \bar{\rho}_a(y_a) \\
&\cdot \prod_{i\in V} \left( \sum_{x\in\mathcal{X}} \omega_p^{x\cdot(\sum_{a\in\partial i} H_{a,i}\cdot w_a - z_i)} \right) \prod_{a\in F} \left( \sum_{v\in\mathcal{X}} \omega_p^{-v\cdot(y_a + w_a)} \right) \\
= &\frac{1}{p^{|F|}} \sum_{\boldsymbol{z}\in\mathcal{X}^N} \sum_{\boldsymbol{y}\in\mathcal{X}^{|F|}} \sum_{\boldsymbol{w}\in\mathcal{X}^{|F|}} \prod_{i\in V} \bar{\lambda}_i(z_i) \prod_{a\in F} \bar{\rho}_a(y_a) \\
&\cdot \prod_{i\in V} \left[ p\,\mathbb{I}\left\{ \sum_{a\in\partial i} H_{a,i}\cdot w_a = z_i \right\} \right] \prod_{a\in F} \left[ p\,\mathbb{I}\{y_a + w_a = 0\} \right]. \qquad \square
\end{aligned}
$$

The dual factor graph obtained by the Fourier transform is described in the right of Fig. 3.1. While the MacWilliams identity was obtained in the context of coding theory [MacWilliams and Sloane, 1977], it is also used in statistical physics as the high temperature expansion and the duality transformation.

## 3.3 Loop calculus for the binary alphabet

Let $\mathcal{Y} = \mathcal{X}$. We now consider the following additional conditions on $((\phi_{i,a}, \hat{\phi}_{i,a}))_{(i,a)\in E}$. For each $i \in V$ and $a \in F$, the additional conditions are

$$\begin{aligned}\sum_{x\in\mathcal{X}} f_i(x) \prod_{a\in\partial i} \phi_{i,a}(x, y_a) = 0, &\quad \exists! b \in \partial i, y_b \neq 0\\ \sum_{\boldsymbol{x}_{\partial a}\in\mathcal{X}^{d_a}} f_a(\boldsymbol{x}_{\partial a}) \prod_{i\in\partial a} \hat{\phi}_{i,a}(y_i, x_i) = 0, &\quad \exists! j \in \partial a, y_j \neq 0.\end{aligned} \tag{3.5}$$

On these conditions, the term in (3.4) corresponding to $\boldsymbol{y}$ is zero, if the subset $\{(i,a) \in E \mid y_{i,a} \neq 0\} \subseteq E$ of edges generates degree-one variable nodes or degree-one factor nodes. Hence, in (3.4), we only have to take the sum over $\boldsymbol{y} \in \mathcal{X}^{|E|}$ satisfying $\{(i,a) \in E \mid y_{i,a} \neq 0\} \in \mathcal{G}$ where $\mathcal{G}$ is the set of generalized loops defined as $\mathcal{G} := \left\{E' \subseteq E \mid d_s(E') \neq 1, \forall s \in V \cup F\right\}$. Here, $d_i(E') := |\{(i,a) \in E' \mid a \in \partial i\}|$ and $d_a(E') := |\{(i,a) \in E' \mid i \in \partial a\}|$. The conditions (3.3) and (3.5) are equivalent to the condition (3.3) together with

$$\begin{aligned}\hat{\phi}_{i,a}(0, x) &= \frac{1}{\hat{f}_i(\mathbf{0})} f_i(x) \prod_{b\in\partial i\setminus\{a\}} \phi_{i,b}(x, 0)\\ \phi_{i,a}(x, 0) &= \frac{1}{\hat{f}_a(\mathbf{0})} \sum_{\substack{\boldsymbol{x}_{\partial a}\in\mathcal{X}^{d_a},\\ x_i = x}} f_a(\boldsymbol{x}_{\partial a}) \prod_{j\in\partial a\setminus\{i\}} \hat{\phi}_{j,a}(0, x_j)\end{aligned} \tag{3.6}$$

where $\mathbf{0}$ denotes the all-zero assignment.

For $((m_{a\to i}, m_{i\to a}))_{(i,a)\in E}$ which satisfies the BP equations (2.10),

$$\phi_{i,a}(x, 0) = c_{i,a} m_{a\to i}(x), \qquad \hat{\phi}_{i,a}(0, x) = \hat{c}_{i,a} m_{i\to a}(x) \tag{3.7}$$

provides a solution of (3.6) where $c_{i,a}\hat{c}_{i,a} = 1/Z_{i,a}(m_{i\to a}, m_{a\to i})$. In this case, the contribution of the all-zero assignment in (3.4) is the Bethe partition function $Z_{\mathrm{Bethe}}((m_{i\to a}), (m_{a\to i}))$ since

$$\hat{f}_a(\mathbf{0}) = Z_a((m_{i\to a})_{i\in\partial a}) \prod_{i\in\partial a} \hat{c}_{i,a}, \qquad \hat{f}_i(\mathbf{0}) = Z_i((m_{a\to i})_{a\in\partial i}) \prod_{a\in\partial i} c_{i,a}.$$

For the binary case, i.e., $\mathcal{X} = \{0, 1\}$, $(\phi_{i,a}(x, 1), \hat{\phi}_{i,a}(1, x))_{x\in\mathcal{X}}$ satisfying the condition (3.3) is uniquely determined up to a constant

$$\phi_{i,a}(x, 1) = (-1)^{\bar{x}} c_{i,a} m_{i\to a}(\bar{x}), \qquad \hat{\phi}_{i,a}(1, x) = (-1)^{\bar{x}} \hat{c}_{i,a} m_{a\to i}(\bar{x})$$

where $\bar{x} := 1 - x$. In this case, one obtains the following lemma by substituting the above values of $(\phi_{i,a}, \hat{\phi}_{i,a})_{(i,a)\in E}$ to (3.4).

**Lemma 3.4** ([Chertkov and Chernyak, 2006a], [Sudderth et al., 2008])**.** *Assume that the alphabet is binary, i.e.,* $\mathcal{X} = \{0, 1\}$*. Let* $\eta_i := \langle X_i \rangle_{b_i} = b_i(1)$*. For any* $((b_i), (b_a)) \in \mathrm{Int}(\mathcal{F}_{\mathrm{Bethe}})$*,*

$$Z(G) = Z_{\mathrm{Bethe}}((b_i)_{i \in V}, (b_a)_{a \in F}) \sum_{E' \subseteq E} \mathcal{Z}_G(E') \tag{3.8}$$

*where*

$$\mathcal{Z}_G(E') := \prod_{a \in F} \left\langle \prod_{i \in \partial a, (i,a) \in E'} \frac{X_i - \eta_i}{\sqrt{\langle (X_i - \eta_i)^2 \rangle_{b_i}}} \right\rangle_{b_a} \prod_{i \in V} \left\langle \left( \frac{X_i - \eta_i}{\sqrt{\langle (X_i - \eta_i)^2 \rangle_{b_i}}} \right)^{d_i(E')} \right\rangle_{b_i}.$$

*Proof.* It holds

$$\begin{aligned}
Z &= \sum_{\mathbf{y} \in \{0,1\}^{|E|}} \prod_{a \in F} \hat{f}_a(\mathbf{y}_{\partial a, a}) \prod_{i \in V} \hat{f}_i(\mathbf{y}_{i, \partial i}) \\
&= Z_{\mathrm{Bethe}} \sum_{\mathbf{y} \in \{0,1\}^{|E|}} \prod_{a \in F} \left\langle \prod_{i \in \partial a} \frac{\hat{\phi}_{i,a}(X_i, y_{i,a})}{\hat{\phi}_{i,a}(X_i, 0)} \right\rangle_{b_a} \prod_{i \in V} \left\langle \prod_{a \in \partial i} \frac{\phi_{i,a}(X_i, y_{i,a})}{\phi_{i,a}(X_i, 0)} \right\rangle_{b_i} \\
&= Z_{\mathrm{Bethe}} \sum_{E' \subseteq E} \prod_{a \in F} \left\langle \prod_{i \in \partial a, (i,a) \in E'} \frac{\hat{\phi}_{i,a}(X_i, 1)}{\hat{\phi}_{i,a}(X_i, 0)} \right\rangle_{b_a} \prod_{i \in V} \left\langle \prod_{a \in \partial i, (i,a) \in E'} \frac{\phi_{i,a}(X_i, 1)}{\phi_{i,a}(X_i, 0)} \right\rangle_{b_i} \\
&= Z_{\mathrm{Bethe}} \sum_{E' \subseteq E} \prod_{a \in F} \left\langle \prod_{i \in \partial a, (i,a) \in E'} \frac{(-1)^{\bar{X}_i} m_{i \to a}(\bar{X}_i) m_{a \to i}(\bar{X}_i)}{m_{i \to a}(0) m_{i \to a}(1)} \right\rangle_{b_a} \\
&\quad \cdot \prod_{i \in V} \left\langle \prod_{a \in \partial i, (i,a) \in E'} \frac{(-1)^{\bar{X}_i} m_{i \to a}(\bar{X}_i) m_{a \to i}(\bar{X}_i)}{m_{a \to i}(0) m_{a \to i}(1)} \right\rangle_{b_i}.
\end{aligned}$$

Finally, equations (2.9), $X_i - \eta_i = (-1)^{\bar{X}_i} b_i(\bar{X}_i)$ and $\langle (X_i - \eta_i)^2 \rangle_{b_i} = b_i(0) b_i(1)$ complete the proof. □

## 3.4 Loop calculus for non-binary finite alphabets

For non-binary finite alphabets, the conditions (3.3) and (3.5) do not fix $\left(\phi_{i,a}, \hat{\phi}_{i,a}\right)_{(i,a) \in E}$ uniquely. In [Chernyak and Chertkov, 2007], it is suggested to use loop calculus iteratively for each $\mathcal{Z}(E')$. Here, a representation of $(\phi_{i,a}, \hat{\phi}_{i,a})$ is given, which includes the full degree of freedom. As shown in Example 2.10, the family of distributions on a finite alphabet can be regarded as an exponential family. Let $\boldsymbol{\theta}_i$ and $\boldsymbol{\eta}_i$ be a natural parameter and an expectation parameter of $b_i$, respectively. Then, $\phi_{i,a}(x, y)$ and $\hat{\phi}_{i,a}(x, y)$ for $x \in \mathcal{X}$ and $y \in \mathcal{X} \setminus \{0\}$ can be represented as

$$\frac{\phi_{i,a}(x, y)}{c_{i,a} m_{a \to i}(x)} = \frac{\partial \log b_i(x)}{\partial \eta_{i,y}}, \qquad \frac{\hat{\phi}_{i,a}(y, x)}{\hat{c}_{i,a} m_{i \to a}(x)} = \frac{\partial \log b_i(x)}{\partial \theta_{i,y}}. \tag{3.9}$$

The partial derivatives in the first and second equations in the above are those with respect to the coordinate systems $(\eta_{i,y})_{y\in\mathcal{X}\setminus\{0\}}$ and $(\theta_{i,y})_{y\in\mathcal{X}\setminus\{0\}}$, respectively. One can easily confirm that these $(\phi_{i,a}, \hat{\phi}_{i,a})$ satisfy the condition (3.3) as follows. For $w \in \mathcal{X}\setminus\{0\}$, it holds

$$\sum_{x\in\mathcal{X}} \hat{\phi}_{i,a}(0,x)\phi_{i,a}(x,w) = \sum_{x\in\mathcal{X}} b_i(x)\frac{\partial \log b_i(x)}{\partial \eta_{i,w}} = 0.$$

Similarly, $\sum_{x\in\mathcal{X}} \hat{\phi}_{i,a}(y,x)\phi_{i,a}(x,0) = 0$ for any $y \in \mathcal{X}\setminus\{0\}$. For $y, w \in \mathcal{X}\setminus\{0\}$, it holds

$$\sum_{x\in\mathcal{X}} \hat{\phi}_{i,a}(y,x)\phi_{i,a}(x,w) = \sum_{x\in\mathcal{X}} b_i(x)\frac{\partial \log b_i(x)}{\partial \eta_{i,y}}\frac{\partial \log b_i(x)}{\partial \theta_{i,w}}$$
$$= \sum_{x\in\mathcal{X}} \frac{\partial b_i(x)}{\partial \eta_{i,y}}\left[t_{i,w}(x) - \eta_{i,w}\right] = \frac{\partial \eta_{i,w}}{\partial \eta_{i,y}} - \eta_{i,w}\sum_{x\in\mathcal{X}}\frac{\partial b_i(x)}{\partial \eta_{i,y}} = \delta(y,w).$$

In this representation, the degree of freedom for $\left(\phi_{i,a}, \hat{\phi}_{i,a}\right)_{(i,a)\in E}$ satisfying (3.3) and (3.5) is regarded as the degree of freedom for choice of sufficient statistic $\left(t_{i,y}(x)\right)_{x\in\mathcal{X}}$. Any function $t : \mathcal{X} \to \mathbb{R}$ can be represented as a linear combination of $(\mathbb{I}\{x = z\})_{z\in\mathcal{X}\setminus\{0\}}$ up to a translation. Hence, in both the cases, $(|\mathcal{X}|-1)\times(|\mathcal{X}|-1)$ invertible matrix represents the degrees of freedom, and hence, the representation (3.9) does not lose the degree of freedom for $\left(\phi_{i,a}, \hat{\phi}_{i,a}\right)_{(i,a)\in E}$ satisfying (3.3) and (3.5). Note that the relationship

$$\left\langle \frac{\partial \log b_i(X_i)}{\partial \eta_{i,y}}\frac{\partial \log b_i(X_i)}{\partial \theta_{i,w}} \right\rangle_{b_i} = \delta(y,w) \tag{3.10}$$

is well known in theory of information geometry [Amari and Nagaoka, 2000]. From this representation, one obtains the following theorem.

**Theorem 3.5** ([Mori and Tanaka, 2012b]). *For any* $((b_i),(b_a)) \in \mathrm{Int}(\mathcal{F}_{\mathrm{Bethe}})$, (3.8) *holds where*

$$\mathcal{Z}_G(E') := \sum_{y\in(\mathcal{X}\setminus\{0\})^{|E'|}} \prod_{a\in F}\left\langle \prod_{i\in\partial a,(i,a)\in E'} \frac{\partial \log b_i(X_i)}{\partial \theta_{i,y_{i,a}}} \right\rangle_{b_a} \prod_{i\in V}\left\langle \prod_{a\in\partial i,(i,a)\in E'} \frac{\partial \log b_i(X_i)}{\partial \eta_{i,y_{i,a}}} \right\rangle_{b_i}. \tag{3.11}$$

*If one chooses a sufficient statistic* $\mathbf{t}_i(x_i)$ *for* $i \in V$ *such that the Fisher information matrix is diagonal at* $b_i$*, it holds*

$$\mathcal{Z}_G(E') = \sum_{y\in(\mathcal{X}\setminus\{0\})^{|E'|}} \prod_{a\in F}\left\langle \prod_{i\in\partial a,(i,a)\in E'} \frac{t_{i,y_{i,a}}(X_i) - \eta_{i,y_{i,a}}}{\sqrt{\left\langle \left(t_{i,y_{i,a}}(X_i) - \eta_{i,y_{i,a}}\right)^2\right\rangle_{b_i}}} \right\rangle_{b_a}$$
$$\cdot \prod_{i\in V}\left\langle \prod_{a\in\partial i,(i,a)\in E'} \frac{t_{i,y_{i,a}}(X_i) - \eta_{i,y_{i,a}}}{\sqrt{\left\langle \left(t_{i,y_{i,a}}(X_i) - \eta_{i,y_{i,a}}\right)^2\right\rangle_{b_i}}} \right\rangle_{b_i}.$$

*Proof.* Similarly to the proof of Lemma 3.4, one obtains

$$Z = Z_{\text{Bethe}} \sum_{y \in \mathcal{X}^{|E|}} \prod_{a \in F} \left\langle \prod_{i \in \partial a} \frac{\hat{\phi}_{i,a}(X_i, y_{i,a})}{\hat{\phi}_{i,a}(X_i, 0)} \right\rangle_{b_a} \prod_{i \in V} \left\langle \prod_{a \in \partial i} \frac{\phi_{i,a}(X_i, y_{i,a})}{\phi_{i,a}(X_i, 0)} \right\rangle_{b_i}.$$

The equation (3.11) is obtained by substituting (3.9) into the above formula. For the second result, it generally holds

$$\frac{\partial \log b_i(X_i)}{\partial \eta_{i,y_{i,a}}} = \sum_{w \in \mathcal{X} \setminus \{0\}} \frac{\partial \theta_{i,w}}{\partial \eta_{i,y_{i,a}}} \frac{\partial \log b_i(X_i)}{\partial \theta_{i,w}}.$$

From Lemma 2.9, $\frac{\partial \theta_{i,w}}{\partial \eta_{i,y_{i,a}}}$ is the $(w, y_{i,a})$-element of the Fisher information matrix $\mathcal{J}(\boldsymbol{\eta})$. When the Fisher information matrix is diagonal, it holds

$$\frac{\partial \log b_i(X_i)}{\partial \eta_{i,y_{i,a}}} = \mathcal{J}(\boldsymbol{\theta})^{-1}_{y_{i,a},y_{i,a}} \frac{\partial \log b_i(X_i)}{\partial \theta_{i,y_{i,a}}} = \frac{t_{i,y_{i,a}}(X_i) - \eta_{i,y_{i,a}}}{\left\langle \left(t_{i,y_{i,a}}(X_i) - \eta_{i,y_{i,a}}\right)^2 \right\rangle_{b_i}}$$

for $(i,a) \in E'$. □

In fact, the weight $\mathcal{Z}_G(E')$ of each generalized loop $E' \in \mathcal{G}$ does not depend on the choice of sufficient statistics.

**Lemma 3.6.** *The weight $\mathcal{Z}_G(E')$ in* (3.11) *of each generalized loop $E' \in \mathcal{G}$ does not depend on the choice of sufficient statistics. In fact, it holds*

$$\begin{aligned}\mathcal{Z}_G(E') = &\sum_{z \in (\mathcal{X})^{2|E'|}} \prod_{a \in F} \left\langle \prod_{i \in \partial a, (i,a) \in E'} \frac{\delta\left(z_{(i,a),a}, X_i\right) - b_i(z_{(i,a),a})}{\sqrt{b_i(z_{(i,a),a})}} \right\rangle_{b_a} \\ &\cdot \prod_{i \in V} \left\langle \prod_{a \in \partial i, (i,a) \in E'} \frac{\delta\left(z_{(i,a),i}, X_i\right) - b_i(z_{(i,a),i})}{\sqrt{b_i(z_{(i,a),i})}} \right\rangle_{b_i} \\ &\cdot \prod_{(i,a) \in E'} \left\langle \frac{\left(\delta\left(z_{(i,a),a}, X_i\right) - b_i(z_{(i,a),a})\right)\left(\delta\left(z_{(i,a),i}, X_i\right) - b_i(z_{(i,a),i})\right)}{\sqrt{b_i(z_{(i,a),a}) b_i(z_{(i,a),i})}} \right\rangle_{b_i}.\end{aligned}$$

*Proof.* This lemma is easily obtained by using the generalized Holant theorem (3.4) based on (3.10). It holds

$$\sum_{y_{i,a} \in \mathcal{X} \setminus \{0\}} \frac{\partial \log b_i(x_i)}{\partial \theta_{i,y_{i,a}}} \frac{\partial \log b_i(z_i)}{\partial \eta_{i,y_{i,a}}} = \delta(x_i, z_i) \frac{1}{b_i(z_i)} - 1$$

for arbitrary choice of sufficient statistics since (3.9) satisfies (3.3). Hence, it holds

$$\begin{aligned}&\sum_{y \in (\mathcal{X} \setminus \{0\})^{d_a(E')}} \left\langle \prod_{i \in \partial a, (i,a) \in E'} \frac{\partial \log b_i(X_i)}{\partial \theta_{i,y_{i,a}}} \right\rangle_{b_a} \prod_{i \in \partial a, (i,a) \in E'} \frac{\partial \log b_i(z_{(i,a),a})}{\partial \eta_{i,y_{i,a}}} \\ &= \left\langle \prod_{i \in \partial a, (i,a) \in E'} \frac{\delta\left(z_{(i,a),a}, X_i\right) - b_i(z_{(i,a),a})}{b_i(z_{(i,a),a})} \right\rangle_{b_a}\end{aligned}$$

and

$$\sum_{\boldsymbol{y}\in(\mathcal{X}\setminus\{0\})^{d_i(E')}} \left\langle \prod_{a\in\partial i,(i,a)\in E'} \frac{\partial \log b_i(X_i)}{\partial \eta_{i,y_{i,a}}} \right\rangle_{b_i} \prod_{a\in\partial i,(i,a)\in E'} \frac{\partial \log b_i(z_{(i,a),i})}{\partial \theta_{i,y_{i,a}}}$$
$$= \left\langle \prod_{a\in\partial i,(i,a)\in E'} \frac{\delta\left(z_{(i,a),i}, X_i\right) - b_i(z_{(i,a),i})}{b_i(z_{(i,a),i})} \right\rangle_{b_i}.$$

Finally, the lemma is obtained from

$$\sum_{y_{i,a}\in\mathcal{X}\setminus\{0\}} b_i(z_{(i,a),a}) \frac{\partial \log b_i(z_{(i,a),a})}{\partial \theta_{i,y_{i,a}}} b_i(z_{(i,a),i}) \frac{\partial \log b_i(z_{(i,a),i})}{\partial \eta_{i,y_{i,a}}}$$
$$= \delta\left(z_{(i,a),a}, z_{(i,a),i}\right) b_i(z_{(i,a),a}) - b_i(z_{(i,a),a}) b_i(z_{(i,a),i})$$
$$= \left\langle \left(\delta\left(z_{(i,a),a}, X_i\right) - b_i(z_{(i,a),a})\right)\left(\delta\left(z_{(i,a),i}, X_i\right) - b_i(z_{(i,a),i})\right)\right\rangle_{b_i}. \qquad \square$$

## 3.5 Loop calculus for arbitrary alphabet

On the other hand, a similar result is known for arbitrary alphabet $\mathcal{X}$.

**Lemma 3.7** ([Xiao and Zhou, 2011]). *Assume that the alphabet $\mathcal{X}$ is not necessarily finite. For any $((b_i),(b_a)) \in \mathrm{Int}(\mathcal{F}_{\mathrm{Bethe}})$,*

$$Z(G) = Z_{\mathrm{Bethe}}((b_i)_{i\in V}, (b_a)_{a\in F}) \sum_{E'\subseteq E} \tilde{\mathcal{Z}}_G(E')$$

*where*

$$\tilde{\mathcal{Z}}_G(E') := \sum_{\boldsymbol{w}\in\mathcal{X}^{|E|}} \prod_{a\in F} b_a(\boldsymbol{w}_{\partial a,a}) \prod_{i\in V} \left\langle \prod_{a\in\partial i,(i,a)\in E'} \frac{\delta(w_{i,a}, X_i) - b_i(w_{i,a})}{b_i(w_{i,a})} \right\rangle_{b_i}.$$

*Here, $\sum_{\boldsymbol{x}\in\mathcal{X}^N}$ and $\delta(x_i, X_i)$ are replaced by $\int_{\mathcal{X}^N} \mathrm{d}\boldsymbol{x}$ and $\delta(x_i - X_i)$ when the alphabet is continuous where $\delta(\cdot)$ denotes the Dirac delta function.*

*Proof.* The proof is almost the same as (2.14).

$$Z(G) = Z_{\mathrm{Bethe}}((b_i)_{i\in V}, (b_a)_{a\in F}) \sum_{\boldsymbol{x}\in\mathcal{X}^N} \prod_{a\in F} \frac{b_a(\boldsymbol{x}_{\partial a})}{\prod_{i\in\partial a} b_i(x_i)} \prod_{i\in V} b_i(x_i)$$
$$= Z_{\mathrm{Bethe}}((b_i)_{i\in V}, (b_a)_{a\in F}) \sum_{\boldsymbol{x}\in\mathcal{X}^N, \boldsymbol{w}\in\mathcal{X}^{|E|}} \prod_{a\in F} b_a(\boldsymbol{w}_{\partial a,a}) \prod_{i\in V} b_i(x_i) \prod_{(i,a)\in E} \frac{\delta(w_{i,a}, x_i)}{b_i(w_{i,a})}$$
$$= Z_{\mathrm{Bethe}}((b_i)_{i\in V}, (b_a)_{a\in F}) \sum_{\boldsymbol{x}\in\mathcal{X}^N, \boldsymbol{w}\in\mathcal{X}^{|E|}} \prod_{a\in F} b_a(\boldsymbol{w}_{\partial a,a}) \prod_{i\in V} b_i(x_i)$$

$$\cdot \prod_{(i,a)\in E} \left[ 1 + \frac{\delta(w_{i,a}, x_i) - b_i(w_{i,a})}{b_i(w_{i,a})} \right]$$

$$= Z_{\mathrm{Bethe}}((b_i)_{i\in V}, (b_a)_{a\in F}) \sum_{\boldsymbol{x}\in\mathcal{X}^N, \boldsymbol{w}\in\mathcal{X}^{|E|}} \prod_{a\in F} b_a(\boldsymbol{w}_{\partial a,a}) \prod_{i\in V} b_i(x_i)$$

$$\cdot \sum_{E'\subseteq E} \prod_{(i,a)\in E'} \frac{\delta(w_{i,a}, x_i) - b_i(w_{i,a})}{b_i(w_{i,a})}$$

$$= Z_{\mathrm{Bethe}}((b_i)_{i\in V}, (b_a)_{a\in F})$$

$$\cdot \sum_{E'\subseteq E} \sum_{\boldsymbol{w}\in\mathcal{X}^{|E|}} \prod_{a\in F} b_a(\boldsymbol{w}_{\partial a,a}) \prod_{i\in V} \left\langle \prod_{a\in\partial i,(i,a)\in E'} \frac{\delta(w_{i,a}, X_i) - b_i(w_{i,a})}{b_i(w_{i,a})} \right\rangle_{b_i}. \qquad \square$$

Lemma 3.7 is generalized for CVM free energies in [Zhou et al., 2011]. It is not obvious whether or not Lemma 3.7 is useful for finite alphabets. While in Theorem 3.5, the alphabet for $\mathcal{Z}_G(E')$ is $\mathcal{X} \setminus \{0\}$, the alphabet for $\tilde{\mathcal{Z}}_G(E')$ is $\mathcal{X}$ in Lemma 3.7. The difference of the alphabets is significant when $\mathcal{X}$ is the binary alphabet since $\mathcal{X} \setminus \{0\}$ is the unary alphabet and the summation disappears as shown in Lemma 3.4. Interestingly, the weight $\tilde{\mathcal{Z}}_G(E')$ in Lemma 3.7 is equal to the weight $\mathcal{Z}_G(E')$ in Theorem 3.5.

**Lemma 3.8.** *Lemma 3.6 holds also for infinite alphabet $\mathcal{X}$. Furthermore, it holds $\mathcal{Z}_G(E') = \tilde{\mathcal{Z}}_G(E')$ for $E' \subseteq E$ for arbitrary alphabet.*

*Proof.* It holds

$$\tilde{\mathcal{Z}}_G(E') = \sum_{\boldsymbol{w}\in\mathcal{X}^{|E|}} \prod_{a\in F} b_a(\boldsymbol{w}_{\partial a,a}) \prod_{i\in V} \left\langle \prod_{a\in\partial i,(i,a)\in E'} \frac{\delta(w_{i,a}, X_i) - b_i(w_{i,a})}{b_i(w_{i,a})} \right\rangle_{b_i}$$

$$= \sum_{\boldsymbol{y}\in\mathcal{X}^{|E|}, \boldsymbol{w}\in\mathcal{X}^{|E|}} \prod_{a\in F} b_a(\boldsymbol{y}_{\partial a,a}) \prod_{i\in V} \left\langle \prod_{a\in\partial i,(i,a)\in E'} \frac{\delta(w_{i,a}, X_i) - b_i(w_{i,a})}{b_i(w_{i,a})} \right\rangle_{b_i} \prod_{(i,a)\in E} \delta(w_{i,a}, y_{i,a})$$

$$= \sum_{\boldsymbol{y}\in\mathcal{X}^{|E|}, \boldsymbol{w}\in\mathcal{X}^{|E|}} \prod_{a\in F} b_a(\boldsymbol{y}_{\partial a,a}) \prod_{i\in V} \left\langle \prod_{a\in\partial i,(i,a)\in E'} \frac{\delta(w_{i,a}, X_i) - b_i(w_{i,a})}{b_i(w_{i,a})} \right\rangle_{b_i} \prod_{(i,a)\in E} b_i(w_{i,a})$$

$$\cdot \prod_{(i,a)\in E} \left[ 1 + \frac{\delta(w_{i,a}, y_{i,a}) - b_i(w_{i,a})}{b_i(w_{i,a})} \right]$$

$$= \sum_{\boldsymbol{y}\in\mathcal{X}^{|E|}, \boldsymbol{w}\in\mathcal{X}^{|E|}} \prod_{a\in F} b_a(\boldsymbol{y}_{\partial a,a}) \prod_{i\in V} \left\langle \prod_{a\in\partial i,(i,a)\in E'} \frac{\delta(w_{i,a}, X_i) - b_i(w_{i,a})}{b_i(w_{i,a})} \right\rangle_{b_i} \prod_{(i,a)\in E} b_i(w_{i,a})$$

$$\cdot \sum_{E''\subseteq E} \prod_{(i,a)\in E''} \frac{\delta(w_{i,a}, y_{i,a}) - b_i(w_{i,a})}{b_i(w_{i,a})}.$$

From

$$\frac{\delta(w_{i,a}, y_{i,a}) - b_i(w_i)}{b_i(w_i)} = \sum_{v_{i,a}\in\mathcal{X}} b_i(v_{i,a}) \left( \frac{\delta(y_{i,a}, v_{i,a})}{b_i(v_{i,a})} - 1 \right) \left( \frac{\delta(w_{i,a}, v_{i,a})}{b_i(v_{i,a})} - 1 \right)$$

it holds

$$\tilde{\mathcal{Z}}_G(E') = \sum_{E'' \subseteq E} \sum_{v \in \mathcal{X}^{|E''|}, w \in \mathcal{X}^{|E|}} \prod_{a \in F} \left\langle \prod_{i \in \partial a, (i,a) \in E''} \frac{\delta(v_{i,a}, X_i) - b_i(v_{i,a})}{b_i(v_{i,a})} \right\rangle_{b_a}$$
$$\cdot \prod_{i \in V} \left\langle \prod_{a \in \partial i, (i,a) \in E'} \frac{\delta(w_{i,a}, X_i) - b_i(w_{i,a})}{b_i(w_{i,a})} \right\rangle_{b_i}$$
$$\cdot \prod_{(i,a) \in E \setminus E''} b_i(w_{i,a}) \prod_{(i,a) \in E''} \left(\delta(w_{i,a}, v_{i,a}) b_i(w_{i,a}) - b_i(w_{i,a}) b_i(v_{i,a})\right).$$

For $E'' \neq E'$, the weight corresponding to $E''$ is zero. Hence, it holds

$$\tilde{\mathcal{Z}}_G(E') = \sum_{v \in \mathcal{X}^{|E'|}, w \in \mathcal{X}^{|E'|}} \prod_{a \in F} \left\langle \prod_{i \in \partial a, (i,a) \in E'} \frac{\delta(v_{i,a}, X_i) - b_i(v_{i,a})}{b_i(v_{i,a})} \right\rangle_{b_a}$$
$$\cdot \prod_{i \in V} \left\langle \prod_{a \in \partial i, (i,a) \in E'} \frac{\delta(w_{i,a}, X_i) - b_i(w_{i,a})}{b_i(w_{i,a})} \right\rangle_{b_i}$$
$$\cdot \prod_{(i,a) \in E'} \left\langle \left(\delta\left(v_{i,a}, X_i\right) - b_i(v_{i,a})\right) \left(\delta\left(w_{i,a}, X_i\right) - b_i(w_{i,a})\right) \right\rangle_{b_i}.$$

This coincides with the expression of $\mathcal{Z}_G(E')$ in Lemma 3.6. □

## 3.6 Loop calculus for marginal distributions

Similarly to the previous section, loop calculus formula for marginal distributions can be obtained as follows.

**Lemma 3.9.** *Assume that the alphabet is binary, i.e., $\mathcal{X} = \{0, 1\}$. Let $C \subseteq V$, $F_C := \{a \in F \mid \partial a \subseteq C\}$, $E(F_C) := \{(i, a) \in E \mid a \in F_C\}$ and $g : \mathcal{X}^{|C|} \to \mathbb{C}$. For any $((b_i), (b_a)) \in \mathrm{Int}(\mathcal{F}_{\mathrm{Bethe}})$,*

$$Z(G)\langle g(\boldsymbol{X}_C)\rangle_p = Z_{\mathrm{Bethe}}((b_i)_{i \in V}, (b_a)_{a \in F}) \sum_{E' \subseteq E \setminus E(F_C)} \mathcal{Z}_G(E') \tag{3.12}$$

*where*

$$\mathcal{Z}_G(E') := \prod_{a \in F \setminus F_C} \left\langle \prod_{i \in \partial a, (i,a) \in E'} \frac{X_i - \eta_i}{\sqrt{\langle (X_i - \eta_i)^2 \rangle_{b_i}}} \right\rangle_{b_a} \prod_{i \in V \setminus C} \left\langle \left( \frac{X_i - \eta_i}{\sqrt{\langle (X_i - \eta_i)^2 \rangle_{b_i}}} \right)^{d_i(E')} \right\rangle_{b_i}$$
$$\cdot \left\langle g(\boldsymbol{X}_C) \prod_{i \in C} \left( \frac{X_i - \eta_i}{\sqrt{\langle (X_i - \eta_i)^2 \rangle_{b_i}}} \right)^{d_i(E')} \right\rangle_{b_C}.$$

*Here, $\langle \cdot \rangle_{b_C}$ is a pseudo-expectation with respect to an un-normalized distribution*

$$b_C(\boldsymbol{x}_C) = \prod_{i \in C} b_i(x_i) \prod_{a \in F_C} \frac{b_a(\boldsymbol{x}_{\partial a})}{\prod_{i \in \partial a} b_i(x_i)}.$$

*Proof.* By transforming $f_a$ only for $a \notin F_C$, one obtains

$$Z\langle g(\boldsymbol{X}_C)\rangle_p = \sum_{\boldsymbol{y} \in \{0,1\}^{|E \setminus E(F_C)|}} \prod_{a \in F \setminus F_C} \hat{f}_a(\boldsymbol{y}_{\partial a}) \prod_{i \in V \setminus C} \hat{f}_i(\boldsymbol{y}_{\partial i}) \cdot \left( \sum_{\boldsymbol{x}_C} g(\boldsymbol{x}_C) \prod_{i \in C} \prod_{a \in \partial i, a \notin F_C} \phi_{i,a}(x_i, y_{i,a}) \prod_{a \in F_C} f_a(\boldsymbol{x}_{\partial a}) \right).$$

One obtains the lemma from

$$\prod_{a \in F_C} f_a(\boldsymbol{x}_{\partial a}) \prod_{i \in C} \prod_{a \in \partial i, a \notin F_C} m_{a \to i}(x_i) = \prod_{i \in C} Z_i \prod_{a \in F_C} \frac{Z_a}{\prod_{i \in \partial a} Z_{i,a}} \prod_{i \in C} b_i(x_i) \prod_{a \in F_C} \frac{b_a(\boldsymbol{x}_{\partial a})}{\prod_{i \in \partial a} b_i(x_i)}.$$

□

In the same way, the following lemma is obtained for non-binary finite alphabets.

**Lemma 3.10.** *Let $C \subseteq V$ and $g : \mathcal{X}^{|C|} \to \mathbb{C}$. For any $((b_i), (b_a)) \in \mathrm{Int}(\mathcal{F}_{\mathrm{Bethe}})$, it holds* (3.12) *where*

$$\mathcal{Z}_G(E') := \sum_{\boldsymbol{y} \in (\mathcal{X} \setminus \{0\})^{|E'|}} \prod_{a \in F \setminus F_C} \left\langle \prod_{i \in \partial a, (i,a) \in E'} \frac{\partial \log b_i(X_i)}{\partial \theta_{i, y_{i,a}}} \right\rangle_{b_a} \cdot \left[ \prod_{i \in V \setminus C} \left\langle \prod_{a \in \partial i, (i,a) \in E'} \frac{\partial \log b_i(X_i)}{\partial \eta_{i, y_{i,a}}} \right\rangle_{b_i} \right] \left\langle g(\boldsymbol{X}_C) \prod_{i \in C, (i,a) \in E'} \frac{\partial \log b_i(X_i)}{\partial \eta_{i, y_{i,a}}} \right\rangle_{b_C}.$$

*If one chooses a sufficient statistic $\boldsymbol{t}_i(x_i)$ for $i \in V$ such that the Fisher information matrix is diagonal at $b_i$, it holds*

$$\mathcal{Z}_G(E') = \sum_{\boldsymbol{y} \in (\mathcal{X} \setminus \{0\})^{|E'|}} \prod_{a \in F \setminus F_C} \left\langle \prod_{i \in \partial a, (i,a) \in E'} \frac{t_{i, y_{i,a}}(X_i) - \eta_{i, y_{i,a}}}{\sqrt{\left\langle \left( t_{i, y_{i,a}}(X_i) - \eta_{i, y_{i,a}} \right)^2 \right\rangle_{b_i}}} \right\rangle_{b_a} \cdot \prod_{i \in V \setminus C} \left\langle \prod_{a \in \partial i, (i,a) \in E'} \frac{t_{i, y_{i,a}}(X_i) - \eta_{i, y_{i,a}}}{\sqrt{\left\langle \left( t_{i, y_{i,a}}(X_i) - \eta_{i, y_{i,a}} \right)^2 \right\rangle_{b_i}}} \right\rangle_{b_i} \cdot \left\langle g(\boldsymbol{X}_C) \prod_{i \in C, (i,a) \in E'} \frac{t_{i, y_{i,a}}(X_i) - \eta_{i, y_{i,a}}}{\sqrt{\left\langle \left( t_{i, y_{i,a}}(X_i) - \eta_{i, y_{i,a}} \right)^2 \right\rangle_{b_i}}} \right\rangle_{b_C}.$$

The weight of the empty set $\emptyset$ of edges is $\mathcal{Z}_G(\emptyset) = \langle g(\boldsymbol{X}_C)\rangle_{b_C}$. Even if a factor graph is tree, $\emptyset$ is not a unique subset of edges with non-zero weight since even if variable nodes in $C$ have degree one the weight can be non-zero. Lemmas 3.6, 3.7 and 3.8 can be generalized in a similar way.

Let $\mathrm{Var}_p[t_i(X_i)]$ and $\mathrm{Cov}_p[t_i(X_i), t_j(X_j)]$ be a $(|\mathcal{X}|-1)\times(|\mathcal{X}|-1)$-matrix whose $(k,l)$-elements are $\left\langle (t_{i,k}(X_i)-\eta_{i,k})(t_{i,l}(X_i)-\eta_{i,l})\right\rangle_p$ and $\left\langle (t_{i,k}(X_i)-\eta_{i,k})(t_{j,l}(X_j)-\eta_{j,l})\right\rangle_p$, respectively. Let $\mathrm{Cor}_p[t_i(X_i), t_j(X_j)] := \mathrm{Var}[t_i(X_i)]^{-1/2}\mathrm{Cor}_p[t_i(X_i), t_j(X_j)]\mathrm{Var}[t_j(X_j)]^{-1/2}$. The matrices $\mathrm{Var}_p[t_i(X_i)]$, $\mathrm{Cov}_p[t_i(X_i), t_j(X_j)]$ and $\mathrm{Cor}_p[t_i(X_i), t_j(X_j)]$ are called *variance matrix*, *covariance matrix* and *correlation matrix*, respectively.

**Corollary 3.11** (Correlation matrix on a tree factor graph [Watanabe, 2010])**.** *For a tree factor graph $G$, the correlation matrix for $i, j \in V$ is decomposed to*

$$\mathrm{Cor}_p[t_i(X_i), t_j(X_j)] = \mathrm{Cor}_p[t_i(X_i), t_{i_1}(X_{i_1})]\mathrm{Cor}_p[t_{i_1}(X_{i_1}), t_{i_2}(X_{i_2})]\cdots\mathrm{Cor}_p[t_{i_l}(X_{i_l}), t_j(X_j)]$$

*where $(i, i_1 \in V, i_2 \in V, \dots, i_l \in V, j)$ is the unique path of variable nodes between $i$ and $j$.*

*Proof.* When $l = 0$, i.e., $i$ and $j$ are adjacent, the lemma is trivial. It is sufficient to prove

$$\begin{aligned}\mathrm{Cov}_p[t_i(X_i), t_j(X_j)] = {}& \mathrm{Cov}_p[t_i(X_i), t_{i_1}(X_{i_1})]\mathrm{Cor}_p[t_{i_1}(X_{i_1}), t_{i_2}(X_{i_2})]\\ &\cdots\mathrm{Cor}_p[t_{i_{l-1}}(X_{i_{l-1}}), t_{i_l}(X_{i_l})]\mathrm{Cov}_p[t_{i_l}(X_{i_l}), t_j(X_j)]\end{aligned}$$

for $l \geq 1$. For a tree factor graph, the stationary point of the Bethe free energy is unique and the pseudo-marginals $((b_i)_{i\in V}, (b_a)_{a\in F})$ on the stationary point are exact marginal distributions. Let $C = \{i, j\}$ and

$$g(x_i, x_j) = (t_{i,k}(x_i) - \eta_{i,k})(t_{j,l}(x_j) - \eta_{j,l}) = \frac{\partial \log b_i(x_i)}{\partial\theta_{i,k}}\frac{\partial \log b_j(x_j)}{\partial\theta_{j,l}}$$

for Lemma 3.10. In this case, $F_C$ is the set of degree-one factor nodes connected to $i$ or $j$. Since $\mathcal{Z}_G(E') = 0$ for $E' \subseteq E$ generating degree-one variable node or degree-one factor node except for $i$ and $j$, we only have to consider the empty set and the set of edges in the unique path between $i$ and $j$. Since $b_C(x_i, x_j) = b_i(x_i)b_j(x_j)$, the weight of the empty set is zero. The lemma is obtained from

$$\begin{aligned}&\left\langle g(\boldsymbol{X}_C)\frac{\partial \log b_i(X_i)}{\partial\eta_{i,y_{i,a}}}\frac{\partial \log b_j(X_j)}{\partial\eta_{j,y_{j,b}}}\right\rangle_{b_C}\\ &= \left\langle \frac{\partial \log b_i(X_i)}{\partial\theta_{i,k}}\frac{\partial \log b_i(X_i)}{\partial\eta_{i,y_{i,a}}}\right\rangle_{b_i}\left\langle \frac{\partial \log b_j(X_j)}{\partial\theta_{j,l}}\frac{\partial \log b_j(X_j)}{\partial\eta_{j,y_{j,b}}}\right\rangle_{b_j} = \delta(k, y_{i,a})\delta(l, y_{j,b})\end{aligned}$$

and Lemma 2.1. □

Lemma 3.10 can be used also for generalizing Corollary 3.11 to general factor graphs. The results are considered to be useful for proving the decay of correlations in factor graphs [Kudekar and Macris, 2011], [Kudekar, 2009], [Weitz, 2006].

## 3.7 Historical remarks on the loop calculus

The loop calculus formula, Lemma 3.4, is obtained in [Chertkov and Chernyak, 2006a], [Chertkov and Chernyak, 2006b]. In [Sudderth et al., 2008], a simple proof and a simple expression are obtained. Furthermore, it is proved that for some Ising model, all weights in the loop calculus formula are non-negative, and hence, the Bethe partition function is a lower bound of the true partition function. A clear derivation using (3.1) is obtained in [Chertkov and Chernyak, 2006c] and [Chernyak and Chertkov, 2007], and mentioned in [Forney and Vontobel, 2011] using the concept of the holographic transformation for normal factor graphs proposed in [Al-Bashabsheh and Mao, 2011]. The formula are generalized for continuous alphabets in [Xiao and Zhou, 2011] and for CVM free energies in [Zhou et al., 2011]. An improvement of the Bethe approximation by taking partial sum among generalized loops is considered in [Chertkov and Chernyak, 2006c], [Gómez et al., 2007]. The loop calculus formula has been used not only for Ising model but also for the permanent problem [Watanabe and Chertkov, 2010] and the independent set problem [Chandrasekaran et al., 2011]. Almost simultaneously with the loop calculus formula, a similar method based on the cavity method is proposed in [Montanari and Rizzo, 2005].

# 4 Characterization of the Bethe Entropy by Graph Covers

**In this chapter, characterization of the Bethe entropy using graph cover is introduced, which is shown by Vontobel. The method of graph covers is also useful for considering relationship between true partition function and its Bethe approximation as shown in the next chapter.**

## 4.1 Method of types

In this section, the *method of types* is introduced, which is an intuitive combinatorial tool invented by Csiszár and Körner [Csiszár, 1998], [Csiszár and Körner, 2011]. Here, the key lemma in the method of types is proved without using Stirling's formula.

**Lemma 4.1.** *Let* $(N(x))_{x=1,\dots,q}$ *be natural numbers satisfying* $\sum_{x=1}^{q} N(x) = N$. *When* $\lim_{N\to\infty} N(x)/N = \nu(x)$, *it holds*

$$\lim_{N\to\infty} \frac{1}{N} \log \binom{N}{N(1)\, N(2)\, \cdots\, N(q)} = \mathcal{H}(\boldsymbol{\nu}).$$

*Proof.* From

$$\begin{aligned}
\log \binom{N}{N(1)\, N(2)\, \cdots\, N(q)} &= \log N! - \sum_{x=1}^{q} \log N(x)! = \sum_{k=1}^{N} \log k - \sum_{x=1}^{q} \sum_{k=1}^{N(x)} \log k \\
&= \sum_{k=1}^{N} \log \frac{k}{N} - \sum_{x=1}^{q} \sum_{k=1}^{N(x)} \log \frac{k}{N(x)} - \sum_{x=1}^{q} N(x) \log \frac{N(x)}{N} \\
&= N \left[ \sum_{k=1}^{N} \frac{1}{N} \log \frac{k}{N} - \sum_{x=1}^{q} \frac{N(x)}{N} \sum_{k=1}^{N(x)} \frac{1}{N(x)} \log \frac{k}{N(x)} - \sum_{x=1}^{q} \frac{N(x)}{N} \log \frac{N(x)}{N} \right]
\end{aligned}$$

one obtains

$$\begin{aligned}
&\lim_{N\to\infty} \frac{1}{N} \log \binom{N}{N(1)\, N(2)\, \cdots\, N(q)} \\
&= \int_0^1 \log z \,\mathrm{d}z - \sum_{x=1}^{q} \nu(x) \int_0^1 \log z \,\mathrm{d}z - \sum_{x=1}^{q} \nu(x) \log \nu(x) = \mathcal{H}(\boldsymbol{\nu}). \qquad \square
\end{aligned}$$

The idea of the method of types is the following. Assume that a function $f_N : \{1,2, \ldots, q\}^N \to \mathbb{R}_{\geq 0}$ depends on the argument $\boldsymbol{x} \in \{1,2,\ldots,q\}^N$ only through the frequency of occurrences $N_{\boldsymbol{x}}(x)$ of the symbols $x \in \{1,2,\ldots,q\}$ in $\boldsymbol{x}$. Here, $(N_{\boldsymbol{x}}(x)/N)_{x=1,\ldots,q}$ is called a *type* of $\boldsymbol{x} \in \{1,2,\ldots,q\}^N$. Then, one obtains an equation

$$\sum_{\boldsymbol{x}\in\{1,2,\ldots,q\}^N} f_N(\boldsymbol{x}) = \sum_{N(1),N(2),\ldots,N(q)} U_N\left((N(x)/N)_{x=1,2,\ldots,q}\right) f\left(\frac{N(1)}{N}, \frac{N(2)}{N}, \ldots, \frac{N(q)}{N}\right)$$

where $U_N((N(x)/N)_{x=1,2,\ldots,q})$ is the number of sequences $\boldsymbol{x}$ with type $(N(x)/N)_{x=1,2,\ldots,q}$, where $f(N(1)/N, N(2)/N, \ldots, N(q)/N)$ is equal to $f_N(\boldsymbol{x})$ for $\boldsymbol{x} \in \{1,2,\ldots,q\}^N$ with type $(N(x)/N)_{x=1,2,\ldots,q}$ and where the sum is taken among the set of $N$-length types $\mathcal{P}_N(\{1,2,\ldots,q\}) := \left\{(N(x)/N)_{x=1,2,\ldots,q} \mid \sum_{x=1}^q N(x) = N\right\}$. From,

$$U_N((N(x)/N)_{x=1,2,\ldots,q}) = \binom{N}{N(1)\, N(2)\, \cdots\, N(q)}$$

one obtains lower and upper bounds

$$\begin{aligned}&\max_{N(1),N(2),\ldots,N(q)} \binom{N}{N(1)\, N(2)\, \cdots\, N(q)} f\left(\frac{N(1)}{N}, \frac{N(2)}{N}, \ldots, \frac{N(q)}{N}\right) \leq \sum_{\boldsymbol{x}\in\{1,2,\ldots,q\}^N} f_N(\boldsymbol{x}) \\ &\leq \left|\mathcal{P}_N(\{1,2,\ldots,q\})\right| \\ &\quad \cdot \max_{N(1),N(2),\ldots,N(q)} \binom{N}{N(1)\, N(2)\, \cdots\, N(q)} f\left(\frac{N(1)}{N}, \frac{N(2)}{N}, \ldots, \frac{N(q)}{N}\right).\end{aligned}$$

Since the number of $N$-length types is $\left|\mathcal{P}_N(\{1,2,\ldots,q\})\right| = \binom{N+q-1}{q-1}$ and hence polynomial in $N$, when $f$ is continuous, it holds that

$$\begin{aligned}&\lim_{N\to\infty} \frac{1}{N} \log \sum_{\boldsymbol{x}\in\{1,2,\ldots,q\}^N} f_N(\boldsymbol{x}) \\ &= \lim_{N\to\infty} \frac{1}{N} \max_{N(1),N(2),\ldots,N(q)} \log\left[\binom{N}{N(1)\, N(2)\, \cdots\, N(q)} f\left(\frac{N(1)}{N}, \frac{N(2)}{N}, \ldots, \frac{N(q)}{N}\right)\right] \\ &= \max_{\nu(1),\nu(2),\ldots,\nu(q)} \left\{\mathcal{H}(\boldsymbol{\nu}) + \log f\left(\nu(1), \nu(2), \ldots, \nu(q)\right)\right\}.\end{aligned}$$

In the last equality, Lemma 4.1 is used. This method is called the *Laplace method* [Flajolet and Sedgewick, 2009].

**Example 4.2** (Sanov's theorem for a finite alphabet)**.** Let $(X_i)_{j=1,2,\ldots,N}$ be i.i.d. random variables on a finite alphabet $\{1,2,\ldots,q\}$. For any open subset $\mathcal{S}$ of $\mathcal{P}(\{1,2,\ldots,q\})$, it holds

$$\begin{aligned}&\Pr\left(\left(\frac{1}{N}\sum_{i=1}^N \mathbb{I}\{X_j = x\}\right)_{x=1,2,\ldots,q} \in \mathcal{S}\right) = \sum_{\substack{(\nu(x))_{x=1,2,\ldots,q}\in\mathcal{S} \\ N\nu(x)\in\mathbb{N} \text{ for } x\in\mathcal{X}}} \binom{N}{(N\nu(x))_{x\in\mathcal{X}}} \prod_{i=1}^q P_X(x)^{N\nu(x)} \\ &\doteq \exp\left\{\sup_{\nu\in\mathcal{S}} - \sum_{x=1}^q \nu(x)\log\frac{\nu(x)}{P_X(x)}\right\} =: \exp\left\{\sup_{\nu\in\mathcal{S}} -\mathcal{D}_{\mathrm{KL}}(\nu \| P_X)\right\}.\end{aligned}$$

Here, $\mathcal{D}_{\mathrm{KL}}(\nu \| P_X)$ is called a *Kullback-Leibler divergence*.

## 4.2 Graph covers

A *graph cover* has been well considered in graph theory [Stark and Terras, 2000]. Let $G = (V, F, E, (f_a)_{a\in F})$ be a factor graph. Then, its $M$-copied factor graph is denoted by $G^{\oplus M}$. Let $i^{(k)}$ be a $i$-th variable node in $k$-th copy of $G$. A notation $a^{(k)}$ is defined in the same way. Then, $G^{\oplus M} = (V^{\oplus M}, F^{\oplus M}, E^{\oplus M}, (f_{a^{(k)}} = f_a)_{a\in F, k\in M})$. Its partition function is

$$Z(G^{\oplus M}) = \sum_{\boldsymbol{x}\in(\mathcal{X}^M)^N} \prod_{k=1}^{M} \prod_{a\in F} f_a(\boldsymbol{x}_{\partial a^{(k)}}) = \left( \sum_{\boldsymbol{x}\in\mathcal{X}^N} \prod_{a\in F} f_a(\boldsymbol{x}_{\partial a}) \right)^M = Z(G)^M.$$

Let $\sigma_{i,a} : \{1, 2, \dots, M\} \mapsto \{1, 2, \dots, M\}$ be a permutation on $\{1, 2, \dots, M\}$ for $(i, a) \in E$. Let $G_{\boldsymbol{\sigma}}$ be a factor graph in which the set of edges $\{(i^{(k)}, a^{(k)}) \mid (i, a) \in E, k = 1, 2, \dots, M\}$ in $G^{\oplus M}$ is replaced by $\{(i^{(k)}, a^{(\sigma_{i,a}(k))}) \mid (i, a) \in E, k = 1, 2, \dots, M\}$ for $\boldsymbol{\sigma} := (\sigma_{i,a})_{(i,a)\in E}$. The factor graph $G_{\boldsymbol{\sigma}}$ is called an $M$-fold graph cover of $G$. Note that there exist $(M!)^{|E|}$ graph covers without any identification among factor graphs.

Since the belief propagation (2.10) for an $M$-fold graph cover $G_{\boldsymbol{\sigma}}$ of $G$ is the same as that for $G$ except for the existence of indices $k \in \{1, \dots, M\}$ for copies in $i^{(k)}$ and $a^{(k)}$, the projection $((b_{i^{(k)}} = b^*_i)_{i\in V^{\oplus M}}, (b_{a^{(k)}} = b^*_a)_{a\in F^{\oplus M}})$ of a fixed point $((b^*_i)_{i\in V}, (b^*_a)_{a\in F})$ for $G$ is a stationary point of the Bethe free energy for $G_{\boldsymbol{\sigma}}$. On the other hand, for large $M$ and small $d$, any depth-$d$ neighborhood of $i^{(k)}$ does not include cycles for almost all $M$-fold graph covers. Hence, the Bethe approximation is considered to be accurate for almost all $M$-fold graph covers for large $M$. From these observations, it is expected that the partition functions of $M$-fold graph covers $G_{\boldsymbol{\sigma}}$ for large $M$ is related to the Bethe partition function for the original factor graph $G$.

## 4.3 Expected number sequences on random graph covers and the Bethe entropy

In this section, characterization of the Bethe entropy by graph covers is introduced, which is recently obtained by [Vontobel, 2010b]. For the identity permutation $\boldsymbol{\sigma}_0$, it holds $Z(G_{\boldsymbol{\sigma}_0}) = Z(G)^M$ as shown in the previous section. For other $\boldsymbol{\sigma}$, generally $Z(G_{\boldsymbol{\sigma}}) \neq Z(G)^M$. The partition function $Z(G_{\boldsymbol{\sigma}})$ of a graph cover $G_{\boldsymbol{\sigma}}$ is

$$Z(G_{\boldsymbol{\sigma}}) = \sum_{\boldsymbol{x}\in(\mathcal{X}^M)^N} \prod_{k=1}^{M} \prod_{a\in F} f_a(\boldsymbol{x}_{\partial a^{(k)}}). \tag{4.1}$$

Now, the idea of the method of types can be used for the calculation of $Z(G_{\sigma})$. Since the weight of $\boldsymbol{x}$ in (4.1) is not determined only by a frequency of occurrences of each symbol in $\boldsymbol{x}$, one has to consider more detailed types for (4.1). Let $N_{i,\boldsymbol{x}}(z)$ be the number of $z \in \mathcal{X}$ in $(x_i^{(1)}, \dots, x_i^{(M)}) \in \mathcal{X}^M$ for $\boldsymbol{x} \in (\mathcal{X}^M)^N$. Furthermore, $N_{a,\boldsymbol{x}}(\boldsymbol{z}_{\partial a})$ be the number of $\boldsymbol{z}_{\partial a} \in \mathcal{X}^{d_a}$ in $(\boldsymbol{x}_{\partial a^{(1)}}, \dots, \boldsymbol{x}_{\partial a^{(M)}}) \in (\mathcal{X}^{d_a})^M$ for $\boldsymbol{x} \in (\mathcal{X}^M)^N$. They must satisfy the condition

$$N_{i,\boldsymbol{x}}(z) = \sum_{\boldsymbol{z}_{\partial a} \in \mathcal{X}^{d_a}, x_i = z} N_{a,\boldsymbol{x}}(\boldsymbol{z}_{\partial a}). \tag{4.2}$$

Here, $((N_{i,\boldsymbol{x}}/M)_{i \in V}, (N_{a,\boldsymbol{x}}/M)_{a \in F})$ is called a type of $\boldsymbol{x} \in (\mathcal{X}^M)^N$ for a factor graph $G$. According to the type, it holds

$$Z(G_{\sigma}) = \sum_{(N_i),(N_a)} U_{M,G_{\sigma}}((N_i/M)_{i \in V}, (N_a/M)_{a \in F}) \prod_{a \in F} \prod_{\boldsymbol{x}_{\partial a}} f_a(\boldsymbol{x}_{\partial a})^{N_a(\boldsymbol{x}_{\partial a})}. \tag{4.3}$$

where $U_{M,G_{\sigma}}((N_i/M)_{i \in V}, (N_a/M)_{a \in F})$ is the number of sequences $\boldsymbol{x} \in (\mathcal{X}^M)^N$ with type $((N_i/M)_{i \in V}, (N_a/M)_{a \in F})$. For each factor graph $G_{\sigma}$, it is difficult to calculate $U_{M,G_{\sigma}}((N_i/M)_{i \in V}, (N_a/M)_{a \in F})$. However, its expectation taken over all graph covers has a simple expression

$$\begin{aligned} \langle U_{M,G_{\Sigma_M}}((N_i/M)_{i \in V}, (N_a/M)_{a \in F}) \rangle_{\Sigma_M} &= \prod_{i \in V} \binom{M}{(N_i(x))_{x \in \mathcal{X}}} \prod_{a \in F} \binom{M}{(N_a(\boldsymbol{x}_{\partial x}))_{\boldsymbol{x}_{\partial x} \in \mathcal{X}^{d_a}}} \\ &\quad \cdot \prod_{(i,a) \in E} \binom{M}{(N_i(x))_{x \in \mathcal{X}}}^{-1} \end{aligned} \tag{4.4}$$

where $\Sigma_M$ is a uniform random permutation among all $(M!)^{|E|}$ possible permutations of edges and where $\langle \cdot \rangle_{\Sigma_M}$ is an expectation with respect to $\Sigma_M$. When $N_i(z)/M \to b_i(z)$ and $N_a(\boldsymbol{z}_{\partial a})/M \to b_a(\boldsymbol{z}_{\partial a})$, it holds

$$\begin{aligned} \langle U_{M,G_{\Sigma_M}}((N_i/M)_{i \in V}, (N_a/M)_{a \in F}) \rangle_{\Sigma_M} &\doteq \exp\left\{ M \left[ \sum_{a \in F} \mathcal{H}(b_a) - \sum_{i \in V} (d_i - 1) \mathcal{H}(b_i) \right] \right\} \\ &= \exp\left\{ M \mathcal{H}_{\mathrm{Bethe}}\big((b_i)_{i \in V}, (b_a)_{a \in F}\big) \right\}. \end{aligned}$$

This is a novel characterization of the Bethe entropy, obtained by [Vontobel, 2010b]. From this result, when $Z(G) > 0$, one obtains

$$\begin{aligned} \langle Z(G_{\Sigma_M}) \rangle_{\Sigma_M} &= \sum_{(N_i)_{i \in V}, (N_a)_{a \in F}} \left\langle U_{M,G_{\Sigma_M}}((N_i/M)_{i \in V}, (N_a/M)_{a \in F}) \right\rangle_{\Sigma_M} \prod_{a \in F} \prod_{\boldsymbol{x}_{\partial a}} f_a(\boldsymbol{x}_{\partial a})^{N_a(\boldsymbol{x}_{\partial a})} \\ &\doteq \max_{(b_i)_{i \in V}, (b_a)_{a \in F}} \exp\left\{ M \left[ \mathcal{H}_{\mathrm{Bethe}}((b_i)_{i \in V}, (b_a)_{a \in F}) + \sum_{a \in F} \sum_{\boldsymbol{x}_{\partial a}} b_a(\boldsymbol{x}_{\partial a}) \log f(\boldsymbol{x}_{\partial a}) \right] \right\} \\ &= \exp\left\{ -M \min_{(b_i)_{i \in V}, (b_a)_{a \in F}} \mathcal{F}_{\mathrm{Bethe}}((b_i)_{i \in V}, (b_a)_{a \in F}) \right\}. \end{aligned}$$

When $Z(G) = 0$, it holds $\langle Z(G_{\Sigma_M})\rangle_{\Sigma_M} = 0$ for some $M$ including 1. In this case,

$$\limsup_{M\to\infty} \frac{1}{M} \log\langle Z(G_{\Sigma_M})\rangle_{\Sigma_M} = -\min_{(b_i)_{i\in V},(b_a)_{a\in F}} \mathcal{F}_{\text{Bethe}}((b_i)_{i\in V},(b_a)_{a\in F})$$

Here, the minimization of the Bethe free energy appears naturally as well. The following theorem is the summary of this chapter.

**Theorem 4.3** ([Vontobel, 2010b]).

$$\langle U_{M,G_{\Sigma_M}}((N_i/M)_{i\in V},(N_a/M)_{a\in F})\rangle_{\Sigma_M} \doteq \exp\left\{M\mathcal{H}_{\text{Bethe}}((b_i)_{i\in V},(b_a)_{a\in F})\right\}$$
$$\limsup_{M\to\infty} \frac{1}{M} \log\langle Z(G_{\Sigma_M})\rangle_{\Sigma_M} = -\min_{(b_i)_{i\in V},(b_a)_{a\in F}} \mathcal{F}_{\text{Bethe}}((b_i)_{i\in V},(b_a)_{a\in F}).$$

## 4.4 Historical remarks on the method of graph covers

The method of graph covers is introduced for analyzing linear programming (LP) decoding for LDPC codes [Vontobel and Köetter, 2005]. The new characterization of the Bethe entropy and the Bethe free energy is obtained in [Vontobel, 2010b]. In [Vontobel, 2011b], it is conjectured that for the permanent problem, defined in (1.5), it holds $Z(G_\sigma) \preceq Z(G_{\sigma_0})$ for arbitrary permutations $\sigma$ of edges where $(i,j)$-element of the matrix is a variable $z_{i,j}$ and where $p\big((z_{i,j})_{(i,j)\in V}\big) \preceq q\big((z_{i,j})_{(i,j)\in V}\big)$ means that the coefficient of an arbitrary monomial in $p$ is equal to or smaller than the coefficient of the same monomial in $q$. If the conjecture is true, it shows $Z_{\text{Bethe}}(G) \leq Z(G)$ for the permanent problem for non-negative matrices, which is proved by [Gurvits, 2011] in a different way. In [Watanabe, 2011], it is conjectured that for the problem of independent set, which is a pairwise binary model defined by

$$f_{i,j}(x_i,x_j) = \begin{cases} 0, & x_i = x_j = 1 \\ 1, & \text{otherwise} \end{cases}, \qquad \text{for } (i,j) \in E$$
$$f_i(x_i) = z_i^{x_i}, \qquad \text{for } i \in V$$

where $(z_i)_{i\in V}$ are variables, it holds $Z(G_\sigma) \preceq Z(G_{\sigma_0})$ for arbitrary permutations $\sigma$ of edges. In [Watanabe, 2011], it is shown that if the conjecture is true, for any binary pairwise attractive model, it holds $Z(G_\sigma) \leq Z(G)^M$ and hence $Z_{\text{Bethe}}(G) \leq Z(G)$, which is conjectured in [Sudderth et al., 2008]. In [Ruozzi, 2012], it is proved that for some class of factor functions $(f_a)_{a\in F}$ including the binary pairwise attractive model, it holds $Z(G_\sigma) \leq Z(G)^M$ for arbitrary permutations $\sigma$ of edges, which proves the conjecture for the binary pairwise attractive model in [Sudderth et al., 2008]. The characterization of the Bethe entropy is generalized to the fractional Bethe entropy and the CVM entropy [Vontobel, 2011a].

# 5 Series of the Generalized Bethe Approximations

**In this chapter, new series of the generalized Bethe approximations is introduced on the basis of an asymptotic expansion of the partition function of graph covers.**

## 5.1 Loop calculus for graph covers

From the discussion in Section 4.2, any $((b_i)_{i\in V}, (b_a)_{a\in F}) \in \mathrm{Int}(\mathcal{F}_{\mathrm{Bethe}})$ for $G$ gives a stationary point of the Bethe free energy for any $M$-fold graph cover $G_\sigma$. On the choice of the stationary point, the Bethe partition function for $G_\sigma$ is $Z_{\mathrm{Bethe}}((b_i)_{i\in V}, (b_a)_{a\in F})^M$. Hence, from Theorem 3.5, it holds

$$
\begin{aligned}
Z(G) &= Z_{\mathrm{Bethe}}((b_i)_{i\in V}, (b_a)_{a\in F}) \sum_{E'\subseteq E(G)} \mathcal{Z}_G(E')\\
Z(G_\sigma) &= Z_{\mathrm{Bethe}}\left((b_{i^{(k)}} = b_i)_{i^{(k)}\in V\oplus M}, (b_{a^{(k)}} = b_a)_{a^{(k)}\in F\oplus M}\right) \sum_{E'\subseteq E(G_\sigma)} \mathcal{Z}_{G_\sigma}(E')\\
&= Z_{\mathrm{Bethe}}((b_i)_{i\in V}, (b_a)_{a\in F})^M \sum_{E'\subseteq E(G_\sigma)} \mathcal{Z}_{G_\sigma}(E').
\end{aligned}
$$

Let $\mathrm{Min}(\mathcal{F}_{\mathrm{Bethe}}) := \arg\min\left\{\mathcal{F}_{\mathrm{Bethe}}((b_i)_{i\in V}, (b_a)_{a\in F})\right\}$. Theorem 4.3 implies that for $((b^*_i)_{i\in V}, (b^*_a)_{a\in F}) \in \mathrm{Min}(\mathcal{F}_{\mathrm{Bethe}}) \cap \mathrm{Int}(\mathcal{F}_{\mathrm{Bethe}})$ for $G$,

$$
\frac{\left\langle Z(G_{\Sigma_M})\right\rangle_{\Sigma_M}}{Z_{\mathrm{Bethe}}((b^*_i)_{i\in V}, (b^*_a)_{a\in F})^M} = \left\langle \sum_{E'\subseteq E(G_{\Sigma_M})} \mathcal{Z}_{G_{\Sigma_M}}(E') \right\rangle_{\Sigma_M} = \exp\{o(M)\} \tag{5.1}
$$

as $M \to \infty$. Let a *circuit rank* $c(E')$ of a connected subset $E' \subseteq E$ of edges be $|E'| - |V(E')| - |F(E')| + 1$ where $V(E')$ and $F(E')$ be the subsets of variable nodes and factor nodes connected to $E'$, respectively. The circuit rank $c(E')$ equals 0 if and only if $E'$ induces a tree and equals 1 if and only if $E'$ includes one cycle. Let $\mathcal{L}_1(G)$ be the set of generalized loops whose connected components are with circuit rank 1, i.e.,

$$
\mathcal{L}_1(G) := \left\{E' \subseteq E(G) \mid E' \neq \emptyset, d_s(E') = 0 \text{ or } 2, \forall s \in V(G) \cup F(G)\right\}.
$$

Let $E' \in \mathcal{L}_1(G)$ be a connected subset of edges with circuit rank 1. The expected number of such structures in uniform $M$-fold graph covers is exactly 1 since among $M$ copies $e^{(1)}, e^{(2)}, \ldots, e^{(M)}$ of edge $e \in E'$ the probability that $e^{(1)}$ participates in the generalized loop is $1/M$. More generally, for any connected generalized loop $E' \subseteq E(G_\sigma)$ in any $M$-fold graph cover $G_\sigma$, the expected number of the same type of generalized loops is $\Theta(M^{1-c(E')})$. Similar analyses have been considered for neighborhood graphs in LDPC codes [Montanari, 2006], [Mori et al., 2013]. On the other hand, the number of types of generalized loops in $M$-fold graph covers grows as $M \to \infty$. If this behavior of growing number of types of generalized loops can be neglected, the contributions of generalized loops with circuit rank greater than 1 is $\Theta(1/M)$ and hence,

$$\left\langle \sum_{E' \subseteq E(G_{\Sigma_M})} \mathcal{Z}_{G_{\Sigma_M}}(E') \right\rangle_{\Sigma_M} = 1 + \left\langle \sum_{E' \in \mathcal{L}_1(G_{\Sigma_M})} \mathcal{Z}_{G_{\Sigma_M}}(E') \right\rangle_{\Sigma_M} + \Theta\left(\frac{1}{M}\right). \tag{5.2}$$

Let $\hat{\mathcal{L}}_1(G)$ be a set of *simple loops* in $G$ which are connected subsets $E' \in \mathcal{L}_1(G)$ of edges with circuit rank 1, i.e., $\hat{\mathcal{L}}_1(G) := \left\{ E' \in \mathcal{L}_1(G) \mid E' \text{ is connected} \right\}$. From Lemmas 2.2 and 2.9, one obtains $\mathcal{Z}_G(E')$ in Theorem 3.5 for any simple loop $E' = \{(i_1, a_1), (i_2, a_1), (i_2, a_2), (i_3, a_2), \ldots, (i_\ell, a_\ell), (i_1, a_\ell)\} \in \hat{\mathcal{L}}_1(G)$ as

$$\mathcal{Z}_G(E') = \mathrm{tr}\left(\mathrm{Cor}_{b_{a_1}}[t_{i_1}(X_{i_1}), t_{i_2}(X_{i_2})]\mathrm{Cor}_{b_{a_2}}[t_{i_2}(X_{i_2}), t_{i_3}(X_{i_3})] \cdots \mathrm{Cor}_{b_{a_\ell}}[t_{i_\ell}(X_{i_\ell}), t_{i_1}(X_{i_1})]\right). \tag{5.3}$$

A backtrackless closed walk $\mathfrak{w} \in \mathfrak{C}/\overset{\mathrm{d}}{\sim}$ on $G$ can be naturally projected onto $E' \in \hat{\mathcal{L}}_1(G_\sigma)$ for some graph cover $G_\sigma$ of $G$ where $\overset{\mathrm{d}}{\sim}$ is an equivalence relation on $\mathfrak{C}$ up to cyclic permutations and reversal of direction. Hence, it is expected that the edge zeta function in Appendix A is related to the above quantity.

**Theorem 5.1.** *Let* $u^a_{i\to j} = \mathrm{Cor}_{b_a}[t_i(X_i), t_j(X_j)]$ *for arbitrary choice of sufficient statistics. If* $|\boldsymbol{u}|$ *is smaller than the radius of convergence of* $\zeta(\boldsymbol{u})$ *at* $\boldsymbol{u} = \boldsymbol{0}$*, it holds*

$$1 + \lim_{M\to\infty} \left\langle \sum_{E' \in \mathcal{L}_1(G_{\Sigma_M})} \mathcal{Z}_{G_{\Sigma_M}}(E') \right\rangle_{\Sigma_M} = \sqrt{\zeta(\boldsymbol{u})}.$$

*Proof.* Let $N_{G_\sigma}\left((\boldsymbol{k}_{\mathfrak{p},t})_{\mathfrak{p}\in\mathfrak{P}/\overset{\mathrm{d}}{\sim},\, t\in\mathbb{N}}\right)$ be the number of generalized loops in $G_\sigma$ which consist of $k_{\mathfrak{p},t}$ connected generalized loops corresponding to the backtrackless closed walk $\mathfrak{p}^t$ for $\mathfrak{p} \in \mathfrak{P}/\overset{\mathrm{d}}{\sim}$ and $t \in \mathbb{N}$. Then, it holds

$$\begin{aligned} &1 + \lim_{M\to\infty} \left\langle \sum_{E' \in \mathcal{L}_1(G_{\Sigma_M})} \mathcal{Z}_{G_{\Sigma_M}}(E') \right\rangle_{\Sigma_M} \\ &= \sum_{k_{\mathfrak{p},t}=0:\, \mathfrak{p}\in\mathfrak{P}/\overset{\mathrm{d}}{\sim},\, t\in\mathbb{N}}^{\infty} \left( \lim_{M\to\infty} \left\langle N_{G_{\Sigma_M}}\left((\boldsymbol{k}_{\mathfrak{p},t})_{\mathfrak{p}\in\mathfrak{P}/\overset{\mathrm{d}}{\sim},\, t\in\mathbb{N}}\right) \right\rangle_{\Sigma_M} \right) \prod_{\mathfrak{p}\in\mathfrak{P}/\overset{\mathrm{d}}{\sim}} \prod_{t=1}^{\infty} \mathcal{Z}(\mathfrak{p}^t)^{k_{\mathfrak{p},t}}. \end{aligned}$$

For fixed $\mathfrak{p} \in \mathfrak{P}/\overset{\mathrm{d}}{\sim}$ and $t \in \mathbb{N}$, when $\mathfrak{p}$ visits each variable node at most once, it holds

$$\begin{aligned}\left\langle N_{G_{\Sigma_M}}(k_{\mathfrak{p},t}) \right\rangle_{\Sigma_M} &= \frac{M!}{k_{\mathfrak{p},t}!(t!)^{k_{\mathfrak{p},t}}(M - k_{\mathfrak{p},t}t)!} \\ &\quad \cdot \prod_{j=1}^{k_{\mathfrak{p},t}} \left[ \frac{1}{M-(j-1)t-(t-1)} \prod_{s=0}^{t-2} \frac{t-1-s}{M-(j-1)t-s} \right] \\ &= \frac{1}{t^{k_{\mathfrak{p},t}} k_{\mathfrak{p},t}!}.\end{aligned}$$

Moreover, it generally holds

$$\lim_{M\to\infty} \left\langle N_{G_{\Sigma_M}}((\boldsymbol{k}_{\mathfrak{p},t})_{\mathfrak{p}\in\mathfrak{P}/\overset{\mathrm{d}}{\sim}, t\in\mathbb{N}}) \right\rangle_{\Sigma_M} = \prod_{\mathfrak{p}\in\mathfrak{P}/\overset{\mathrm{d}}{\sim}} \prod_{t=1}^{\infty} \frac{1}{t^{k_{\mathfrak{p},t}} k_{\mathfrak{p},t}!}.$$

Hence, it holds

$$\begin{aligned}1 + \lim_{M\to\infty} \left\langle \sum_{E'\in\mathcal{L}_1(G_{\Sigma_M})} \mathcal{Z}_{G_{\Sigma_M}}(E') \right\rangle_{\Sigma_M} &= \prod_{\mathfrak{p}\in\mathfrak{P}/\overset{\mathrm{d}}{\sim}} \prod_{t=1}^{\infty} \left( \sum_{k=0}^{\infty} \frac{\mathcal{Z}(\mathfrak{p}^t)^k}{t^k k!} \right) \\ &= \prod_{\mathfrak{p}\in\mathfrak{P}/\overset{\mathrm{d}}{\sim}} \prod_{t=1}^{\infty} \exp\left\{ \frac{\mathcal{Z}(\mathfrak{p}^t)}{t} \right\}.\end{aligned} \qquad \square$$

In coding theory, the edge zeta function is used in [Köetter et al., 2004] and [Vontobel, 2010a]. In [Köetter et al., 2004], it is shown that the existence of monomial in the edge zeta function implies the existence of the corresponding pseudo-codeword. In [Vontobel, 2010a], it is shown that the growth rate of coefficients of monomial (i.e., the logarithm of inverse of the radius of convergence) in the edge zeta function is equal to the slope at zero of the exponent of the number of codewords in a cycle code. In my knowledge, Theorem 5.1 for the first time uses the value of $\zeta(\boldsymbol{u})$ in this area. One may regard

$$1 + \lim_{M\to\infty} \left\langle \sum_{E'\in\mathcal{L}_1(G_{\Sigma_M})} \mathcal{Z}_{G_{\Sigma_M}}(E') \right\rangle_{\Sigma_M} = \sqrt{\zeta(\boldsymbol{u})}$$

as an approximation of

$$1 + \sum_{E'\in\mathcal{L}_1(G)} \mathcal{Z}_G(E').$$

On the basis of this idea, a new series of generalized Bethe approximations will be proposed.

## 5.2 Asymptotic analysis of the partition function of graph covers

In this section, (5.1) is evaluated. First, useful tools for detailed analysis are introduced.

**Lemma 5.2** (Stirling's formula)**.**

$$N! = \sqrt{2\pi N}\left(\frac{N}{\mathrm{e}}\right)^N\left(1+\Theta\left(\frac{1}{N}\right)\right).$$

**Lemma 5.3** (Local approximation)**.** *Let $(N(x))_{x\in\mathcal{X}}$ be natural numbers satisfying $\sum_{x\in\mathcal{X}} N(x) = N$. Assume $\lim_{N\to\infty} N(x)/N = \nu(x)$ where $\nu(x)$ is a probability measure on $\mathcal{X}$ satisfying $\nu(x) > 0$ for all $x \in \mathcal{X}$. For a function $n(x)$ satisfying $\sum_{x\in\mathcal{X}} n(x) = 0$ and $n(x) = o(N^{\frac{2}{3}})$,*

$$\binom{N}{(N(x)+n(x))_{x\in\mathcal{X}}} = \frac{\sqrt{2\pi N}}{\prod_{x\in\mathcal{X}}\sqrt{2\pi N(x)}}\exp\{N\mathcal{H}(\nu)\}$$
$$\cdot \exp\left\{-\sum_{x\in\mathcal{X}}\left[n(x)\log\nu(x) + \frac{n(x)^2}{2N\nu(x)}\right]\right\}\left(1+\sum_{x\in\mathcal{X}}\Theta\left(\frac{n(x)}{N}\right)+\Theta\left(\frac{n(x)^3}{N^2}\right)\right)$$

*Proof.* From Stirling's formula, it holds

$$\frac{N!}{\prod_{x\in\mathcal{X}}(N(x)+n(x))!}$$
$$= \frac{\sqrt{2\pi N}N^N}{\prod_{x\in\mathcal{X}}\sqrt{2\pi\,(N(x)+n(x))}(N(x)+n(x))^{N(x)+n(x)}}\left(1+\Theta\left(\frac{1}{N}\right)\right).$$

The lemma is obtained from

$$\log\frac{N^N}{\prod_{x\in\mathcal{X}}(N(x)+n(x))^{N(x)+n(x)}} = -\sum_{x\in\mathcal{X}}(N(x)+n(x))\log\frac{N(x)+n(x)}{N}$$
$$= -\sum_{x\in\mathcal{X}}N(x)\log\frac{N(x)}{N} - \sum_{x\in\mathcal{X}}\left[n(x)\log\frac{N(x)}{N} + \frac{(N(x)+n(x))n(x)}{N(x)} - \frac{n(x)^2}{2N(x)}\right.$$
$$\left. + \Theta\left(\frac{n(x)^3}{N^2}\right)\right]. \qquad \square$$

Let $A(M) \approx B(M) \overset{\text{def}}{\iff} A(M) = B(M)(1+o(1))$ as $M\to\infty$. Let $\nabla^2\mathcal{F}_{\text{Bethe}}\big((b_i^*)_{i\in V}, (b_a^*)_{a\in F}\big)$ be the Hessian matrix of the Bethe free energy with respect to an arbitrary sub-set $\mathcal{B}$ of $\left\{b_a(\boldsymbol{x}_{\partial a}) \mid \boldsymbol{x}_{\partial a} \in \mathrm{Supp}(f_a), a \in F\right\}$ which forms a basis of the linear space

$$\begin{aligned}\mathcal{V} := \Bigg\{ & b_i(x_i) \in \mathbb{R}, b_a(\boldsymbol{x}_{\partial a}) \in \mathbb{R} \mid i \in V, a \in F, x_i \in \mathcal{X}, \boldsymbol{x}_{\partial a} \in \mathrm{Supp}(f_a), \\ & \sum_{x_{i'} \in \mathcal{X}} b_{i'}(x_{i'}) = 0, \quad \forall i' \in V, \qquad \sum_{\boldsymbol{x}_{\partial a'} \in \mathrm{Supp}(f_{a'})} b_{a'}(\boldsymbol{x}_{\partial a'}) = 0, \quad \forall a' \in F, \\ & \sum_{\boldsymbol{x}_{\partial a'} \in \mathrm{Supp}(f_{a'}), x_{i'} = z} b_{a'}(\boldsymbol{x}_{\partial a'}) = b_{i'}(z), \quad \forall (i', a') \in E, \forall z \in \mathcal{X} \setminus \{0\} \Bigg\} \end{aligned} \tag{5.4}$$

at $((b_i^*)_{i \in V}, (b_a^*)_{a \in F})$. The dimension of the linear space $\mathcal{V}$, which is equal to $|\mathcal{B}|$, is at least $\sum_{a \in F}(|\mathrm{Supp}(f_a)| - 1) - (|E| - |V|)(|\mathcal{X}| - 1)$ which is achieved when all constraints in (5.4) are linearly independent. The following theorem gives a more detailed asymptotic result than Theorem 4.3.

**Theorem 5.4** ([Mori and Tanaka, 2012a], [Mori and Tanaka, 2012b]). *Assume that* $|\mathrm{Min}(\mathcal{F}_{\mathrm{Bethe}})| < \infty$ *and* $\mathrm{Min}(\mathcal{F}_{\mathrm{Bethe}}) \subseteq \mathrm{Int}(\mathcal{F}_{\mathrm{Bethe}})$. *Furthermore, assume* $|\mathcal{B}| = \sum_{a \in F} (|\mathrm{Supp}(f_a)| - 1) - (|E| - |V|)(|\mathcal{X}| - 1)$ *and* $\det\left(\nabla^2 \mathcal{F}_{\mathrm{Bethe}}((b_i^*)_{i \in V}, (b_a^*)_{a \in F})\right) > 0$ *for all* $((b_i^*)_{i \in V}, (b_a^*)_{a \in F}) \in \mathrm{Min}(\mathcal{F}_{\mathrm{Bethe}})$. *Then, for* $M \in \mathbb{N}$ *such that* $\langle Z(G_{\Sigma_M})\rangle_{\Sigma_M} > 0$, *it holds*

$$\begin{aligned}\langle Z(G_{\Sigma_M})\rangle_{\Sigma_M} &\approx \exp\left\{-M \min \mathcal{F}_{\mathrm{Bethe}}((b_i)_{i \in V}, (b_a)_{a \in F})\right\} \\ &\cdot \sum_{((b_i^*)_{i \in V}, (b_a^*)_{a \in F}) \in \mathrm{Min}(\mathcal{F}_{\mathrm{Bethe}})} \sqrt{\frac{\det\left(\nabla^2 \mathcal{F}_{\mathrm{Bethe}}((b_i^*)_{i \in V}, (b_a^*)_{a \in F})\right)^{-1}}{\prod_{i \in V} \prod_{x_i \in \mathcal{X}} b_i^*(x_i)^{1-d_i} \prod_{a \in F} \prod_{\boldsymbol{x}_{\partial a} \in \mathrm{Supp}(f_a)} b_a^*(\boldsymbol{x}_{\partial a})}}.\end{aligned}$$

*Proof.* The proof is based on the Laplace method and central approximation [Flajolet and Sedgewick, 2009]. Here, the proof is given for the case $|\mathrm{Min}(\mathcal{F}_{\mathrm{Bethe}})| = 1$. For other cases, a similar proof works. From (4.3) and (4.4),

$$\begin{aligned}\langle Z(G_{\Sigma_M})\rangle_{\Sigma_M} &= \sum_{\left((N_i(x))_{i \in V}, (N_a(\boldsymbol{x}_{\partial a}))_{a \in F}\right)} \left\langle U_{M, G_{\Sigma_M}}\left((N_i/M)_{i \in V}, (N_a/M)_{a \in F}\right)\right\rangle_{\Sigma_M} \\ &\quad \cdot \prod_{a \in F} \prod_{\boldsymbol{x}_{\partial a}} f_a(\boldsymbol{x}_{\partial a})^{N_a(\boldsymbol{x}_{\partial a})} \\ &= \sum_{\left((N_i(x))_{i \in V}, (N_a(\boldsymbol{x}_{\partial a}))_{a \in F}\right)} \prod_{i \in V} \binom{M}{(N_i(x))_{x \in \mathcal{X}}} \prod_{a \in F} \binom{M}{(N_a(\boldsymbol{x}_{\partial x}))_{\boldsymbol{x}_{\partial x} \in \mathcal{X}^{d_a}}} \\ &\quad \cdot \prod_{(i,a) \in E} \binom{M}{(N_i(x))_{x \in \mathcal{X}}}^{-1} \prod_{a \in F} \prod_{\boldsymbol{x}_{\partial a}} f_a(\boldsymbol{x}_{\partial a})^{N_a(\boldsymbol{x}_{\partial a})}.\end{aligned}$$

Let $\alpha \in (1/2, 2/3)$. From the assumption $\mathrm{Min}(\mathcal{F}_{\mathrm{Bethe}}) = 1$, from Lemma 5.3, it holds

$$
\begin{aligned}
&\sum_{\substack{((n_i(x_i)),(n_a(\boldsymbol{x}_{\partial a})))\in\mathcal{V} \\ \|n_i(x_i)\|_2 \geq M^\alpha, \|n_a(\boldsymbol{x}_{\partial a})\|_2 \geq M^\alpha}} \prod_{i\in V} \binom{M}{(N_i(x)+n_i(x))_{x\in\mathcal{X}}} \prod_{a\in F} \binom{M}{(N_a(\boldsymbol{x}_{\partial x})+n_a(\boldsymbol{x}_{\partial x}))_{\boldsymbol{x}_{\partial x}\in\mathcal{X}^{d_a}}} \\
&\quad \cdot \prod_{(i,a)\in E} \binom{M}{(N_i(x)+n_i(x))_{x\in\mathcal{X}}}^{-1} \prod_{a\in F}\prod_{\boldsymbol{x}_{\partial a}} f_a(\boldsymbol{x}_{\partial a})^{N_a(\boldsymbol{x}_{\partial a})+n_a(\boldsymbol{x}_{\partial a})} \\
&\quad = \exp\left\{-M\mathcal{F}_{\mathrm{Bethe}}((b_i^*),(b_a^*)) - CM^{2\alpha-1} + \Theta(M^{3\alpha-2})\right\}.
\end{aligned}
$$

On the other hand, it holds

$$
\begin{aligned}
&\sum_{\substack{((n_i(x_i)),(n_a(\boldsymbol{x}_{\partial a})))\in\mathcal{V} \\ \|n_i(x_i)\|_2 < M^\alpha, \|n_a(\boldsymbol{x}_{\partial a})\|_2 < M^\alpha}} \prod_{i\in V} \binom{M}{(N_i(x)+n_i(x))_{x\in\mathcal{X}}} \prod_{a\in F} \binom{M}{(N_a(\boldsymbol{x}_{\partial x})+n_a(\boldsymbol{x}_{\partial x}))_{\boldsymbol{x}_{\partial x}\in\mathcal{X}^{d_a}}} \\
&\quad \cdot \prod_{(i,a)\in E} \binom{M}{(N_i(x)+n_i(x))_{x\in\mathcal{X}}}^{-1} \prod_{a\in F}\prod_{\boldsymbol{x}_{\partial a}} f_a(\boldsymbol{x}_{\partial a})^{N_a(\boldsymbol{x}_{\partial a})+n_a(\boldsymbol{x}_{\partial a})} \\
&\approx \sum_{\substack{((n_i(x_i)),(n_a(\boldsymbol{x}_{\partial a})))\in\mathcal{V} \\ \|n_i(x_i)\|_2 < M^\alpha, \|n_a(\boldsymbol{x}_{\partial a})\|_2 < M^\alpha}} \exp\left\{-M\mathcal{F}_{\mathrm{Bethe}}((b_i^*),(b_a^*))\right\} \\
&\quad \cdot \prod_{i\in V}\left(\frac{\sqrt{2\pi M}}{\prod_{x_i\in\mathcal{X}}\sqrt{2\pi M b_i^*(x_i)}} \exp\left\{-\sum_{x_i\in\mathcal{X}} n_i(x_i)\log b_i^*(x_i) + \frac{n_i(x_i)^2}{2Mb_i^*(x_i)}\right\}\right)^{1-d_i} \\
&\quad \cdot \prod_{a\in F}\left(\frac{\sqrt{2\pi M}}{\prod_{\boldsymbol{x}_{\partial a}\in\mathrm{Supp}(f_a)}\sqrt{2\pi M b_a^*(\boldsymbol{x}_{\partial a})}}\right. \\
&\quad \left. \cdot \exp\left\{-\sum_{\boldsymbol{x}_{\partial a}\in\mathrm{Supp}(f_a)} n_a(\boldsymbol{x}_{\partial a})\log b_a^*(\boldsymbol{x}_{\partial a}) + \frac{n_a(\boldsymbol{x}_{\partial a})^2}{2Mb_a^*(\boldsymbol{x}_{\partial a})}\right\}\right) \\
&\approx \prod_{i\in V}\left(\frac{\sqrt{2\pi}}{\prod_{x_i\in\mathcal{X}}\sqrt{2\pi b_i^*(x_i)}}\right)^{1-d_i} \prod_{a\in F}\left(\frac{\sqrt{2\pi}}{\prod_{\boldsymbol{x}_{\partial a}\in\mathrm{Supp}(f_a)}\sqrt{2\pi b_a^*(\boldsymbol{x}_{\partial a})}}\right) \\
&\quad \cdot \exp\left\{-M\mathcal{F}_{\mathrm{Bethe}}((b_i^*),(b_a^*))\right\} \\
&\quad \cdot \int \exp\left\{-\frac{1}{2}\boldsymbol{\epsilon}^t\nabla^2\mathcal{F}_{\mathrm{Bethe}}((b_i^*),(b_a^*))\boldsymbol{\epsilon}\right\} \prod_{\{a\in F, \boldsymbol{x}_{\partial a}\in\mathrm{Supp}(f_a) | b_a(\boldsymbol{x}_{\partial a})\in\mathcal{B}\}} d\epsilon_a(\boldsymbol{x}_{\partial a}) \\
&= \prod_{i\in V}\prod_{x_i\in\mathcal{X}}\left(\frac{1}{\sqrt{b_i^*(x_i)}}\right)^{1-d_i} \prod_{a\in F}\prod_{\boldsymbol{x}_{\partial a}\in\mathrm{Supp}(f_a)}\left(\frac{1}{\sqrt{b_a^*(\boldsymbol{x}_{\partial a})}}\right) \\
&\quad \cdot \exp\left\{-M\mathcal{F}_{\mathrm{Bethe}}((b_i^*),(b_a^*))\right\}\sqrt{\det\left(\nabla^2\mathcal{F}_{\mathrm{Bethe}}((b_i^*),(b_a^*))\right)^{-1}}
\end{aligned}
$$

where $\boldsymbol{\epsilon}$ denotes the column vector $\left[\epsilon_a(\boldsymbol{x}_{\partial a})\right]_{a\in F, \boldsymbol{x}_{\partial a}\in\mathrm{Supp}(f_a), b_a(\boldsymbol{x}_{\partial a})\in\mathcal{B}}$. □

From the Watanabe-Fukumizu formula, Corollary A.10, the constant coefficient can be represented by using the edge zeta function.

**Corollary 5.5.** *Assume the conditions in Theorem 5.4. Let*

$$\mathcal{V}_a := \left\{ n_a(\boldsymbol{x}_{\partial a}) \in \mathbb{R} \,\middle|\, \boldsymbol{x}_{\partial a} \in \mathrm{Supp}(f_a), \sum_{\boldsymbol{x}_{\partial a} \in \mathrm{Supp}(f_a)} n_a(\boldsymbol{x}_{\partial a}) = 0 \right\}.$$

*Assume that variables in*

$$\left\{ n_i(z) := \sum_{\boldsymbol{x}_{\partial a} \in \mathrm{Supp}(f_a), x_i = z} n_a(\boldsymbol{x}_{\partial a}) \,\middle|\, i \in \partial a, z \in \mathcal{X} \setminus \{0\} \right\}$$

*are independent on the linear space $\mathcal{V}_a$. Then, it holds*

$$\begin{aligned} \langle Z(G_{\boldsymbol{\Sigma}_M}) \rangle_{\Sigma_M} &\approx \exp\left\{ -M \min \mathcal{F}_{\mathrm{Bethe}}((b_i)_{i \in V}, (b_a)_{a \in F}) \right\} \\ &\quad \cdot K(G) \left( \sum_{((b_i^*)_{i \in V}, (b_a^*)_{a \in F}) \in \mathrm{Min}(\mathcal{F}_{\mathrm{Bethe}})} \sqrt{\zeta(\boldsymbol{u})} \right) \end{aligned}$$

*where $K(G)$ is a positive constant determined by the structure of a factor graph $G$ and $(\mathrm{Supp}(f_a))_{a \in F}$, and $u^a_{i \to j} = \mathrm{Cor}_{b_a}[t_i(X_i), t_j(X_j)]$ for arbitrary choice of sufficient statistics. When $\mathrm{Supp}(f_a) = \mathcal{X}^{d_a}$ for all $a \in F$, it holds $K(G) = 1$.*

*Proof.* From Corollary A.10, it holds

$$\begin{aligned} \zeta(\boldsymbol{u})^{-1} &= \det\left( \nabla^2 \left( -\mathcal{H}_{\mathrm{Bethe}}((\eta_i)_{i \in V}, (\eta_{\langle a \rangle})_{a \in F}) \right) \right) \\ &\quad \cdot \prod_{i \in V} \det(\mathrm{Var}_{b_i}[t_i(X_i)])^{1-d_i} \prod_{a \in F} \det(\mathrm{Var}_{b_a}[t_a(X_{\partial a})]). \end{aligned}$$

When a sufficient statistic for $b_i$ is $\left(t_{i,z_i}(x_i) = \mathbb{I}\{x_i = z_i\}\right)_{z_i \in \mathcal{X} \setminus \{0\}}$ for $i \in V$, it holds

$$\begin{aligned} \mathrm{Var}_{b_i}[t_i(X_i)] &= \begin{bmatrix} b_i(1) & & 0 \\ & \ddots & \\ 0 & & b_i(|\mathcal{X}|-1) \end{bmatrix} - \begin{bmatrix} b_i(1) \\ \vdots \\ b_i(|\mathcal{X}|-1) \end{bmatrix} \begin{bmatrix} b_i(1) & \cdots & b_i(|\mathcal{X}|-1) \end{bmatrix} \\ &= \det(\mathrm{Var}_{b_i}[t_i(X_i)]) = \prod_{x_i \in \mathcal{X}} b_i(x_i). \end{aligned}$$

On the other hand, it holds that

$$\det(J)^2 \det\left( \nabla^2 \left( -\mathcal{H}_{\mathrm{Bethe}}((\eta_i)_{i \in V}, (\eta_{\langle a \rangle})_{a \in F}) \right) \right) = \det\left( \nabla^2 \left( \mathcal{F}_{\mathrm{Bethe}}((b_i)_{i \in V}, (b_a)_{a \in F}) \right) \right)$$

for sufficient statistics satisfying $((\eta_i)_{i \in V}, (\eta_{\langle a \rangle})_{a \in F}) = J[b_a(\boldsymbol{x}_{\partial a})]_{b_a(\boldsymbol{x}_{\partial a}) \in \mathcal{B}}$, and that

$$\det(\mathrm{Var}_{b_a}[t_a(X_{\partial a})]) = \det(J_a)^2 \prod_{\boldsymbol{x}_{\partial a} \in \mathrm{Supp}(f_a)} b_a(\boldsymbol{x}_{\partial a})$$

for sufficient statistics satisfying $(\eta_i, \eta_{\langle a\rangle}) = J_a[b_a(\boldsymbol{x}_{\partial a})]_{\boldsymbol{x}_{\partial a} \in \mathrm{Supp}(f_a)\setminus\{\boldsymbol{z}_{\partial a}\}}$ for any $\boldsymbol{z}_{\partial a} \in \mathrm{Supp}(f_a)$. Hence, one obtains

$$\zeta(\boldsymbol{u})^{-1} = K(G)^2 \det\left(\nabla^2 \left(\mathcal{F}_{\mathrm{Bethe}}((b_i)_{i\in V}, (b_a)_{a\in F})\right)\right) \prod_{i\in V}\prod_{x_i\in\mathcal{X}} b_i(x_i)^{1-d_i} \prod_{a\in F}\prod_{\boldsymbol{x}_{\partial a}\in\mathrm{Supp}(f_a)} b_a(\boldsymbol{x}_{\partial a})$$

for a constant $K(G) := \left|\det(J)^{-1}\prod_{a\in F}\det(J_a)\right|$. When $\mathrm{Supp}(f_a) = \mathcal{X}^{d_a}$ for all $a \in F$, one can choose $t_{\langle a\rangle}(\boldsymbol{x}_{\partial a})$ such that $\det(J_a) = 1$ for all $a \in F$, e.g., $t_{\langle a\rangle}(\boldsymbol{x}_{\partial a}) = (b_a(\boldsymbol{x}_{\partial a}))_{\boldsymbol{x}_{\partial a}\in\mathcal{X}^{d_a}\setminus(\boldsymbol{0}\cup\mathcal{S})}$ where $\mathcal{S} := \{\boldsymbol{x}_{\partial a} \in \mathcal{X}^{d_a} \mid \exists! i \in \partial a, x_i \neq 0\}$. In this case, it also holds $\det(J) = 1$. □

When the minimum of the Bethe free energy is unique and $K(G) = 1$, one can expect that (5.2) is correct. For permanents of positive matrix [Vontobel, 2011b], the factor graph satisfies the condition $|\mathcal{B}| = \sum_{a\in F}(|\mathrm{Supp}(f_a)| - 1) - (|E| - |V|)(|\mathcal{X}| - 1)$ in Theorem 5.4, but does not satisfy the condition in Corollary 5.5. In this case, Theorem 5.4 should be used directly.

## 5.3 Series of approximations via asymptotic expansion

### 5.3.1 Asymptotic Bethe approximations

From Theorem 5.4, on the same conditions, it holds

$$\log\langle Z(G_{\Sigma_M})\rangle_{\Sigma_M} = -M\mathcal{F}_{\mathrm{Bethe}}((b_i^*)_{i\in V}, (b_a^*)_{a\in F}) + \log\left(K(G)\sum_{((b_i^*)_{i\in V},(b_a^*)_{a\in F})\in\mathrm{Min}(\mathcal{F}_{\mathrm{Bethe}})} \sqrt{\zeta(\boldsymbol{u})}\right) + o(1).$$

By considering complete asymptotic expansion [Flajolet and Sedgewick, 2009] [Boyd, 1999] [Butler, 2007], one obtains

$$\log\langle Z(G_{\Sigma_M})\rangle_{\Sigma_M} \sim -M\mathcal{F}_{\mathrm{Bethe}}((b_i^*)_{i\in V}, (b_a^*)_{a\in F}) + \sum_{k=0}^{\infty}\frac{g_k}{M^k}.$$

where $g_0 := \log\left(K(G)\sum_{(\{b_i^*\},\{b_a^*\})\in\mathrm{Min}(\mathcal{F}_{\mathrm{Bethe}})}\sqrt{\zeta(\boldsymbol{u})}\right)$ and where $(g_k)_{k=1,2,\ldots}$ are some constants unless the problem includes a kind of singularity. From the discussion in Section 5.1, we propose the following series of approximations

**Definition 5.6** (Asymptotic Bethe approximation). For $m = 0, 1, \ldots,$ the asymptotic Bethe approximation of order $m$ is defined as

$$Z_{\mathrm{AB}}^{(m)}(G) := Z_{\mathrm{Bethe}}(G)\exp\left\{\sum_{k=0}^{m-1} g_k\right\}.$$

Here, one can guess that $Z_{\mathrm{AB}}^{(m)}(G)$ takes account of the contributions of generalized loops in graph covers whose connected components have circuit rank at most $m$. When $f_a(\boldsymbol{x}_{\partial a})$ represents a linear constraint for all $a \in F$, the Bethe free energy is minimized by the uniform messages. In this case, if all degrees of factor nodes are greater than two, it holds $\sqrt{\zeta(\boldsymbol{u})} = 1$ and $K(G) = |\mathcal{X}|^{|F|-r}$ which represents the rate loss where $r$ is the rank of the linear constraints while the Bethe approximation gives the design rate, i.e., $Z_{\mathrm{Bethe}}(G) = |\mathcal{X}|^{N-|F|}$. Hence, $Z_{\mathrm{AB}}^{(1)}(G) = Z(G) = |\mathcal{X}|^{N-r}$ when there exists at least one solution for the linear constraints.

### 5.3.2 Asymptotic exactness of the asymptotic Bethe approximation of order 1

In this section, examples of factor graphs are given, in which the asymptotic Bethe approximation $Z_{\mathrm{AB}}^{(1)}(G)$ of order 1 is asymptotically better in some limit than the Bethe approximation $Z_{\mathrm{Bethe}}(G)$.

**Example 5.7** (Single-cycle factor graph)**.** Single-cycle graphs are considered to be the simplest non-trivial example. Assume that $f_a(\boldsymbol{x}_{\partial a}) > 0$ for all $a \in F$ and $\boldsymbol{x}_{\partial a} \in \mathcal{X}^{d_a}$. In this case, the Bethe free energy is convex with respect to the expectation parameters and hence the stationary point is unique [Watanabe, 2010]. For the unique solution $((b_i^*)_{i\in V}, (b_a^*)_{a\in F})$, one obtains from Theorem 3.5 and (5.3) that

$$Z(G) = Z_{\mathrm{Bethe}}(G)\,(1 + \mathrm{tr}(A))$$

where

$$A := \mathrm{Cor}_{b_{a_1}^*}[t_1(X_1), t_2(X_2)]\mathrm{Cor}_{b_{a_2}^*}[t_2(X_2), t_3(X_3)] \cdots \mathrm{Cor}_{b_{a_N}^*}[t_N(X_N), t_1(X_1)]$$

since the set of generalized loops for a single-cycle factor graph only includes the empty set and the unique cycle. On the other hand, the square root of the edge zeta function is

$$\sqrt{\zeta(\boldsymbol{u})} = \frac{1}{\det\left(I_{|\mathcal{X}|-1} - A\right)}.$$

From $\det(I_{|\mathcal{X}|-1} - A) = 1 - \mathrm{tr}(A) + O(\rho(A)^2)$ as $A \to 0$, where $\rho(A)$ denotes the spectral radius of $A$, one obtains the following asymptotic equality

$$\begin{aligned}\sqrt{\zeta(\boldsymbol{u})} &= \frac{1}{1 - \mathrm{tr}(A) + O(\rho(A)^2)}\\ &= 1 + \mathrm{tr}(A) + O(\rho(A)^2) = \frac{Z(G)}{Z_{\mathrm{Bethe}}(G)} + O(\rho(A)^2).\end{aligned}$$

Hence, $\sqrt{\zeta(\boldsymbol{u})}$ is an accurate approximation for $Z(G)/Z_{\mathrm{Bethe}}(G)$ whenever the matrix $A$ is close to zero. □

The asymptotic Bethe approximation of order 1 is asymptotically better than the Bethe approximation for the Ising model (1.2) in the high-temperature limit.

**Lemma 5.8.** *For the Ising model* (1.2)*, it holds*

$$Z(G) = Z_{\text{Bethe}}(G)\left(1 + \left(\sqrt{\zeta(\boldsymbol{u})} - 1\right) + o\left(\left(\sqrt{\zeta(\boldsymbol{u})} - 1\right)\right)\right), \qquad \textit{as } \beta \to 0.$$

*Proof.* The correlation coefficient evaluated by the pseudo-marginal $b_a$ at an arbitrary saddle point of the Bethe free energy is represented as

$$\mathrm{Cor}_{b_a}[X_i, X_j] = \frac{\sinh(2\beta J_{i,j})}{\sqrt{\cosh(2h_{i\to a}) + \cosh(2\beta J_{i,j})}\sqrt{\cosh(2h_{j\to a}) + \cosh(2\beta J_{i,j})}}$$

where $m_{i\to a}(x) \propto \exp\{h_{i\to a}x\}$ and $m_{j\to a}(x) \propto \exp\{h_{j\to a}x\}$ [Watanabe, 2010]. Since $|\mathrm{Cor}_{b_a}[X_i, X_j]|$ takes the maximum $|\tanh(\beta J_{i,j})|$ at $h_{i\to a} = h_{j\to a} = 0$, $\mathrm{Cor}_{b_a}[X_i, X_j] \to 0$ as $\beta \to 0$. From Lemma 3.4, it holds $\mathcal{Z}_G(E') = o(\mathcal{Z}_G(E''))$ for $E'' \subsetneq E'$ where $E', E'' \in \mathcal{G}$. Since an arbitrary generalized loop includes some simple loops, it holds

$$Z(G) = Z_{\text{Bethe}}(G)\left(1 + \sum_{E'\in\hat{\mathcal{L}}_1(G)} \mathcal{Z}_G(E') + o\left(\sum_{E'\in\hat{\mathcal{L}}_1(G)} \mathcal{Z}_G(E')\right)\right), \qquad \text{as } \beta \to 0.$$

On the other hand, from Lemma A.5, it holds

$$\begin{aligned}\sqrt{\zeta(\boldsymbol{u})} &= \exp\left\{\sum_{\mathfrak{w}=(e_1\rightharpoonup e_2\cdots\rightharpoonup e_n\rightharpoonup e_1)\in\mathfrak{C}} \frac{1}{2n}\mathrm{tr}\left(u_{e_1,e_2}u_{e_2,e_3}\cdots u_{e_n,e_1}\right)\right\} \\ &= \left(1 + \sum_{E'\in\hat{\mathcal{L}}_1(G)} \mathcal{Z}_G(E') + o\left(\sum_{E'\in\hat{\mathcal{L}}_1(G)} \mathcal{Z}_G(E')\right)\right), \qquad \text{as } \beta \to 0.\end{aligned}$$

□

On the basis of the proof of Lemma 5.8, $1 + \log\sqrt{\zeta(\boldsymbol{u})}$ is also considered as good approximation for $Z(G)/Z_{\text{Bethe}}(G)$. In Figure 5.1, the Bethe approximation $Z_{\text{Bethe}}(G)$ and the asymptotic Bethe approximation $Z^{(1)}_{\text{AB}}(G)$ are compared on the Ising model defined on a randomly generated graph. The errors $\left(\log Z(G) - \log Z_{\text{Bethe}}(G)\right)/N$ and $\left(\log Z(G) - \log Z^{(1)}_{\text{AB}}(G)\right)/N$ of approximations are plotted. The stationary point of the Bethe free energy is obtained by simple BP iterations. Hence, it is not necessarily the exact minimum of the Bethe free energy. As shown in Lemma 5.8, the asymptotic Bethe approximation is accurate in high temperature region. Furthermore, it can be confirmed that approximations are improved for the whole region of $\beta \geq 0$ from Figure 5.1. Even if each $\mathrm{Cor}_{b_a}[X_i, X_j]$ does not go to 0, if the product of them along a loop goes to 0, the Bethe approximation can be accurate. Some factor graphs with diverging girth satisfy this condition in the large-size limit $N \to \infty$ [Chandrasekaran et al., 2011]. Justification of the approximation $Z^{(1)}_{\mathrm{AB}}(G)$ for some factor graphs with diverging girth is an open problem. In [Sudderth et al., 2008] and [Ruozzi, 2012], it is proved that the Bethe approximation gives an exact lower bound of the partition function of Ising model with $\beta \geq 0$. However, the experimental results show that $Z^{(1)}_{\mathrm{AB}}(G)$ is neither lower bound nor upper bound.

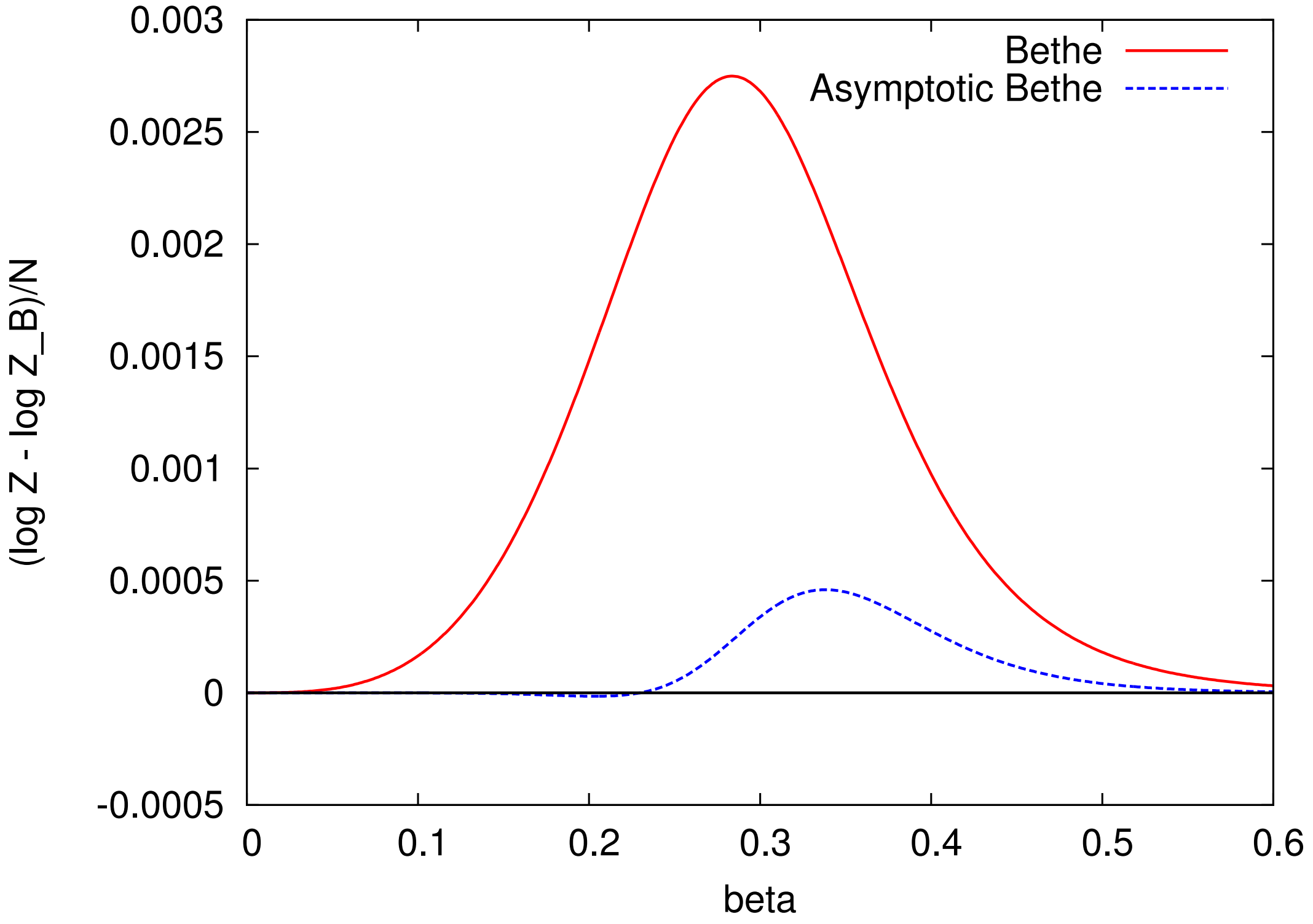

Figure 5.1: Comparisons of $Z_{\mathrm{Bethe}}(G)$ and $Z^{(1)}_{\mathrm{AB}}(G)$ for the ferromagnetic Ising model by numerical experiment. The graph is generated by Erdős-Rényi law. The average degree is 3.25. The number $N$ of variables is 24, $J_{i,j} = 1$ for all $(i, j) \in E$ and $h_i = 0.5$ for all $i \in V$.

When the Hessian of the Bethe free energy is not positive definite at some critical temperature $\beta_{\mathrm{c}}$, the edge zeta function diverges. This situation is considered as a finite-size analogue of the second-order phase transition. Similarly, if the minimum point of the Bethe free energy discontinuously jumps at $\beta_{\mathrm{c}}$, $Z^{(m)}_{\mathrm{AB}}(G)$ is discontinuous at $\beta_{\mathrm{c}}$ for $m \geq 1$. This situation is considered as a finite-size analogue of the first-order phase transition. In these cases, it is better to consider another limit for $\beta$ around $\beta_{\mathrm{c}}$, e.g., $\delta = (\beta - \beta_{\mathrm{c}})/M^c$ is fixed for some $c > 0$ [Parisi et al., 1993].

# Part II

# The Replica Method

# 6 New Derivations of the Replica Free Energies

**In this chapter, new derivations of expected log partition function is proposed using the replica method. In the derivation, an analogue of the minimization problem of the Bethe free energy appears. From this fact, it is easy to understand why the replica symmetric assumption implies asymptotic exactness of the Bethe approximation.**

## 6.1 Random factor graph and the replica method

In this part, we analyze typical behaviors of a probability distribution defined by a random factor graph. For the purpose, the *replica method* is introduced, which is a non-rigorous method invented in statistical physics. This second part of this thesis gives a new understanding of the replica method based on the method of types for factor graphs similarly to Chapter 4. In this part, we also deal with the factor graph model (2.8). The replica method is a non-rigorous method for the derivation of $\lim_{N\to\infty} \mathbb{E}[\log Z(G)]/N$ where $\mathbb{E}[\cdot]$ denotes the expectation with respect to a probability measure on a random factor graph $G$. In the replica method, the following equality is used.

$$\lim_{N\to\infty} \frac{1}{N}\mathbb{E}[\log Z(G)] = \lim_{N\to\infty} \frac{1}{N} \lim_{n\to 0} \frac{1}{n} \log \mathbb{E}[Z(G)^n] = \lim_{n\to 0} \frac{1}{n} \lim_{N\to\infty} \frac{1}{N} \log \mathbb{E}[Z(G)^n].$$

In the last equation, the exchange of limits is assumed to be valid. This is the first ansatz in the replica method. The second ansatz is concerned with $n$ in the above equation. Here, one only derives $\mathbb{E}[Z(G)^n]$ for $n \in \mathbb{N}$. Then, the result is analytically extended for $n \in \mathbb{R}$ (or more generally $n \in \mathbb{C}$) in a *natural way*. This interpolation is similar to the analytic continuation in complex analysis. However, in this case the analytic continuation is not rigorous. The third ansatz is about computation of $\lim_{N\to\infty}(1/N)\log \mathbb{E}[Z(G)^n]$. Here, the ansatz called *replica symmetry assumption* is used. The free energies $\lim_{N\to\infty}(1/N)\log \mathbb{E}[Z(G)]$ and $\lim_{N\to\infty}(1/N)\mathbb{E}[\log Z(G)]$ are called *annealed free energy* and *quenched free energy*, respectively.

## 6.2 Random sparse factor graph

### 6.2.1 Annealed free energy

In this section, we deal with random regular sparse factor graph. Let $l$ and $r$ denote degrees of variable and factor nodes. Connection of edges is chosen uniformly from all $(Nl)!$ connections. For simplicity, it is assumed that $f_a(\boldsymbol{x})$ does not depend on a factor node $a \in F$ and is denoted by $f(\boldsymbol{x})$. Similarly, it is also assumed that $f_i(x_i)$ does not depend on a variable node $i \in V$ and is denoted by $h(x_i)$. The basic idea of the calculation is type classification of $\boldsymbol{x} \in \mathcal{X}^N$ as in Chapter 4. Let $v(x)$ denote the number of variable nodes of value $x \in \mathcal{X}$. Let $u(\boldsymbol{x})$ denote the number of factor nodes connecting to $r$ variable nodes of value $\boldsymbol{x} \in \mathcal{X}^r$. Then, $(v(x)/N)_{x\in\mathcal{X}}$ and $(u(\boldsymbol{x})/((l/r)N))_{\boldsymbol{x}\in\mathcal{X}^r}$ are called a type of $\boldsymbol{x} \in \mathcal{X}^N$ for a factor graph $G$. Let $U_N(v/N, u/N; G)$ denote the number of assignments with type $v/N$ and $u/N$ on factor graph $G$. We can consider the classification according to the type of $\boldsymbol{x} \in \mathcal{X}^N$ in the partition function, namely,

$$Z(G) = \sum_{\boldsymbol{x}\in\mathcal{X}^N} \prod_{a\in F} f(\boldsymbol{x}_{\partial a}) \prod_{i\in V} h(x_i) = \sum_{v,u} U_N(v/N, u/N; G) \prod_{\boldsymbol{x}\in\mathcal{X}^r} f(\boldsymbol{x})^{u(\boldsymbol{x})} \prod_{x\in\mathcal{X}} h(x)^{v(x)}.$$

In the above equation, $v$ and $u$ must satisfy the condition for consistency

$$\sum_{\boldsymbol{z}\in\mathcal{X}^r} N_x(\boldsymbol{z}) u(\boldsymbol{z}) = l v(x) \tag{6.1}$$

where $N_x(\boldsymbol{z})$ denotes the number of occurrences of $x \in \mathcal{X}$ in $\boldsymbol{z} \in \mathcal{X}^r$. Both of the sides count the number of edges connected to variable nodes with value $z \in \mathcal{X}$. The expected number of assignments with type $v$ and $u$ is

$$\mathbb{E}[U_N(v/N, u/N; G)] = \binom{N}{(v(x))_{x\in\mathcal{X}}} \binom{\frac{l}{r}N}{(u(\boldsymbol{x}))_{\boldsymbol{x}\in\mathcal{X}^r}} \frac{\prod_{x\in\mathcal{X}} (v(x)l)!}{(Nl)!}.$$

Now, we consider the exponent of the contribution of types $\nu$ and $\mu$ where $\nu(x) := v(x)/N$ and $\mu(x) := u(x)/((l/r)N)$, respectively. From the Laplace method, it holds

$$\begin{aligned}\lim_{N\to\infty} \frac{1}{N} \log \mathbb{E}[Z(\nu, \mu)] &= \frac{l}{r}\mathcal{H}(\mu) - (l-1)\mathcal{H}(\nu) + \frac{l}{r} \sum_{\boldsymbol{x}\in\mathcal{X}^r} \mu(\boldsymbol{x}) \log f(\boldsymbol{x}) + \sum_{x\in\mathcal{X}} \nu(x) \log h(x) \\ &=: -F_{\mathrm{Bethe}}(\nu, \mu).\end{aligned}$$

Note that $F_{\mathrm{Bethe}}$ has a form similar to the Bethe free energy. It holds

$$\lim_{N\to\infty} \frac{1}{N} \log \mathbb{E}[Z(G)] = \max_{\nu,\mu} \left\{ -F_{\mathrm{Bethe}}(\nu, \mu) \right\}$$

where, $\nu$ and $\mu$ have to satisfy the following conditions.

$$\nu(x) \geq 0, \quad \forall x \in \mathcal{X}, \qquad \mu(\boldsymbol{x}) \geq 0, \quad \forall \boldsymbol{x} \in \mathcal{X}^r$$
$$\sum_{x \in \mathcal{X}} \nu(x) = 1, \qquad \sum_{\boldsymbol{x} \in \mathcal{X}^r} \mu(\boldsymbol{x}) = 1,$$

$$\frac{1}{r} \sum_{t=1}^{r} \sum_{\substack{\boldsymbol{z} \in \mathcal{X}^r \\ z_t = x}} \mu(\boldsymbol{z}) = \nu(x), \quad \forall z \in \mathcal{X}.$$

The last condition comes from (6.1). The above maximization problem is similar to the minimization problem of Bethe free energy in Definition 2.14. More precisely, $F_{\text{Bethe}}$ is the Bethe free energy of a small factor graph divided by $r$ which is the complete bipartite factor graph including $r$ variable nodes and $l$ factor nodes, in which all pseudo-marginals for variable nodes and factor nodes do not depend on indices of variable nodes and indices of factor nodes, respectively.

In the same way, the exponent of the $n$-th moment $\mathbb{E}[Z(G)^n]$ can be calculated for $n \in \mathbb{N}$ since $Z(G)^n$ can be regarded as a partition function of a factor graph on alphabet $\mathcal{X}^n$ and factors $\prod_{k=1}^n f(\boldsymbol{x}^{(i)})$ and $\prod_{k=1}^n h(x)$. Here, $\boldsymbol{x}^{(i)} \in \mathcal{X}^r$ denotes vector $(x_1^{(i)}, \dots, x_r^{(i)})$ where $\boldsymbol{x}_j$ is $j$-th elements of $\boldsymbol{x} \in (\mathcal{X}^n)^r$ and $x_j^{(i)}$ denotes $i$-th element of $\boldsymbol{x}_j \in \mathcal{X}^n$. For generality, from now on, it is assumed that $f_a(\boldsymbol{x})$ and $h_i(x)$ are i.i.d. drawings of random functions $f(\boldsymbol{x})$ and $h(x)$, respectively. Furthermore, the parameter of the inverse temperature $\beta$ is also introduced. Then, we now consider the $n$-th moment $\mathbb{E}[Z(G,\beta)^n]$ in which $\mathbb{E}[\cdot]$ denotes the expectation with respect to both connections of edges and random functions. Since the random functions are i.i.d., in the derivation $\prod_{k=1}^n f(\boldsymbol{x}^{(i)})$ and $\prod_{k=1}^n h(x)$ are simply replaced by $\left[\prod_{k=1}^n f(\boldsymbol{x}^{(i)})^\beta\right]_f$ and $\left[\prod_{k=1}^n h(x)^\beta\right]_h$, respectively where $[\cdot]_f$ and $[\cdot]_h$ denote the expectations for $f$ and $h$, respectively. Similarly to the derivation of (2.10) and Definition 2.16, the following theorem is obtained.

**Theorem 6.1** ([Mori, 2011])**.**

$$\lim_{N\to\infty} \frac{1}{N} \log \mathbb{E}[Z(G,\beta)^n] = \max_{(m_{\mathrm{v}\to\mathrm{f}}(\boldsymbol{x}), m_{\mathrm{f}\to\mathrm{v}}(\boldsymbol{x})) \in \mathcal{S}} \left\{ \frac{l}{r} \log Z_{\mathrm{f}} + \log Z_{\mathrm{v}} - l \log Z_{\mathrm{e}} \right\}.$$

*where $\mathcal{S}$ denotes the set of saddle points of the function for which the maximization is taken and where*

$$Z_{\mathrm{f}} := \sum_{\boldsymbol{x} \in (\mathcal{X}^n)^r} \left[\prod_{k=1}^n f(\boldsymbol{x}^{(k)})^\beta\right]_f \prod_{j=1}^r m_{\mathrm{v}\to\mathrm{f}}(\boldsymbol{x}_j), \quad Z_{\mathrm{v}} := \sum_{\boldsymbol{x} \in \mathcal{X}^n} \left[\prod_{k=1}^n h(x^{(k)})^\beta\right]_h m_{\mathrm{f}\to\mathrm{v}}(\boldsymbol{x})^l$$
$$Z_{\mathrm{e}} := \sum_{\boldsymbol{x} \in \mathcal{X}^n} m_{\mathrm{f}\to\mathrm{v}}(\boldsymbol{x}) m_{\mathrm{v}\to\mathrm{f}}(\boldsymbol{x}).$$

*The conditions of saddle point are*

$$m_{\mathrm{v}\to\mathrm{f}}(\boldsymbol{x}) \propto \left[\prod_{k=1}^{n} h(x^{(k)})^{\beta}\right]_h m_{\mathrm{f}\to\mathrm{v}}(\boldsymbol{x})^{l-1} \tag{6.2}$$

$$m_{\mathrm{f}\to\mathrm{v}}(\boldsymbol{x}) \propto \sum_{t=1}^{r} \sum_{\substack{z\in\mathcal{X}^r \\ z_t=\boldsymbol{x}}} \left[\prod_{k=1}^{n} f(\boldsymbol{z}^{(k)})^{\beta}\right]_f \prod_{j=1, j\neq t}^{r} m_{\mathrm{v}\to\mathrm{f}}(\boldsymbol{z}_j). \tag{6.3}$$

The equations (6.2), (6.3) are the BP equation for the small factor graph. From the proof of Theorem 6.1, we can easily understand why BP equation appears in the calculation of the exponent of moments since the problem is formulated as an analogue of the minimization of Bethe free energy. This result can be generalized straightforwardly for irregular factor graphs [Mori, 2011].

### 6.2.2 Replica symmetric free energy

From Theorem 6.1, $\lim_{N\to\infty}(1/N)\log\mathbb{E}[Z(G,\beta)^n]$ can be computed for $n\in\mathbb{N}$. For $n\in(0,1)$ (or more generally $n\in\mathbb{C}$), the replica symmetric (RS) assumption, $k$-step replica symmetry breaking (RSB) assumption or full-step replica symmetry breaking assumption is introduced. Replica symmetric assumption is the simplest assumption in which solutions $m_{\mathrm{f}\to\mathrm{v}}(x^{(1)},\dots,x^{(n)})$ and $m_{\mathrm{v}\to\mathrm{f}}(x^{(1)},\dots,x^{(n)})$ of the maximization problem are invariant under permutations on $(x^{(1)},\dots,x^{(n)})$. In that case, there are the representations

$$\begin{aligned} m_{\mathrm{v}\to\mathrm{f}}(\boldsymbol{x}) &= \int \mathrm{d}\Psi(M_{\mathrm{v}\to\mathrm{f}}) \prod_{k=1}^{n} M_{\mathrm{v}\to\mathrm{f}}(x^{(k)}) \\ m_{\mathrm{f}\to\mathrm{v}}(\boldsymbol{x}) &= \int \mathrm{d}\hat{\Psi}(M_{\mathrm{f}\to\mathrm{v}}) \prod_{k=1}^{n} M_{\mathrm{f}\to\mathrm{v}}(x^{(k)}) \end{aligned} \tag{6.4}$$

where $\Psi$ and $\hat{\Psi}$ denote probability measures on $\mathcal{P}(\mathcal{X})$, i.e., $\Psi$ and $\hat{\Psi}$ are elements of $\mathcal{P}(\mathcal{P}(\mathcal{X}))$ [Mottishaw and de Dominicis, 1987], [Wong and Sherrington, 1988]. By substituting them to Theorem 6.1, one obtains the following definition of $\phi_{\mathrm{RS}}(n,\beta)$ for $(1/n)\lim_{N\to\infty}(1/N)\log\mathbb{E}[Z(G,\beta)^n]$ on the RS assumption. Let $\langle\cdot\rangle$ be the expectation with respect to $\Psi$ or $\hat{\Psi}$. Let $\mathrm{extr}_{\Psi,\hat{\Psi}}\{\}$ denote an appropriately chosen extremal point. In this thesis, criterion of choice of extremal point is not discussed.

**Definition 6.2.** For $n\in\mathbb{C}$,

$$\phi_{\mathrm{RS}}(n,\beta) := \frac{1}{\beta}\operatorname*{extr}_{\Psi,\hat{\Psi}}\left\{\frac{1}{n}\left(\frac{l}{r}\log\langle[\mathcal{Z}_{\mathrm{f}}^n]_f\rangle + \log\langle[\mathcal{Z}_{\mathrm{v}}^n]_h\rangle - l\log\langle\mathcal{Z}_{\mathrm{e}}^n\rangle\right)\right\}$$

where

$$\mathcal{Z}_{\mathrm{f}} := \sum_{\boldsymbol{x}\in\mathcal{X}^r} f(\boldsymbol{x})^\beta \prod_{k=1}^{r} M^{(k)}_{\mathrm{v}\to\mathrm{f}}(x_k), \qquad \mathcal{Z}_{\mathrm{v}} := \sum_{x\in\mathcal{X}} h(x)^\beta \prod_{k=1}^{l} M^{(k)}_{\mathrm{f}\to\mathrm{v}}(x)$$

$$\mathcal{Z}_{\mathrm{e}} := \sum_{x\in\mathcal{X}} M_{\mathrm{v}\to\mathrm{f}}(x) M_{\mathrm{f}\to\mathrm{v}}(x).$$

Here, $(M^{(k)}_{\mathrm{v}\to\mathrm{f}})_{k=1}^{l}$ and $(M^{(k)}_{\mathrm{f}\to\mathrm{v}})_{k=1}^{r}$ are i.i.d. random messages obeying $\Psi$ and $\hat{\Psi}$, respectively. The saddle point equations are

$$\psi(m_{\mathrm{v}\to\mathrm{f}}) = \frac{\langle \mathcal{Z}_{\mathrm{e}}^n \rangle}{\langle [\mathcal{Z}_{\mathrm{v}}^n]_h \rangle} \Bigg\langle \Bigg[ \left( \sum_{x\in\mathcal{X}} h(x)^\beta \prod_{k=1}^{l-1} M^{(k)}_{\mathrm{f}\to\mathrm{v}}(x) \right)^n \cdot \delta\left( m_{\mathrm{v}\to\mathrm{f}}, \frac{h(x)^\beta \prod_{k=1}^{l-1} M^{(k)}_{\mathrm{f}\to\mathrm{v}}(x)}{\sum_{x\in\mathcal{X}} h(x)^\beta \prod_{k=1}^{l-1} M^{(k)}_{\mathrm{f}\to\mathrm{v}}(x)} \right) \Bigg]_h \Bigg\rangle \tag{6.5}$$

$$\hat{\psi}(m_{\mathrm{f}\to\mathrm{v}}) = \frac{\langle \mathcal{Z}_{\mathrm{e}}^n \rangle}{\langle [\mathcal{Z}_{\mathrm{f}}^n]_f \rangle} \frac{1}{r} \sum_{t=1}^{r} \Bigg\langle \Bigg[ \left( \sum_{\boldsymbol{x}\in\mathcal{X}^r} f(\boldsymbol{x})^\beta \prod_{k=1,k\neq t}^{r} M^{(k)}_{\mathrm{v}\to\mathrm{f}}(x_k) \right)^n \cdot \delta\left( m_{\mathrm{f}\to\mathrm{v}}, \frac{\sum_{\boldsymbol{x}\in\mathcal{X}^r, x_t=x} f(\boldsymbol{x})^\beta \prod_{k=1,k\neq t}^{r} M^{(k)}_{\mathrm{v}\to\mathrm{f}}(x_k)}{\sum_{\boldsymbol{x}\in\mathcal{X}^r} f(\boldsymbol{x})^\beta \prod_{k=1,k\neq t}^{r} M^{(k)}_{\mathrm{v}\to\mathrm{f}}(x_k)} \right) \Bigg]_f \Bigg\rangle. \tag{6.6}$$

Now, one can substitute any $n \in \mathbb{C}$ to $\phi_{\mathrm{RS}}(n,\beta)$. By letting $n \to 0$, one obtains the following definition.

**Definition 6.3.**

$$\phi_{\mathrm{RS}}(\beta) := \lim_{n\to 0} \phi_{\mathrm{RS}}(n,\beta) = \frac{1}{\beta} \operatorname*{extr}_{\Phi,\hat{\Phi}} \left\{ \frac{l}{r} \langle [\log \mathcal{Z}_{\mathrm{f}}]_f \rangle + \langle [\log \mathcal{Z}_{\mathrm{v}}]_h \rangle - l \langle \log \mathcal{Z}_{\mathrm{e}} \rangle \right\}$$

The saddle point conditions are

$$\psi(m_{\mathrm{v}\to\mathrm{f}}) = \left\langle \left[ \delta\left( m_{\mathrm{v}\to\mathrm{f}}, \frac{h(x)^\beta \prod_{k=1}^{l-1} M^{(k)}_{\mathrm{f}\to\mathrm{v}}(x)}{\sum_{x\in\mathcal{X}} h(x)^\beta \prod_{k=1}^{l-1} M^{(k)}_{\mathrm{f}\to\mathrm{v}}(x)} \right) \right]_h \right\rangle \tag{6.7}$$

$$\hat{\psi}(m_{\mathrm{f}\to\mathrm{v}}) = \frac{1}{r} \sum_{t=1}^{r} \left\langle \left[ \delta\left( m_{\mathrm{f}\to\mathrm{v}}, \frac{\sum_{\boldsymbol{x}\in\mathcal{X}^r, x_t=x} f(\boldsymbol{x})^\beta \prod_{k=1,k\neq k}^{r} M^{(k)}_{\mathrm{v}\to\mathrm{f}}(x_k)}{\sum_{\boldsymbol{x}\in\mathcal{X}^r} f(\boldsymbol{x})^\beta \prod_{k=1,k\neq t}^{r} M^{(k)}_{\mathrm{v}\to\mathrm{f}}(x_k)} \right) \right]_f \right\rangle. \tag{6.8}$$

This derivation of the RS solution is simpler than previously known ones in which complicated tools are used [Condamin, 2002], [Montanari, 2001] e.g., integral expression of the delta function. Another advantage of this proof is that we can understand why the saddle point equation in the RS solution is equal to the density evolution equation. Note that the RS assumption is correct for many problems [Montanari, 2001], [Tanaka, 2002].

### 6.2.3 One-step replica symmetry breaking free energy

The RS assumption is not generally correct. The next simplest assumption is the one-step replica symmetry breaking (1RSB) assumption. Let $m \in \mathbb{N}$ be a divisor of $n$. On the 1RSB assumption, $n$ variables $x^{(1)}, \dots, x^{(n)}$ is classified to $n/m$ groups each of which includes $m$ variables. Then, it is assumed that $m_{\mathrm{v}\to\mathrm{f}}(x^{(1)}, \dots, x^{(n)})$ and $m_{\mathrm{f}\to\mathrm{v}}(x^{(1)}, \dots, x^{(n)})$ are invariant under any permutation among $m$ variables in each group and under any group-wise permutation for $n/m$ groups. In that case, there are the representations

$$m_{\mathrm{v}\to\mathrm{f}}(\boldsymbol{x}) = \int \mathrm{d}\Psi_1(\Psi_0) \prod_{k_1=1}^{n/m} \int \mathrm{d}\Psi_0(M_{\mathrm{v}\to\mathrm{f}}) \prod_{k_2=1}^{m} M_{\mathrm{v}\to\mathrm{f}}(x^{(k_1)(k_2)})$$

$$m_{\mathrm{f}\to\mathrm{v}}(\boldsymbol{x}) = \int \mathrm{d}\hat{\Psi}_1(\hat{\Psi}_0) \prod_{k_1=1}^{n/m} \int \mathrm{d}\hat{\Psi}_0(M_{\mathrm{f}\to\mathrm{v}}) \prod_{k_2=1}^{m} M_{\mathrm{f}\to\mathrm{v}}(x^{(k_1)(k_2)})$$

where $\Psi_1$ and $\hat{\Psi}_1$ are elements of $\mathcal{P}(\mathcal{P}(\mathcal{P}(\mathcal{X})))$. Let $\langle\cdot\rangle_0$ be expectations with respect to $\Psi_0$ or $\hat{\Psi}_0$ and $\langle\cdot\rangle_1$ be expectations with respect to $\Psi_1$ or $\hat{\Psi}_1$.

**Definition 6.4.**

$$\begin{aligned}\phi_{\mathrm{1RSB}}(n, m, \beta) :=& \frac{1}{\beta} \operatorname*{extr}_{\Psi_1, \hat{\Psi}_1} \left\{ \frac{1}{n}\left( \frac{l}{r} \log \left\langle \left[ \langle \mathcal{Z}_{\mathrm{f}}^m \rangle_0^{\frac{n}{m}} \right]_f \right\rangle_1 + \log \left\langle \left[ \langle \mathcal{Z}_{\mathrm{v}}^m \rangle_0^{\frac{n}{m}} \right]_h \right\rangle_1 \right.\right. \\ &\left.\left. - l \log \left\langle \langle \mathcal{Z}_{\mathrm{e}}^m \rangle_0^{\frac{n}{m}} \right\rangle_1 \right) \right\}.\end{aligned}$$

On the 1RSB assumption, one has to take infimum for $m \in (0, 1]$ after taking $n \to 0$.

**Definition 6.5.**

$$\begin{aligned}\phi_{\mathrm{1RSB}}(\beta) :=& \inf_{m\in(0,1]} \lim_{n\to 0} \phi_{\mathrm{1RSB}}(n, m, \beta) \\ =& \frac{1}{\beta} \inf_{m\in(0,1]} \operatorname*{extr}_{\Psi_1, \hat{\Psi}_1} \left\{ \frac{1}{m}\left( \frac{l}{r} \langle [\log \langle \mathcal{Z}_{\mathrm{f}}^m \rangle_0]_f \rangle_1 + \langle \log \langle \mathcal{Z}_{\mathrm{v}}^m \rangle_0 \rangle_1 - l \langle \log \langle \mathcal{Z}_{\mathrm{e}}^m \rangle_0 \rangle_1 \right) \right\}.\end{aligned}$$

The saddle point equations are

$$\begin{aligned}\psi_1(\Psi_0) =& \left\langle \left[ \delta\left( \psi_0, \frac{1}{\left\langle \left( \sum_{x\in\mathcal{X}} h(x)^\beta \prod_{i=1}^{l-1} M_{\mathrm{f}\to\mathrm{v}}^{(i)}(x) \right)^m \right\rangle_0} \right.\right.\right. \\ &\left.\left.\left. \cdot \left\langle \left( \sum_{x\in\mathcal{X}} h(x)^\beta \prod_{i=1}^{l-1} M_{\mathrm{f}\to\mathrm{v}}^{(i)}(x) \right)^m \delta\left( m_{\mathrm{v}\to\mathrm{f}}, \frac{h(x)^\beta \prod_{i=1}^{l-1} M_{\mathrm{f}\to\mathrm{v}}^{(i)}(x)}{\sum_{x\in\mathcal{X}} h(x)^\beta \prod_{i=1}^{l-1} M_{\mathrm{f}\to\mathrm{v}}^{(i)}(x)} \right) \right\rangle_0 \right) \right]_h \right\rangle_1 \end{aligned} \tag{6.9}$$

$$\hat{\psi}_1(\hat{\Psi}_0) = \frac{1}{r}\sum_{k=1}^{r}\left\langle\left[\delta\left(\hat{\psi}_0, \frac{1}{\left\langle\left(\sum_{x\in\mathcal{X}^r} f(x)\prod_{i=1,i\neq k}^{r} M^{(i)}_{\mathrm{v}\to\mathrm{f}}(x_i)\right)^m\right\rangle_0}\right.\right.\right.$$
$$\cdot\left\langle\left(\sum_{x\in\mathcal{X}^r} f(x)\prod_{i=1,i\neq k}^{r} M^{(i)}_{\mathrm{v}\to\mathrm{f}}(x_i)\right)^m\right.$$
$$\left.\left.\left.\cdot\,\delta\left(m_{\mathrm{f}\to\mathrm{v}}, \frac{\sum_{x\in\mathcal{X}^r\setminus x_k} f(x)\prod_{i=1,i\neq k}^{r} M^{(i)}_{\mathrm{v}\to\mathrm{f}}(x_i)}{\sum_{x\in\mathcal{X}^r} f(x)\prod_{i=1,i\neq k}^{r} M^{(i)}_{\mathrm{v}\to\mathrm{f}}(x_i)}\right)\right\rangle_0\right)\right]_f\right\rangle. \tag{6.10}$$

When $m = 0$ or $m = 1$, the above equations can be further simplified [Montanari et al., 2008] [Mézard and Montanari, 2009]. The above results about regular random factor graphs can be generalized to irregular random factor graphs [Mori, 2011].

### 6.2.4 Trivial solutions of the saddle point equations

#### 6.2.4.1 Fixed points for the replica symmetry free energy

In this section, two types of trivial solutions of (6.5) and (6.6) are considered. Let $\phi_A(\beta) := \lim_{N\to\infty}(1/N)\log\mathbb{E}[Z(G,\beta)]$. In the derivation of $\phi_A(\beta)$, let $m^{*(\beta)}_{\mathrm{v}\to\mathrm{f}}$ and $m^{*(\beta)}_{\mathrm{f}\to\mathrm{v}}$ be the solution of (6.2) and (6.3) for $n = 1$ and inverse temperature $\beta$. Assume that for any $t = 1,\dots,r$ and $(x_k \in \mathcal{X})_{k\in\{1,2,\dots,r\}\setminus\{t\}}$, it holds $|\{x_t \in \mathcal{X} \mid f(x) > 0\}| \le 1$. The condition is called *hard constraints* in [Martin et al., 2004], [Martin et al., 2005]. In this case, there is a solution $(\psi, \hat{\psi})$ of (6.5) and (6.6) whose support is restricted to deterministic messages. More precisely, there is a solution $\psi(\delta_x) = m^{*(n\beta)}_{\mathrm{v}\to\mathrm{f}}(x)$ and $\hat{\psi}(\delta_x) = m^{*(n\beta)}_{\mathrm{f}\to\mathrm{v}}(x)$ where $\delta_x$ is the deterministic message for $x \in \mathcal{X}$. If the extremization can be regarded as maximization, $\phi_{\mathrm{RS}}(n,\beta) = \phi_{\mathrm{A}}(n\beta)$.

If $f(x)$ is deterministic and invariant under permutations on $x \in \mathcal{X}^r$, there is another type of solutions for (6.5) and (6.6), namely $\psi(m_{\mathrm{v}\to\mathrm{f}}) = \delta(m^{*(\beta)}_{\mathrm{v}\to\mathrm{f}}, m_{\mathrm{v}\to\mathrm{f}})$ and $\hat{\psi}(m_{\mathrm{f}\to\mathrm{v}}) = \delta(m^{*(\beta)}_{\mathrm{f}\to\mathrm{v}}, m_{\mathrm{f}\to\mathrm{v}})$. For this solution, $\phi_{\mathrm{RS}}(n,\beta) = \phi_{\mathrm{A}}(\beta)$ for any $n \in \mathbb{C}$. This result is well known for regular LDPC codes [Condamin, 2002].

#### 6.2.4.2 Fixed points for the one-step replica symmetry breaking free energy

Similarly to the previous section, in this section, three types of trivial solutions for 1RSB free energy in Definition 6.4 are considered. Let $(\psi^{*(\beta)}, \hat{\psi}^{*(\beta)})$ be the RS solution (6.7) and (6.8). There exists a trivial solution $\psi_1(\delta_{m_{\mathrm{f}\to\mathrm{v}}}) = \psi^{*(\beta)}(m_{\mathrm{f}\to\mathrm{v}})$, $\hat{\psi}_1(\delta_{m_{\mathrm{v}\to\mathrm{f}}}) = \hat{\psi}^{*(\beta)}(m_{\mathrm{v}\to\mathrm{f}})$. In this case, $\phi_{\mathrm{1RSB}}(n,m,\beta) = \phi_{\mathrm{RS}}(n,\beta)$ for any $m \in (0,1]$.

Assume that $f(x)$ is a hard constraint. There exists a trivial fixed point $\Psi_1(\Psi_0) = \Psi_0^{*(m\beta)}(m_{\mathrm{v}\to\mathrm{f}})$, $\hat{\Psi}_1(\hat{\Psi}_0) = \hat{\Psi}_0^{*(m\beta)}(m_{\mathrm{f}\to\mathrm{v}})$ where $\Psi_0(\delta_x) = m_{\mathrm{v}\to\mathrm{f}}(x)$ and $\hat{\Psi}_0(\delta_x) = m_{\mathrm{f}\to\mathrm{v}}(x)$.

This type of solutions is called a *frozen solution* [Krauth and Mézard, 1989], [Montanari, 2001], [Martin et al., 2004]. For this type of solutions, it holds $\phi_{1\mathrm{RSB}}(n, m, \beta) = \phi_{\mathrm{RS}}(n/m, m\beta)$. Hence, if this type of solution is appropriate, one has to take extremal value of the RS free energy with respect to the inverse temperature.

If $f(\boldsymbol{x})$ is deterministic and invariant under permutation on $\boldsymbol{x} \in \mathcal{X}^r$, there is a solution $\psi_1(\psi_0) = \delta(\psi_0^{*(\beta)}, \psi_0)$, $\hat{\psi}_1(\hat{\psi}_0) = \delta(\hat{\psi}_0^{*(\beta)}, \hat{\psi}_0)$. This type of solution is called a *factorized solution* [Wong and Sherrington, 1988], [Franz et al., 2001], [Nakajima and Hukushima, 2009]. For this type of solutions, it holds $\phi_{1\mathrm{RSB}}(n, m, \beta) = \phi_{\mathrm{RS}}(m, \beta)$. Hence, if this type of solution is appropriate, one has to take extremal value of the RS free energy with respect to the replica number.

## 6.3 Poisson model

In this section, another type of random factor graph ensemble is introduced, which is called *Poisson model*. There are $N$ variable nodes and $\alpha N$ factor nodes. The degree of factor nodes is $p$. For each factor node, a list of variable nodes $(x_1, \dots, x_p)$ is chosen independently and uniformly from $N(N-1)\cdots(N-(p-1))$ choices. Degrees of variable nodes are distributed according to a Poisson distribution under this rule, and hence the name Poisson model. For a type of variable nodes $v(x)$ and a type of factor nodes $u(\boldsymbol{x})$, one obtains

$$\mathbb{E}[U_N(v/N, u/N; G)] = \binom{N}{\left(v(x)\right)_{x\in\mathcal{X}}} \binom{\alpha N}{\left(u(\boldsymbol{x})\right)_{\boldsymbol{x}\in\mathcal{X}^p}} \cdot \prod_{\boldsymbol{x}\in\mathcal{X}^p} \left( \frac{\prod_{x\in\mathcal{X}} v(x)(v(x)-1)\cdots(v(x)-(N_x(\boldsymbol{x})-1))}{N(N-1)\cdots(N-(p-1))} \right)^{u(\boldsymbol{x})}$$

where $N_x(\boldsymbol{x})$ denotes the number of occurrences of $x$ in $\boldsymbol{x}$. Hence, one obtains

$$\begin{aligned}
&\lim_{N\to\infty} \frac{1}{N} \log \mathbb{E}[Z(\nu,\mu)] \\
&= \alpha\mathcal{H}(\mu) + \mathcal{H}(\nu) + \alpha \sum_{\boldsymbol{x}\in\mathcal{X}^p} \mu(\boldsymbol{x}) \log\left(\prod_{k=1}^{p} \nu(x_k)\right) + \alpha \sum_{\boldsymbol{x}\in\mathcal{X}^p} \mu(\boldsymbol{x}) \log f(\boldsymbol{x}) \\
&= -\alpha \sum_{\boldsymbol{x}\in\mathcal{X}^p} \mu(\boldsymbol{x}) \log \frac{\mu(\boldsymbol{x})}{\prod_{k=1}^{p} \nu(x_k)} + \mathcal{H}(\nu) + \alpha \sum_{\boldsymbol{x}\in\mathcal{X}^p} \mu(\boldsymbol{x}) \log f(\boldsymbol{x}).
\end{aligned}$$

The Lagrangian is

$$L(\nu, \mu; \lambda, \rho) = -\alpha \sum_{\boldsymbol{x}\in\mathcal{X}^p} \mu(\boldsymbol{x}) \log \frac{\mu(\boldsymbol{x})}{\prod_{k=1}^{p} \nu(x_k)} + \mathcal{H}(\nu) + \alpha \sum_{\boldsymbol{x}\in\mathcal{X}^p} \mu(\boldsymbol{x}) \log f(\boldsymbol{x})$$

$$+\lambda\left(\sum_{x\in\mathcal{X}}\nu(x)-1\right)+\alpha\rho\left(\sum_{x\in\mathcal{X}^p}\mu(\boldsymbol{x})-1\right).$$

Since the partial derivatives are

$$\frac{\partial L(\nu,\mu;\lambda,\rho)}{\partial\mu(\boldsymbol{x})}=-\alpha(\log\mu(\boldsymbol{x})+1)+\alpha\log\prod_{k=1}^{p}\nu(x_k)+\alpha\log f(\boldsymbol{x})+\alpha\rho$$
$$\frac{\partial L(\nu,\mu;\lambda,\rho)}{\partial\nu(\boldsymbol{x})}=-(\log\nu(x)+1)+\alpha\sum_{k=1}^{p}\sum_{\substack{\boldsymbol{x}\in\mathcal{X}^p\\x_k=x}}\mu(\boldsymbol{x})\frac{1}{\nu(x)}+\lambda$$

the stationary conditions are

$$\mu(\boldsymbol{x})=f(\boldsymbol{x})\prod_{k=1}^{p}\nu(x_k)\exp\{-1+\rho\} \tag{6.11}$$
$$\alpha\sum_{k=1}^{p}\sum_{\substack{\boldsymbol{x}\in\mathcal{X}^p\\x_k=x}}\mu(\boldsymbol{x})=\nu(x)\log\nu(x)+(1-\lambda)\nu(x). \tag{6.12}$$

By taking summations for both of the sides, $\rho$ and $\lambda$ are determined uniquely as

$$\rho=1-\log\left(\sum_{\boldsymbol{x}\in\mathcal{X}^p}f(\boldsymbol{x})\prod_{k=1}^{p}\nu(x_k)\right),\qquad\lambda=1-\alpha p-\mathcal{H}(\nu).$$

By substituting (6.11) to (6.12),

$$\alpha\sum_{k=1}^{p}\sum_{\substack{\boldsymbol{x}\in\mathcal{X}^p\\x_k=x}}f(\boldsymbol{x})\prod_{j=1,j\neq k}^{p}\nu(x_j)\exp\{-1+\rho\}=\log\nu(x)+(1-\lambda)$$

By defining the variables,

$$m_{\mathrm{v}\to\mathrm{f}}(x):=\nu(x),\qquad m_{\mathrm{f}\to\mathrm{v}}(x):=\frac{1}{d}\big(\log\nu(x)+\alpha p+\mathcal{H}(\nu)\big)$$
$$d:=|\mathcal{X}|(\alpha p+\mathcal{H}(\nu))+\sum_{x\in\mathcal{X}}\log m_{\mathrm{v}\to\mathrm{f}}(x)$$

the stationary condition can be written as

$$m_{\mathrm{f}\to\mathrm{v}}(x)=\frac{\alpha}{dZ_{\mathrm{f}}}\sum_{k=1}^{p}\sum_{\substack{\boldsymbol{x}\in\mathcal{X}^p\\x_k=x}}f(\boldsymbol{x})\prod_{j=1,j\neq k}^{p}m_{\mathrm{v}\to\mathrm{f}}(x_j)$$
$$m_{\mathrm{v}\to\mathrm{f}}(x)=\frac{1}{Z_{\mathrm{v}}}\exp\{dm_{\mathrm{f}\to\mathrm{v}}(x)\}$$

where

$$Z_{\mathrm{f}}:=\sum_{x\in\mathcal{X}^p}f(\boldsymbol{x})\prod_{k=1}^{p}m_{\mathrm{v}\to\mathrm{f}}(x_k),\qquad Z_{\mathrm{v}}:=\sum_{x\in\mathcal{X}}\exp\{dm_{\mathrm{f}\to\mathrm{v}}(x)\}.$$

Here, $d$ can be regarded as the mean of the Poisson distribution expressing the degree distribution of variable nodes. Similarly to Theorem 6.1, the $n$-th moment is obtained in the following lemma.

**Lemma 6.6.**

$$\lim_{N\to\infty}\frac{1}{N}\log\mathbb{E}[Z^n] = \max_{(m_{\mathrm{f}\to\mathrm{v}}(x),m_{\mathrm{v}\to\mathrm{f}}(x),d)\in\mathcal{S}}\left\{\alpha\log Z_{\mathrm{f}}+\log Z_{\mathrm{v}}-dZ_{\mathrm{e}}\right\}$$

*where $\mathcal{S}$ denotes the set of saddle points of the function for which the maximization is taken, and where*

$$Z_{\mathrm{f}} := \sum_{\boldsymbol{x}\in(\mathcal{X}^n)^p}\left(\prod_{t=1}^{n}f(\boldsymbol{x}^{(t)})\right)\prod_{k=1}^{p}m_{\mathrm{v}\to\mathrm{f}}(\boldsymbol{x}_k),\qquad Z_{\mathrm{v}} := \sum_{\boldsymbol{x}\in\mathcal{X}^n}\exp\{dm_{\mathrm{f}\to\mathrm{v}}(\boldsymbol{x})\}$$

$$Z_{\mathrm{e}} := \sum_{\boldsymbol{x}\in\mathcal{X}^n}m_{\mathrm{v}\to\mathrm{f}}(\boldsymbol{x})m_{\mathrm{f}\to\mathrm{v}}(\boldsymbol{x}).$$

*The conditions of saddle point are*

$$m_{\mathrm{f}\to\mathrm{v}}(\boldsymbol{x}) = \frac{\alpha}{dZ_{\mathrm{f}}}\sum_{k=1}^{p}\sum_{\substack{\boldsymbol{x}\in(\mathcal{X}^n)^p\\ \boldsymbol{x}_k=\boldsymbol{x}}}\left(\prod_{t=1}^{n}f(\boldsymbol{x}^{(t)})\right)\prod_{j=1,j\neq k}^{p}m_{\mathrm{v}\to\mathrm{f}}(\boldsymbol{x}_j)$$

$$m_{\mathrm{v}\to\mathrm{f}}(\boldsymbol{x}) = \frac{1}{Z_{\mathrm{v}}}\exp\{dm_{\mathrm{f}\to\mathrm{v}}(\boldsymbol{x})\},\qquad d=\frac{\alpha p}{Z_{\mathrm{e}}}.$$

On the RS assumption (6.4), one obtains

$$\begin{aligned}Z_{\mathrm{v}} &= \sum_{\boldsymbol{x}\in\mathcal{X}^n}\left(1+dm_{\mathrm{f}\to\mathrm{v}}(\boldsymbol{x})+\frac{(dm_{\mathrm{f}\to\mathrm{v}}(\boldsymbol{x}))^2}{2!}+\cdots\right)\\ &= |\mathcal{X}|^n+d+d^2\frac{\left\langle\left(\sum_{x\in\mathcal{X}}\prod_{k=1}^{2}M_{\mathrm{f}\to\mathrm{v}}^{(k)}(x)\right)^n\right\rangle}{2!}+d^3\frac{\left\langle\left(\sum_{x\in\mathcal{X}}\prod_{k=1}^{3}M_{\mathrm{f}\to\mathrm{v}}^{(k)}(x)\right)^n\right\rangle}{3!}+\cdots\\ &= (\exp d)\left\langle\left\langle\left(\sum_{x\in\mathcal{X}}\prod_{k=1}^{D}M_{\mathrm{f}\to\mathrm{v}}^{(k)}(x)\right)^n\right\rangle\right\rangle_{D\sim\mathrm{Poisson}(d)}.\end{aligned}$$

In the following, we use a simple notation $\langle\cdot\rangle_{D\sim d}$ instead of $\langle\cdot\rangle_{D\sim\mathrm{Poisson}(d)}$ Then, the following lemma is obtained.

**Lemma 6.7.**

$$\phi_{\mathrm{RS}}(n) := \operatorname*{extr}_{(d,m_{\mathrm{f}\to\mathrm{v}},m_{\mathrm{v}\to\mathrm{f}})}\left\{\alpha\log\left\langle\mathcal{Z}_{\mathrm{f}}^n\right\rangle+\log\left\langle\left\langle\mathcal{Z}_{\mathrm{v}}^n\right\rangle\right\rangle_{D\sim d}-d\left(\left\langle\mathcal{Z}_{\mathrm{e}}^n\right\rangle-1\right)\right\}$$

*where*

$$\mathcal{Z}_{\mathrm{f}} = \sum_{\boldsymbol{x}\in\mathcal{X}^p}f(\boldsymbol{x})\prod_{k=1}^{p}M_{\mathrm{v}\to\mathrm{f}}(\boldsymbol{x}_k),\qquad \mathcal{Z}_{\mathrm{v}} = \sum_{x\in\mathcal{X}}\prod_{k=1}^{D}M_{\mathrm{f}\to\mathrm{v}}^{(k)}(x)$$

$$\mathcal{Z}_{\mathrm{e}} = \sum_{x\in\mathcal{X}}M_{\mathrm{f}\to\mathrm{v}}(x)M_{\mathrm{v}\to\mathrm{f}}(x).$$

# 6.4 $p$-spin model

## 6.4.1 The $p$-spin model

In this section, other examples of the replica method are introduced. The $p$-spin model is represented by a factor graph including $N$ variable nodes and $N(N-1)\cdots(N-(p-1))$ factor nodes. Each factor node connects to $p$ variable nodes $(x_1,\dots,x_p)$, in which order of variable nodes is distinguished. In the $p$-spin model, the factor function depends on $N$ in order to obtain finite normalized free energy $\lim_{N\to\infty}(1/N)\log Z_N$. The $p$-spin model is defined by

$$p_{p\text{-spin}}(\boldsymbol{x}) := \frac{1}{Z_N}\prod_{(i_1,\dots,i_p)} f_{N,(i_1,\dots,i_p)}(\boldsymbol{x}_{i_1,\dots,i_p})$$

$$Z_N := \sum_{\boldsymbol{x}\in\mathcal{X}^N}\prod_{(i_1,\dots,i_p)} f_N(\boldsymbol{x}_{i_1,\dots,i_p})$$

where $f_{N,(i_1,\dots,i_p)}$ is i.i.d. random function for all tuples $(i_1,\dots,i_p)$. For the $p$-spin model, $\mathbb{E}[Z_N]$ is determined only by the type of variable nodes. More precisely, it holds

$$\mathbb{E}[Z_N] = \sum_{v}\binom{N}{(v(x))_{x\in\mathcal{X}}}\prod_{\boldsymbol{x}\in\mathcal{X}^p} f_N(\boldsymbol{x})^{\prod_{x\in\mathcal{X}} v(x)(v(x)-1)\dots(v(x)-N_x(\boldsymbol{x})+1)}.$$

Let $\phi(\boldsymbol{x}) := \lim_{N\to\infty}(1/N)\log\mathbb{E}[f_N(\boldsymbol{x})]^{N^p}$. Then, it holds

$$\lim_{N\to\infty}\frac{1}{N}\log\mathbb{E}[Z_N] = \sup_{\nu\in\mathcal{P}(\mathcal{X})}\left\{\mathcal{H}(\nu) + \sum_{\boldsymbol{x}\in\mathcal{X}^p}\prod_{i=1}^{p}\nu(x_i)\phi(\boldsymbol{x})\right\}.$$

The Lagrangian for this problem is

$$L(\nu;\lambda) := \mathcal{H}(\nu) + \sum_{\boldsymbol{x}\in\mathcal{X}^p}\prod_{i=1}^{p}\nu(x_i)\phi(\boldsymbol{x}) + \lambda\left(\sum_{x\in\mathcal{X}}\nu(x) - 1\right).$$

The stationary condition is

$$\frac{\partial L(\nu;\lambda)}{\partial\nu(x)} = -\log\nu(x) - 1 + \sum_{t=1}^{p}\sum_{\boldsymbol{x}\in\mathcal{X}^p, x_t=x}\phi(\boldsymbol{x})\prod_{k\neq t}\nu(x_k) + \lambda = 0.$$

Hence, $\nu(x)\in\mathcal{P}(x)$ must satisfy

$$\nu(x) \propto \exp\left\{\sum_{i=1}^{p}\sum_{\boldsymbol{x}\in\mathcal{X}^p, x_i=x}\phi(\boldsymbol{x})\prod_{j\neq i}\nu(x_j)\right\}. \tag{6.13}$$

Let $m_{\mathrm{v}\to\mathrm{f}}(x) := \nu(x)$ and

$$
\begin{aligned}
m_{\mathrm{f}\to\mathrm{v}}(x) &:= \frac{1}{d}\sum_{i=1}^{p}\sum_{\boldsymbol{x}\in\mathcal{X}^p, x_i=x}\phi(\boldsymbol{x})\prod_{j\neq i} m_{\mathrm{v}\to\mathrm{f}}(x_j)\\
d &:= \sum_{i=1}^{p}\sum_{\boldsymbol{x}\in\mathcal{X}^p}\phi(\boldsymbol{x})\prod_{j\neq i} m_{\mathrm{v}\to\mathrm{f}}(x_j).
\end{aligned}
\tag{6.14}
$$

Then, (6.13) is rewritten as

$$
m_{\mathrm{v}\to\mathrm{f}}(x) \propto \exp\{d m_{\mathrm{f}\to\mathrm{v}}(x)\}. \tag{6.15}
$$

Similarly to Theorem 6.1, one obtains the following lemma.

**Lemma 6.8.** *Let $\phi_n(\boldsymbol{x}) := \lim_{N\to\infty}(1/N)\log\mathbb{E}\left[\prod_{k=1}^{n} f_N(\boldsymbol{x}^{(k)})\right]^{N^p}$ for $\boldsymbol{x}\in(\mathcal{X}^n)^p$.*

$$
\lim_{N\to\infty}\frac{1}{N}\log\mathbb{E}[Z_N^n] = \max_{(d, m_{\mathrm{f}\to\mathrm{v}}, m_{\mathrm{v}\to\mathrm{f}})\in\mathcal{S}}\left\{Z_{\mathrm{f}} + \log Z_{\mathrm{v}} - dZ_{\mathrm{e}}\right\}
$$

*where $\mathcal{S}$ denotes the set of saddle points of the function for which the maximization is taken and where*

$$
Z_{\mathrm{f}} = \sum_{\boldsymbol{x}\in(\mathcal{X}^n)^p}\phi_n(\boldsymbol{x})\prod_{k=1}^{p} m_{\mathrm{v}\to\mathrm{f}}(\boldsymbol{x}_k), \qquad Z_{\mathrm{v}} = \sum_{\boldsymbol{x}\in\mathcal{X}^n}\exp\left\{d m_{\mathrm{f}\to\mathrm{v}}(\boldsymbol{x})\right\}
$$

$$
Z_{\mathrm{e}} = \sum_{\boldsymbol{x}\in\mathcal{X}^n} m_{\mathrm{f}\to\mathrm{v}}(\boldsymbol{x}) m_{\mathrm{v}\to\mathrm{f}}(\boldsymbol{x}).
$$

*The stationary conditions are* (6.14) *and* (6.15).

Then, similarly to Lemma 6.7, the following general result for the $p$-spin model is obtained.

**Lemma 6.9.**

$$
\phi_{\mathrm{RS}}(n) := \operatorname*{extr}_{d, m_{\mathrm{f}\to\mathrm{v}}, m_{\mathrm{v}\to\mathrm{f}}}\left\{\left\langle\bar{\mathcal{Z}}_{\mathrm{f}}^n\right\rangle + \log\left\langle\left\langle\mathcal{Z}_{\mathrm{v}}^n\right\rangle\right\rangle_{D\sim d} - d\left(\left\langle\mathcal{Z}_{\mathrm{e}}^n\right\rangle - 1\right)\right\} \tag{6.16}
$$

*where*

$$
\bar{\mathcal{Z}}_{\mathrm{f}} = \sum_{\boldsymbol{x}\in\mathcal{X}^p}\phi_n(\boldsymbol{x})\prod_{k=1}^{p} M_{\mathrm{v}\to\mathrm{f}}(\boldsymbol{x}_k), \qquad \mathcal{Z}_{\mathrm{v}} = \sum_{x\in\mathcal{X}}\prod_{k=1}^{D} M_{\mathrm{f}\to\mathrm{v}}^{(k)}(x)
$$

$$
\mathcal{Z}_{\mathrm{e}} = \sum_{x\in\mathcal{X}} M_{\mathrm{f}\to\mathrm{v}}(x) M_{\mathrm{v}\to\mathrm{f}}(x).
$$

### 6.4.2 The Viana-Bray model

The Viana-Bray model is a diluted Ising $p$-spin model [Viana and Bray, 1985]. For each factor node, $f_{N,(i_1,\ldots,i_p)}(\boldsymbol{x}) = \exp\{J_{N,(i_1,\ldots,i_p)} \prod_{k=1}^p x_k\}$ where $J_{N,(i_1,\ldots,i_p)}$ is an i.i.d. random variable for all $p$-tuples obeying

$$J_N = \begin{cases} 0, & \text{with probability } 1 - \alpha/N^{p-1}. \\ J_0, & \text{with probability } \alpha/N^{p-1}. \end{cases}$$

Here, $J_0$ is some random variable independent of $N$. Then, it holds

$$\begin{aligned}
\phi_n(\boldsymbol{x}) &= \lim_{N\to\infty} N^{p-1} \log \left\langle \exp\left\{ J_N \sum_{k=1}^{n} \prod_{i=1}^{p} x_i^{(k)} \right\} \right\rangle_{J_N} \\
&= \lim_{N\to\infty} N^{p-1} \log \left( 1 - \frac{\alpha}{N^{p-1}} + \frac{\alpha}{N^{p-1}} \left\langle \exp\left\{ J_0 \sum_{k=1}^{n} \prod_{i=1}^{p} x_i^{(k)} \right\} \right\rangle_{J_0} \right) \\
&= \alpha \left( \left\langle \exp\left\{ J_0 \sum_{k=1}^{n} \prod_{i=1}^{p} x_i^{(k)} \right\} \right\rangle_{J_0} - 1 \right).
\end{aligned}$$

Hence,

$$\begin{aligned}
\langle \bar{\mathcal{Z}}_{\mathrm{f}}^n \rangle &= \alpha \left( \left\langle \left\langle \left( \sum_{x\in\mathcal{X}^p} \exp\left\{ J_0 \prod_{i=1}^{p} x_i \right\} \prod_{i=1}^{p} M_{\mathrm{v}\to\mathrm{f}}(x_i) \right)^n \right\rangle \right\rangle_{J_0} - 1 \right) \\
&=: \alpha \left( \left\langle \langle \mathcal{Z}_{\mathrm{f}}^n \rangle \right\rangle_{J_0} - 1 \right)
\end{aligned}$$

Then, for the Viana-Bray model, (6.16) is

$$\lim_{N\to\infty} \frac{1}{N} \log \mathbb{E}[Z_N^n] = \operatorname*{extr}_{d,\Psi,\hat{\Psi}} \left\{ \alpha \left( \langle \mathcal{Z}_{\mathrm{f}}^n \rangle - 1 \right) + \log \langle \langle \mathcal{Z}_{\mathrm{v}}^n \rangle \rangle_{D\sim d} - d \left( \langle \mathcal{Z}_{\mathrm{e}}^n \rangle - 1 \right) \right\}.$$

This result is similar to but different from the results for irregular random factor graphs

$$\lim_{N\to\infty} \frac{1}{N} \log \mathbb{E}[Z_N^n] = \operatorname*{extr}_{\Psi,\hat{\Psi}} \left\{ \alpha \log \langle \mathcal{Z}_{\mathrm{f}}^n \rangle + \langle \log \langle \mathcal{Z}_{\mathrm{v}}^n \rangle \rangle_{D\sim \alpha p} - \alpha p \log \langle \mathcal{Z}_{\mathrm{e}}^n \rangle \right\}$$

and the result for Poisson random factor graphs in Lemma 6.7

$$\lim_{N\to\infty} \frac{1}{N} \log \mathbb{E}[Z_N^n] = \operatorname*{extr}_{d,\Psi,\hat{\Psi}} \left\{ \alpha \log \langle \mathcal{Z}_{\mathrm{f}}^n \rangle + \log \langle \langle \mathcal{Z}_{\mathrm{v}}^n \rangle \rangle_{D\sim d} - d \left( \langle \mathcal{Z}_{\mathrm{e}}^n \rangle - 1 \right) \right\}.$$

In the limit $n \to 0$, however, the three models yield the same result.

$$\begin{aligned}
\lim_{n\to 0} \frac{1}{n} \lim_{N\to\infty} \frac{1}{N} \log \mathbb{E}[Z_N^n] &= \operatorname*{extr}_{d,\Psi,\hat{\Psi}} \left\{ \langle \langle \log \mathcal{Z}_{\mathrm{v}} \rangle \rangle_{D\sim d} + \alpha \langle \log \mathcal{Z}_{\mathrm{f}} \rangle - d \langle \log \mathcal{Z}_{\mathrm{e}} \rangle \right\} \\
&= \operatorname*{extr}_{\Psi,\hat{\Psi}} \left\{ \langle \langle \log \mathcal{Z}_{\mathrm{v}} \rangle \rangle_{D\sim \alpha p} + \alpha \langle \log \mathcal{Z}_{\mathrm{f}} \rangle - \alpha p \langle \log \mathcal{Z}_{\mathrm{e}} \rangle \right\}.
\end{aligned}$$

This derivation is clearer than that in [Monasson, 1998].

### 6.4.3 The TAP model

In this section, the replica method for the TAP model, which includes the SK model as a special case, is discussed.

**Definition 6.10** (TAP model)**.** The $p$-spin model satisfying

$$\lim_{N\to\infty} N^{p-1}\mathbb{E}[\log f_N(\boldsymbol{x})] = g_0(\boldsymbol{x})$$
$$\lim_{N\to\infty} N^{p-1}\mathbb{E}\left[\left|\log f_N(\boldsymbol{x}) - \mathbb{E}[\log f_N(\boldsymbol{x})]\right|^2\right] = g(\boldsymbol{x})$$
$$\lim_{N\to\infty} N^{p-1}\mathbb{E}\left[\left|\log f_N(\boldsymbol{x}) - \mathbb{E}[\log f_N(\boldsymbol{x})]\right|^k\right] = 0, \qquad \text{for } k = 3, 4, \dots .$$

is called a *TAP model.*

**Lemma 6.11.** *For the TAP model, it holds*

$$\lim_{N\to\infty} \mathbb{E}[f_N(\boldsymbol{x})]^{N^{p-1}} = \exp\left\{g_0(\boldsymbol{x}) + \frac{1}{2}g(\boldsymbol{x})\right\} = \mathbb{E}_{X(\boldsymbol{x})\sim N(g_0(\boldsymbol{x}),g(\boldsymbol{x}))}[\exp\{X(\boldsymbol{x})\}].$$

*for $\boldsymbol{x} \in \mathcal{X}^p$ where $N(a, b)$ denotes the normal distribution with mean $a$ and variance $b$.*

*Proof.*

$$\begin{aligned}\mathbb{E}[f_N(\boldsymbol{x})]^{N^{p-1}} &= \exp\left\{g_0(\boldsymbol{x})\right\}\mathbb{E}\left[\exp\left\{\log f_N(\boldsymbol{x}) - \frac{1}{N^{p-1}}g_0(\boldsymbol{x})\right\}\right]^{N^{p-1}} \\ &= \exp\left\{g_0(\boldsymbol{x})\right\}\left(1 + \frac{1}{2}\mathbb{E}\left[\left(\log f_N(\boldsymbol{x}) - \frac{1}{N^{p-1}}g_0(\boldsymbol{x})\right)^2\right] + o\left(\frac{1}{N^{p-1}}\right)\right)^{N^{p-1}} \\ &= \exp\left\{g_0(\boldsymbol{x}) + \frac{1}{2}g(\boldsymbol{x})\right\} + o(1).\end{aligned}$$

□

**Example 6.12.** For $\mathcal{X} = \{+1, -1\}$,

$$f_N(\boldsymbol{x}) = \exp\left\{\left(\frac{J_0}{N^{p-1}} + \frac{J}{\sqrt{N^{p-1}}}Y\right)\prod_{i=1}^{p} x_i\right\}$$

where $J_0$ and $J$ are constants and where $Y$ is a random variable which is independent of $N$ and has zero mean and unit variance. In this case, the Ising model is the TAP model with $g_0(\boldsymbol{x}) = J_0\prod_{i=1}^{p} x_i$, $g(\boldsymbol{x}) = J^2$.

**Counterexample 6.13.** For the $p$-spin Ising model,

$$f_N(\boldsymbol{x}) = \exp\left\{J_N\prod_{i=1}^{p} x_i\right\}.$$

When $J_N$ obeys Poisson$(\alpha/N^{p-1})$ and Gamma$(\theta, \alpha/N^{p-1})$

$$\lim_{N\to\infty} \mathbb{E}[f_N(\boldsymbol{x})]^{N^{p-1}} = \exp\left\{\alpha\left(\exp\left\{\prod_{i=1}^{p} x_i\right\} - 1\right)\right\} \qquad \text{and}$$

$$\lim_{N\to\infty} \mathbb{E}[f_N(\boldsymbol{x})]^{N^{p-1}} = \left(1-\theta\prod_{i=1}^{p} x_i\right)^{-\alpha}$$

respectively. These are not the TAP model. The former case is equivalent to the Viana-Bray model.

For the TAP model, it holds

$$\lim_{N\to\infty} N^{p-1}\mathbb{E}\left[\log\prod_{k=1}^{n} f_N(\boldsymbol{x}^{(k)})\right] = \sum_{k=1}^{n} g_0(\boldsymbol{x}^{(k)})$$

$$\begin{aligned}
&\lim_{N\to\infty} N^{p-1}\mathbb{E}\left[\left(\log\prod_{k=1}^{n} f_N(\boldsymbol{x}^{(k)}) - \sum_{k=1}^{n} g_0(\boldsymbol{x}^{(k)})\right)^2\right] \\
&= \sum_{k=1}^{n}\sum_{l=1}^{n} \mathbb{E}\left[\left(\log f_N(\boldsymbol{x}^{(k)}) - g_0(\boldsymbol{x}^{(k)})\right)\left(\log f_N(\boldsymbol{x}^{(l)}) - g_0(\boldsymbol{x}^{(l)})\right)\right] \\
&=: \sum_{k=1}^{n}\sum_{l=1}^{n} g_2\left(\boldsymbol{x}^{(k)}, \boldsymbol{x}^{(l)}\right).
\end{aligned}$$

Hence,

$$\begin{aligned}
\lim_{N\to\infty} \mathbb{E}\left[\prod_{k=1}^{n} f_N(\boldsymbol{x}^{(k)})\right]^{N^{p-1}} &= \exp\left\{\sum_{k=1}^{n} g_0(\boldsymbol{x}^{(k)}) + \sum_{k=1}^{n}\sum_{l=1}^{n}\frac{1}{2}g(\boldsymbol{x}^{(k)}, \boldsymbol{x}^{(l)})\right\} \\
&= \mathbb{E}_{[X(\boldsymbol{x})]\sim N(g_0,g_2)}\left[\exp\left\{\sum_{k=1}^{n} X(\boldsymbol{x}^{(k)})\right\}\right].
\end{aligned}$$

For the TAP model, it holds on the RS assumption that

$$\begin{aligned}
Z_{\mathrm{f}} &= \sum_{\boldsymbol{x}\in(\mathcal{X}^n)^p}\prod_{j=1}^{p} m_{\mathrm{v}\to\mathrm{f}}(\boldsymbol{x}_j)\log\mathbb{E}\left[\prod_{k=1}^{n} f_N(\boldsymbol{x}^{(k)})\right]^{N^{p-1}} \\
&= \lim_{y\to 0}\frac{1}{y}\log\left(\sum_{\boldsymbol{x}\in(\mathcal{X}^n)^p}\prod_{j=1}^{p} m_{\mathrm{v}\to\mathrm{f}}(\boldsymbol{x}_j)\mathbb{E}\left[\prod_{k=1}^{n} f_N(\boldsymbol{x}^{(k)})\right]^{yN^{p-1}}\right) \\
&= \lim_{y\to 0}\frac{1}{y}\log\left(\sum_{\boldsymbol{x}\in(\mathcal{X}^n)^p}\prod_{j=1}^{p} m_{\mathrm{v}\to\mathrm{f}}(\boldsymbol{x}_j)\mathbb{E}_{[X(\boldsymbol{x})]\sim N(g_0,g_2)}\left[\exp\left\{\sum_{k=1}^{n} X(\boldsymbol{x}^{(k)})\right\}\right]^{y}\right) \\
&= \lim_{y\to 0}\frac{1}{y}\log\left(\sum_{\boldsymbol{x}\in(\mathcal{X}^n)^p}\prod_{j=1}^{p} m_{\mathrm{v}\to\mathrm{f}}(\boldsymbol{x}_j)\mathbb{E}_{[X(\boldsymbol{x})]\sim N(yg_0,yg_2)}\left[\exp\left\{\sum_{k=1}^{n} X(\boldsymbol{x}^{(k)})\right\}\right]\right)
\end{aligned}$$

$$
\begin{aligned}
&= \lim_{y\to 0} \frac{1}{y} \log \left\langle \mathbb{E}_{[X(\boldsymbol{x})]\sim N(yg_0, yg_2)} \left[ \left( \sum_{\boldsymbol{x}\in\mathcal{X}^p} \prod_{j=1}^{p} M_{\mathrm{v}\to\mathrm{f}}(x_j) \exp\{X(\boldsymbol{x})\} \right)^n \right] \right\rangle \\
&=: \lim_{y\to 0} \frac{1}{y} \log \left\langle \mathbb{E}_y \left[ \mathcal{Z}_{\mathrm{f}}^n \right] \right\rangle .
\end{aligned}
$$

Hence, it holds

$$
\lim_{N\to\infty} \frac{1}{N} \log \mathbb{E}[Z_N^n] = \lim_{y\to 0} \operatorname*{extr}_{d,\Psi,\hat{\Psi}} \left\{ \frac{1}{y} \log\langle \mathbb{E}_y[\mathcal{Z}_{\mathrm{f}}^n]\rangle + \log\langle\langle \mathcal{Z}_{\mathrm{v}}^n\rangle\rangle_{D\sim d} - d\left( \left\langle \mathcal{Z}_{\mathrm{e}}^n \right\rangle - 1 \right) \right\}
$$

if one can exchange the order of the extremization and the limit $y \to 0$. This result can be regarded as the dense limit of the free energy of the Viana-Bray or Poisson model. This relationship is rigorously proved in [Guerra and Toninelli, 2004].

# 6.5 Markov model

## 6.5.1 The method of types for Markov chain

In this section, the method of types for Markov chain is introduced. Let $M_{xy}(\boldsymbol{x})$ be the number of pairs $(x_i, x_{i+1}) = (x, y) \in \mathcal{X}^2$ for $i = 1, \dots, N-1$. The second-order type $P_{\boldsymbol{x}}$ of a sequence $\boldsymbol{x} \in \mathcal{X}^N$ is defined as the empirical distribution of a pair, i.e., $P_{\boldsymbol{x}} = \left( M_{xy}(\boldsymbol{x})/(N-1) \right)$. Let $\mathcal{P}_N^{(2)}$ be the number of second-order types of length $N$. Let $U_N(x_1, P_{X,Y})$ be the number of $N$ length sequence of second-order type $P_{X,Y} \in \mathcal{P}_N^{(2)}$ with the first element $x_1 \in \mathcal{X}$. In [Whittle, 1955], [Billingsley, 1961], it is shown that

$$
U_N(x_1, P_{X,Y}) = F^{x_1,x_n}(P_{X,Y}) \prod_{x\in\mathcal{X}} \binom{NP_X(x)}{(NP_{X,Y}(x,y))_{y\in\mathcal{X}}} \doteq \exp\{N\mathcal{H}(Y \mid X)\}
$$

where $F^{x_1,x_n}(P_{X,Y})$ is a subexponential factor. For an irreducible Markov source, it holds

$$
\begin{aligned}
\sum_{\boldsymbol{x}\in\mathcal{X}^N, P_{\boldsymbol{x}}^M \in\Sigma} Q(\boldsymbol{x}) &= \sum_{P_{X,Y}\in\mathcal{P}_N^M\cap\Sigma} U_N(P_{X,Y}) Q_0(x_0) \prod_{(x,y)\in\mathcal{X}^2} Q(y \mid x)^{NP_{X,Y}(x,y)} \\
&\doteq \exp\left\{ N \sup_{P_{X,Y}\in\Sigma} \{\mathcal{H}(Y \mid X) + \sum_{(x,y)\in\mathcal{X}^2} P_{X,Y}(x,y) \log Q(y \mid x)\} \right\} \\
&=: \exp\left\{ -N \inf_{P_{X,Y}\in\Sigma} D(Y \| Q \mid X) \right\}.
\end{aligned}
$$

This is Sanov's theorem for Markov chain.

### 6.5.2 One-dimensional Ising model and classical derivation

The one-dimensional Ising model is defined as

$$p_{\text{1d-Ising}}(\boldsymbol{x}) :\propto \exp\left\{-J\sum_{i=1}^{N-1} x_i x_{i+1} - h\sum_{i=1}^{N} x_i\right\}, \qquad \text{for } \boldsymbol{x} \in \{+1,-1\}^N$$
$$Z_N := \sum_{\boldsymbol{x}\in\{+1,-1\}^N} \exp\left\{-J\sum_{i=1}^{N-1} x_i x_{i+1} - h\sum_{i=1}^{N} x_i\right\}$$

for $J, h \in \mathbb{R}$. First, the classical derivation using the transfer matrix is introduced in order to compare it with the derivation using the method of types for Markov chain. Let

$$Z_N(x_1, x_N) := \sum_{(x_2,\ldots,x_{N-1})\in\{+1,-1\}^{N-2}} \exp\left\{-J\sum_{i=1}^{N} x_i x_{i+1} - h\sum_{i=1}^{N} x_i\right\}.$$

Then, from Lemma 2.1, one obtains

$$\begin{bmatrix} Z_N(+1,+1) & Z_N(+1,-1) \\ Z_N(-1,+1) & Z_N(-1,-1) \end{bmatrix} = \begin{bmatrix} \exp\{-h\} & \exp\{0\} \\ \exp\{0\} & \exp\{+h\} \end{bmatrix} \begin{bmatrix} \exp\{-J-h\} & \exp\{+J-h\} \\ \exp\{+J+h\} & \exp\{-J+h\} \end{bmatrix}^{N-1}.$$

In the above equation, the rightmost matrix is called a *transfer matrix*. Then, one obtains

$$\lim_{N\to\infty} \frac{1}{N}\log Z_N = \lim_{N\to\infty} \frac{1}{N}\log \sum_{(x_1,x_N)\in\{+1,-1\}^2} Z_N(x_0, x_N) = \log \lambda_{\max} \tag{6.17}$$

where $\lambda_{\max}$ denotes the largest eigenvalue of the transfer matrix.

### 6.5.3 Derivation from the method of types for Markov chain

In this section, we introduce the derivation using the method of types for Markov chain. From the method of types for Markov chain,

$$\begin{aligned}\lim_{N\to\infty} \frac{1}{N}\log Z_N &= \sup_{P_{ST}\in\mathcal{P}(\{+1,-1\}^2), P_S=P_T} \{H(S \mid T) - J\mathbb{E}[ST] - h\mathbb{E}[T]\} \\ &= \sup_{P_{ST}\in\mathcal{P}(\{+1,-1\}^2), P_S=P_T} \{H(S,T) - H(T) - J\mathbb{E}[ST] - h\mathbb{E}[T]\}\end{aligned}$$

The Lagrangian of the supremum problem is

$$
\begin{aligned}
&L(\omega, \nu; \lambda_S, \lambda_T, \rho_\omega, \rho_\nu)\\
&= - \sum_{(s,t)\in\{+1,-1\}^2} \omega(s,t)\log\omega(s,t) + \sum_{t\in\{+1,-1\}} \nu(t)\log\nu(t)\\
&\quad - J \sum_{(s,t)\in\{+1,-1\}^2} \omega(s,t)st - h \sum_{t\in\{+1,-1\}} \nu(t)t\\
&\quad + \sum_{t\in\{+1,-1\}} \lambda_T(t)\left(\sum_{s\in\{+1,-1\}} \omega(s,t) - \nu(t)\right) + \sum_{s\in\{+1,-1\}} \lambda_S(s)\left(\sum_{t\in\{+1,-1\}} \omega(s,t) - \nu(s)\right)\\
&\quad + \rho_\omega\left(\sum_{s,t}\omega(s,t) - 1\right) + \rho_\nu\left(\sum_t \nu(t) - 1\right).
\end{aligned}
$$

By solving $\partial L/\partial\omega(s,t) = 0$ and $\partial L/\partial\nu(t) = 0$, one respectively obtains

$$
\begin{aligned}
\omega(s,t) &\propto \exp\left\{-Jst + \lambda_T(t) + \lambda_S(s)\right\}\\
\nu(t) &\propto \exp\left\{ht + \lambda_T(t) + \lambda_S(t)\right\}.
\end{aligned}
$$

From the equality of marginals, we can let $\lambda_T = \lambda_S =: \lambda$. By letting a distribution $m_{\mathrm{LR}\to\mathrm{v}}(t) \propto \exp\{\lambda(t)\}$, the normalization constants are

$$
\begin{aligned}
Z_{\mathrm{w}} &= \sum_{(s,t)\in\{+1,-1\}^2} m_{\mathrm{LR}\to\mathrm{v}}(t) m_{\mathrm{LR}\to\mathrm{v}}(s)\exp\{-Jst\}\\
Z_{\mathrm{v}} &= \sum_{t\in\{+1,-1\}} m_{\mathrm{LR}\to\mathrm{v}}(t)^2\exp\{ht\}.
\end{aligned}
$$

Then, one obtains the following lemma.

**Lemma 6.14.**

$$
\lim_{N\to\infty}\frac{1}{N}\log Z_N = \max_{m_{\mathrm{LR}\to\mathrm{v}}\in\mathcal{S}}\left\{\log Z_{\mathrm{w}} - \log Z_{\mathrm{v}}\right\}.
$$

*where $\mathcal{S}$ denotes the set of saddle points of the function for which the maximization is taken.*

The saddle point condition is

$$
\frac{2\sum_{s\in\{+1,-1\}} m_{\mathrm{LR}\to\mathrm{v}}(s)\exp\{-Jst\}}{\sum_{(s,t)\in\{+1,-1\}^2} m_{\mathrm{LR}\to\mathrm{v}}(t) m_{\mathrm{LR}\to\mathrm{v}}(s)\exp\{-Jst\}} = \frac{2 m_{\mathrm{LR}\to\mathrm{v}}(t)\exp\{ht\}}{\sum_{t\in\{+1,-1\}} m_{\mathrm{LR}\to\mathrm{v}}(t)^2\exp\{ht\}}
$$

which is equivalent to the condition of the consistency between $\omega$ and $\nu$. The saddle point condition can be read as

$$
m_{\mathrm{LR}\to\mathrm{v}}(t) \propto \sum_{s\in\{+1,-1\}} m_{\mathrm{LR}\to\mathrm{v}}(s)\exp\{-Jst - ht\} \tag{6.18}
$$

which means that $m_{\mathrm{LR}\to\mathrm{v}}$ is the eigenvector of the matrix

$$\begin{bmatrix} \exp\{-J-h\} & \exp\{J-h\} \\ \exp\{J+h\} & \exp\{-J+h\} \end{bmatrix}. \tag{6.19}$$

Equation (6.18) can be also regarded as the forward-backward algorithm for the model. When $m_{\mathrm{LR}\to\mathrm{v}}$ satisfies (6.18),

$$\begin{aligned} \log\frac{Z_{\mathrm{w}}}{Z_{\mathrm{v}}} &= \log\frac{\sum_{(s,t)\in\{+1,-1\}^2} m_{\mathrm{LR}\to\mathrm{v}}(t)m_{\mathrm{LR}\to\mathrm{v}}(s)\exp\{-Jst\}}{\sum_{t\in\{+1,-1\}} m_{\mathrm{LR}\to\mathrm{v}}(t)^2\exp\{ht\}} \\ &= \log\frac{\sum_{(s,t)\in\{+1,-1\}^2} m_{\mathrm{LR}\to\mathrm{v}}(t)m_{\mathrm{LR}\to\mathrm{v}}(s)\exp\{-Jst\}}{\frac{1}{Z_{\mathrm{LR}\to\mathrm{v}}}\sum_{(s,t)\in\{+1,-1\}^2} m_{\mathrm{LR}\to\mathrm{v}}(s)m_{\mathrm{LR}\to\mathrm{v}}(t)\exp\{-Jst\}} = \log Z_{\mathrm{LR}\to\mathrm{v}} \end{aligned}$$

where $Z_{\mathrm{LR}\to\mathrm{v}}$ is the normalization constant for $m_{\mathrm{LR}\to\mathrm{v}}$, which is the eigenvalue of (6.19) corresponding to the eigenvector $m_{\mathrm{LR}\to\mathrm{v}}$. Hence, Lemma 6.14 gives the largest eigenvalue of the transfer matrix, which is the same result (6.17). From the Perron-Frobenius theorem, the condition $m_{\mathrm{LR}\to\mathrm{v}}(x) \geq 0$ is not restrictive. Note that the simple iteration algorithm obtained by (6.18) is nothing but the power iteration with $\ell_1$ normalization. Similarly to Theorem 6.1, (6.18) is the forward-backward algorithm for the small factor graph. The benefit of the use of the method of types is much clearer for more complicated problems including randomness which cannot be solved without the non-rigorous methods such as replica method and cavity method [Mori and Tanaka, 2011]. The method of types and large deviation for two-dimensional Markov models are open problems [Touchette, 2009].

## 6.6 Detailed asymptotic analysis

In the previous sections, we consider the exponent of the partition function. In the following sections, we consider more detailed analysis similarly to Chapter 5 for some simple model. Let $\mathcal{X} \subsetneq \mathbb{R}$ be a finite set. In this chapter, it is assumed that $n$-th moment of the partition function has the form

$$\begin{aligned} \mathbb{E}[Z^n] := \sum_{\boldsymbol{x}\in(\mathcal{X}^n)^N} \exp\Biggl\{ &\sum_{i=1}^N f\Bigl(\bigl(x_i^{(a)}\bigr)_{a\in\{1,\dots,n\}}\Bigr) \\ &+ Ng\Biggl(\Biggl(\frac{1}{N}\sum_{i=1}^N x_i^{(a)}x_i^{(b)}\Biggr)_{a\in\{1,\dots,n\},\, b\in\{a,\dots,n\}}\Biggr)\Biggr\} \end{aligned} \tag{6.20}$$

where $f$ and $g$ are bounded continuous functions taking $n$ and $n(n+1)/2$ arguments, respectively. The function $g$ is assumed to be invariant under permutations of the replica indices $a$, $b$, and to have a Hessian matrix $\nabla^2 g$. This model includes as special cases various models often studied in statistical physics and information theory, e.g., the SK model [Mézard and Montanari, 2009], random matrices [Edwards and Jones, 1976], CDMA channels [Tanaka, 2002], etc. By using the method of types, one obtains

$$\mathbb{E}[Z^n] = \sum_{v(\boldsymbol{x})} \binom{N}{(v(\boldsymbol{x}))_{x\in\mathcal{X}}} \exp\Bigg\{ \sum_{\boldsymbol{x}\in\mathcal{X}^n} v(\boldsymbol{x}) f\left(\left(x^{(a)}\right)_{a\in\{1,\dots,n\}}\right) + Ng\left(\left(\frac{1}{N}\sum_{\boldsymbol{x}\in\mathcal{X}^n} v(\boldsymbol{x}) x^{(a)} x^{(b)}\right)_{a\in\{1,\dots,n\},\, b\in\{a,\dots,n\}}\right)\Bigg\}$$

where $v(\boldsymbol{x})$ is a type of length $N$ on the alphabet $\mathcal{X}^n$ [Monasson, 1998]. From Laplace's method, it holds

$$F := \lim_{N\to\infty} \frac{1}{N} \log \mathbb{E}[Z^n] = \max_{\nu(\boldsymbol{x})\in\mathcal{P}(\mathcal{X})} \Bigg\{ \mathcal{H}(\nu) + \left\langle f\left(\left(X^{(a)}\right)_{a\in\{1,\dots,n\}}\right)\right\rangle_\nu + g\left(\left(\left\langle X^{(a)} X^{(b)}\right\rangle_\nu\right)_{a\in\{1,\dots,n\},\, b\in\{a,\dots,n\}}\right)\Bigg\} \tag{6.21}$$

where $\nu(\boldsymbol{x})$ denotes a probability measure on $\mathcal{X}^n$ and where $\langle h(\boldsymbol{X})\rangle_\nu := \sum_{\boldsymbol{x}\in\mathcal{X}^n} \nu(\boldsymbol{x}) h(\boldsymbol{x})$ for any function $h(\boldsymbol{x})$. First, we consider the following maximization problem with respect to $\nu \in \mathcal{P}(\mathcal{X}^n)$ for given $(q_{ab} \in \mathbb{R})_{a\in\{1,\dots,n\},b\in\{a,\dots,n\}}$

$$\begin{aligned} &\text{maximize } : \mathcal{H}(\nu) + f\left(\left(X^{(a)}\right)_{a\in\{1,\dots,n\}}\right) \\ &\text{subject to } : \left\langle X^{(a)} X^{(b)}\right\rangle_\nu = q_{ab}, \qquad \text{for } a \in \{1,\dots,n\}, b \in \{a,\dots,n\}. \end{aligned}$$

From the method of Lagrange multiplier, the solution of the maximization problem has the following form for Lagrange multiplier $(\lambda_{ab} \in \mathbb{R})_{a\in\{1,\dots,n\},b\in\{a,\dots,n\}}$

$$\nu(\boldsymbol{x}) = \frac{1}{Z_\nu} \exp\left\{ f\left((x^{(a)})_{a\in\{1,\dots,n\}}\right) - \frac{1}{2}\sum_{a,b} \lambda_{ab} x^{(a)} x^{(b)} \right\}$$

$$Z_\nu := \sum_{\boldsymbol{x}\in(\mathcal{X}^n)^N} \exp\left\{ f\left((x^{(a)})_{a\in\{1,\dots,n\}}\right) - \frac{1}{2}\sum_{a,b} \lambda_{ab} x^{(a)} x^{(b)} \right\}$$

where $(\lambda_{ab})_{a\in\{1,\dots,n\},b\in\{a,\dots,n\}}$ must satisfy

$$\frac{1}{Z_\nu} \sum_{\boldsymbol{x}\in(\mathcal{X}^n)^N} x^{(a)} x^{(b)} \exp\left\{ f\left((x^{(a)})_{a\in\{1,\dots,n\}}\right) - \frac{1}{2}\sum_{a,b} \lambda_{ab} x^{(a)} x^{(b)} \right\} = q_{ab}.$$

By substituting this form of $\nu(\boldsymbol{x})$ to (6.21), one obtains

$$F = \max_{(\lambda_{ab}, q_{ab})\in\mathcal{S}} \left\{ \log Z_\nu + \frac{1}{2}\sum_{a,b} \lambda_{ab} q_{ab} + g\left((q_{ab})_{a\in\{1,\dots,n\},b\in\{a,\dots,n\}}\right) \right\}$$

where $\mathcal{S}$ denotes the set of saddle points of the function for which the maximization is taken.

**Theorem 6.15** (Central approximation for the dense model [Mori and Tanaka, 2012a])**.** *Assume that the solution of the maximization problem* (6.21) *is unique and is denoted by $v^*(\boldsymbol{x})$. Furthermore, assume $v^*(\boldsymbol{x}) > 0$ for all $\boldsymbol{x} \in \mathcal{X}^n$ and*

$$\det\left(I_{n(n+1)/2} - \nabla^2 g(v^*)(U' - U)\right) > 0$$

*where $U'$ and $U$ are $n(n+1)/2 \times n(n+1)/2$ matrices defined by*

$$U'_{(a,b),(c,d)} = \langle X^{(a)} X^{(b)} X^{(c)} X^{(d)} \rangle_{v^*}$$
$$U_{(a,b),(c,d)} = \langle X^{(a)} X^{(b)} \rangle_{v^*} \langle X^{(c)} X^{(d)} \rangle_{v^*}.$$

*Then,*

$$\mathbb{E}[Z^n] \approx \mathrm{e}^{NF} \det\left(I_{n(n+1)/2} - \nabla^2 g(v^*)(U' - U)\right)^{-\frac{1}{2}}$$

*where $F$ is given by* (6.21).

*Proof.* Similarly to the proof of Theorem 5.4, one obtains

$$\begin{aligned}\mathbb{E}[Z^n] &\approx \mathrm{e}^{NF} \frac{1}{\prod_{\boldsymbol{x} \in \mathcal{X}^n} \sqrt{v^*(\boldsymbol{x})}} \det\left(H^t(B - J\nabla^2 g(v^*) J^t) H\right)^{-\frac{1}{2}} \\ &= \mathrm{e}^{NF} \det\left(I_{|\mathcal{X}|^n - 1} - H^t J(\nabla^2 g(v^*)) J^t H (H^t B H)^{-1}\right)^{-\frac{1}{2}}\end{aligned}$$

where $H$ is the $|\mathcal{X}^n| \times (|\mathcal{X}^n| - 1)$ matrix defined by

$$H_{\boldsymbol{x},\boldsymbol{x}'} = \begin{cases} -1, & \text{if } \boldsymbol{x} = \boldsymbol{x}_0 \\ 1, & \text{if } \boldsymbol{x} = \boldsymbol{x}' \\ 0, & \text{otherwise} \end{cases} \qquad \text{for } \boldsymbol{x} \in \mathcal{X}^n,\ \boldsymbol{x}' \in \mathcal{X}^n \setminus \{\boldsymbol{x}_0\}$$

for any fixed $\boldsymbol{x}_0 \in \mathcal{X}^n$, $B$ is the $|\mathcal{X}^n| \times |\mathcal{X}^n|$ diagonal matrix defined by $B_{\boldsymbol{x},\boldsymbol{x}} = 1/v^*(\boldsymbol{x})$, and $J$ is the $|\mathcal{X}^n| \times n(n+1)/2$ matrix defined by $J_{\boldsymbol{x},(a,b)} = x^{(a)} x^{(b)}$. One obtains

$$\det\left(I_{|\mathcal{X}|^n - 1} - H^t J(\nabla^2 g(v^*)) J^t H (H^t B H)^{-1}\right) = \det\left(I_{n(n+1)/2} - \nabla^2 g(v^*)(U' - U)\right)$$

by using Sylvester's determinant theorem (Lemma A.4) and the following equations, which can be verified easily

$$H(H^t B H)^{-1} H^t = S' - S, \qquad\qquad J^t(S' - S)J = U' - U$$

where $S' = B^{-1}$ and $S_{\boldsymbol{x},\boldsymbol{x}'} = v^*(\boldsymbol{x}) v^*(\boldsymbol{x}')$. □

Note that if the solution of the maximization problem (6.21) is not unique but finite, the constant factor is

$$\sum_{v^*(\boldsymbol{x})} \det\left(I_{n(n+1)/2} - \nabla^2 g(v^*)(U' - U)\right)^{-\frac{1}{2}}$$

where the contributions from all solutions $v^*(\boldsymbol{x})$ of the maximization problem (6.21) are summed up. For the $p$-spin model [Mézard and Montanari, 2009], $\nabla^2 g(v^*)$ is a diagonal matrix whose diagonal elements are $\nabla^2 g(v^*)_{(a,b),(a,b)} = \beta^2 \binom{p}{2} \langle X^{(a)} X^{(b)} \rangle_{v^*}^{p-2}$ where $\beta > 0$ is the inverse temperature. The positive definiteness of the matrix for which the determinant is taken is equivalent to the Almedia-Thouless (AT) condition [Almeida and Thouless, 1978], which is a condition for local convexity of a RS solution.

## 6.7 The replica symmetry assumption and $n \to 0$

In the replica theory, we often assume the RS assumption, i.e., $v^*(\boldsymbol{x})$ is invariant under permutations of the $n$ variables in $\boldsymbol{x} \in \mathcal{X}^n$. In this section, for simplicity, it is assumed that the alphabet is $\mathcal{X} = \{+1, -1\}$. The matrices $\nabla^2 g(v^*)$, $U'$ and $U$ can thus be reduced to $n(n-1)/2 \times n(n-1)/2$ matrices since $x^{(a)} x^{(a)} = 1$ always holds. It is known that $\nabla^2 g(v^*)$ and $U' - U$ share the same eigenspaces [Almeida and Thouless, 1978], [Nishimori, 2001], [Tanaka, 2002], [Mézard and Montanari, 2009]. Let $A$ be the $n(n-1)/2 \times n(n-1)/2$ matrix with elements

$$A_{(a,b),(c,d)} = \begin{cases} P, & \text{if } |\{a,b\} \cap \{c,d\}| = 2 \\ Q, & \text{if } |\{a,b\} \cap \{c,d\}| = 1 \\ R, & \text{if } |\{a,b\} \cap \{c,d\}| = 0. \end{cases} \tag{6.22}$$

Both $U' - U$ and $\nabla^2 g(v^*)$ are of this form on the RS assumption. The eigenvectors of $A$ does not depend on $P$, $Q$ and $R$. From this observation, one obtains

$$\begin{aligned} &\det\left(I_{n(n-1)/2} - \nabla^2 g(v^*)(U' - U)\right) = \\ &\left(1 - \left(1 - q^2 + 2(n-2)q(1-q) + \frac{(n-2)(n-3)}{2}(r - q^2)\right)\right. \\ &\qquad \left. \cdot \left(P + 2(n-2)Q + \frac{(n-2)(n-3)}{2} R\right)\right) \\ &\cdot \left(1 - \left(1 - q^2 + (n-4)q(1-q) - (n-3)(r-q^2)\right)\left(P + (n-4)Q - (n-3)R\right)\right)^{n-1} \\ &\cdot \left(1 - \left(1 - q^2 - 2q(1-q) + r - q^2\right)(P - 2Q + R)\right)^{\frac{n(n-3)}{2}} \end{aligned}$$

where $P$, $Q$ and $R$ are (6.22) for $\nabla^2 g(\nu^*)$ and where $q := \langle X^{(a)} X^{(b)} \rangle_{\nu^*}$, $r := \langle X^{(a)} X^{(b)} X^{(c)} X^{(d)} \rangle_{\nu^*}$. In the definitions of $q$ and $r$, the indices $a$, $b$, $c$ and $d$ are all different. In the limit $n \to 0$, the finite-size correction term of the RS free energy $\mathbb{E}[\log Z]/N$ is

$$\begin{aligned}
&\lim_{n\to 0} \frac{1}{n} \frac{1}{N} \log\det \left( I_{n(n-1)/2} - \nabla^2 g(\nu^*)(U' - U) \right)^{-\frac{1}{2}} \\
&= -\frac{1}{2N} \Bigg[ \log\left(1 - (1 - 4q + 3r)\left(P - 4Q + 3R\right)\right) \\
&\quad - \frac{3}{2} \log\left(1 - (1 - 2q + r)\left(P - 2Q + R\right)\right) \Bigg]
\end{aligned}$$

where the variables $q$, $r$, $P$, $Q$ and $R$ are to be determined by the saddle point condition of the RS free energy [Mézard and Montanari, 2009]. For the SK model where $P = \beta^2$, $Q = R = 0$, in the paramagnetic phase $\beta < 1$ where $q = r = 0$, the finite-size correction term is $(1/(4N)) \log(1 - \beta^2)$. This result is known in [Parisi et al., 1993]. For the SK model, at the critical temperature $\beta = \beta_{\mathrm{c}} := 1$, eigenvalues of the Hessian include zero, i.e., the phase transition is the second order. For $\beta > 1$ where the full-step replica symmetry breaking must be considered, the Hessian also includes zero eigenvalue. Hence, for $\beta \geq 1$, the second derivative analysis is not sufficient and the analysis of third or higher-order derivative is needed [Flajolet and Sedgewick, 2009]. For $\beta = 1$, the results are partially obtained in [Parisi et al., 1993].

## 6.8 Perturbation of the empirical joint distribution from the i.i.d. Boltzmann distributions

For $\boldsymbol{x} \in (\mathcal{X}^m)^N$, let the $m$-joint empirical distribution be

$$\nu_m^{\boldsymbol{x}}(\boldsymbol{z}) := \frac{1}{N} \sum_{i=1}^{N} \prod_{a=1}^{m} \mathbb{I}\left\{ x_i^{(a)} = z^{(a)} \right\}, \qquad \text{for } \boldsymbol{z} \in \mathcal{X}^m.$$

For a probability distribution $p(\boldsymbol{x}) \propto \exp\{-E(\boldsymbol{x})\}$, the probability distribution of the empirical joint distribution is defined as

$$P_E((\nu(\boldsymbol{z}))) := \sum_{\boldsymbol{x} \in (\mathcal{X}^m)^N} \frac{\prod_{a=1}^{m} \exp\{-E(\boldsymbol{x}^{(a)})\}}{Z^m} \prod_{\boldsymbol{z} \in \mathcal{X}^m} \mathbb{I}\left\{ \nu_m^{\boldsymbol{x}}(\boldsymbol{z}) \leq \nu(\boldsymbol{z}) \right\}.$$

Here, we consider randomness of the energy function and the expectation of $P_E((\nu(\boldsymbol{z})))$ with respect to it, i.e., $P((\nu(\boldsymbol{z}))) := \mathbb{E}[P_E((\nu(\boldsymbol{z})))]$. By the replica method, it can be calculated as [Mézard and Montanari, 2009]

$$P((\nu(\boldsymbol{z}))) = \lim_{n\to 0} \sum_{\boldsymbol{x} \in (\mathcal{X}^n)^N} \mathbb{E}\left[ \prod_{a=1}^{n} \exp\{-E(\boldsymbol{x}^{(a)})\} \right] \frac{1}{\binom{n}{m}} \sum_{\substack{\mathcal{A} \subseteq \{1,\ldots,n\},\ \boldsymbol{z} \in \mathcal{X}^n \\ |\mathcal{A}| = m}} \prod \mathbb{I}\left\{ \nu_m^{\boldsymbol{x}^{(\mathcal{A})}}(\boldsymbol{z}) \leq \nu(\boldsymbol{z}) \right\}.$$

Almost the same calculation as that of $\mathbb{E}[Z^n]$ shows that it tends to the delta distribution on the RS assumption [Mézard and Montanari, 2009]

$$\lim_{N\to\infty} P((\nu(z))) = \prod_{z\in\mathcal{X}^n} \mathbb{I}\left\{\nu_m^{\mathrm{RS}}(z) \le \nu(z)\right\}$$

where $\nu_m^{\mathrm{RS}}(\boldsymbol{x})$ is the $m$-joint distribution determined from the RS solution. For the dense model, i.e., $\mathbb{E}[Z^n]$ is of the form of (6.20), by the same calculation as that of $\mathbb{E}[Z^n]$, the distribution around the expectation can be obtained from

$$P'((s(z))) := \lim_{n\to 0} \sum_{\boldsymbol{x}\in(\mathcal{X}^n)^N} \mathbb{E}\left[\prod_{a=1}^{n} \exp\{-E(\boldsymbol{x}^{(a)})\}\right] \cdot \prod_{z\in\mathcal{X}^n} \mathbb{I}\left\{ \sqrt{N}(\nu_m^{\boldsymbol{x}^{(1,\dots,m)}}(z) - \nu_m^{\mathrm{RS}}(z)) \le s(z)\right\}.$$

**Theorem 6.16** (Central limit theorem for the dense model [Mori and Tanaka, 2012a])**.** *It is assumed that the replica method on the RS assumption gives a correct result. On the assumption of Theorem 6.15,* $\left\{\sqrt{N}\left(\nu_m^{\boldsymbol{x}}(z) - \nu_m^{\mathrm{RS}}(z)\right)\right\}_{z\in\mathcal{X}^m}$ *weakly converges to the degenerate Gaussian distribution of zero mean and the covariance matrix* $(S'-S)(I_{|\mathcal{X}^m|} - J\nabla^2 g(\nu^*)J^t(S'-S))^{-1}$ *evaluated at the RS solution.*

Let the overlaps $q_{ab}^{\boldsymbol{x}} := \langle X^{(a)}X^{(b)}\rangle_{\nu_m^{\boldsymbol{x}}}$. As a consequence of Theorem 6.16, $\{\sqrt{N}(q_{ab}^{\boldsymbol{x}} - q^{\mathrm{RS}})\}_{a\in\{1,\dots,m\},\, b\in\{a,\dots,m\}}$ weakly converges to the Gaussian distribution of zero mean and the covariance matrix $(U'-U)\left(I_{m(m+1)/2} - \nabla^2 g(\nu^*)(U'-U)\right)^{-1}$. This result is known for SK model at high temperature $\beta < 1$ rigorously (without replica method nor cavity method) [Comets and Neveu, 1995] where the covariance matrix is $1/(1-\beta^2)I_{m(m-1)/2}$. Obviously, a local limit theorem also holds although it is not explicitly stated here. Furthermore, finite-size scaling can be generally obtained on the basis of this analysis by choosing $\beta$ dependently on $N$ for problems including the second-order phase transition similarly to the SK model [Parisi et al., 1993], [Billoire, 2008]. At the critical temperature of the first-order phase transition, i.e., there are two solutions for (6.21), $\left\{\sqrt{N}\left(\nu_m^{\boldsymbol{x}}(z) - \nu_m^{\mathrm{RS}}(z)\right)\right\}_{z\in\mathcal{X}^m}$ converges to mixture of two Gaussian distributions. The weight of each Gaussian distribution is determined by the determinants of the variance–covariance matrices. Similarly to the second-order phase transition, by choosing $\beta$ dependently on $N$, one can control the weights of Gaussian distributions. This idea gives finite-size scaling for the first-order phase transition.

# A Edge Zeta Function and Hessian of the Bethe Entropy

**In this appendix, the Watanabe-Fukumizu formula is introduced, which shows that the determinant of the Hessian of the Bethe entropy can be expressed by using the edge zeta function.**

## A.1 Edge zeta function

**Definition A.1.** For $(i, a) \in E$ and $(j, b) \in E$, $(i \to a) \rightharpoonup (j \to b) \overset{\text{def}}{\iff} j \in \partial a, i \neq j, a \neq b$.

The set of backtrackless closed walks is defined as $\mathfrak{C} := \{(e_1, \ldots, e_n) \in E^n \mid e_1 \rightharpoonup e_2 \rightharpoonup \ldots \rightharpoonup e_n \rightharpoonup e_1\}$. We say $w_1 \sim w_2$ for $w_1, w_2 \in \mathfrak{C}$ if and only if $w_1$ is a cyclic permutation of $w_2$. Let $\mathfrak{D}/\sim$ be the set of all equivalence classes for $\mathfrak{D} \subseteq \mathfrak{C}$.

**Definition A.2** (Prime cycle)**.** The backtrackless closed walk $e_1 \rightharpoonup e_2 \rightharpoonup \ldots \rightharpoonup e_n \rightharpoonup e_1$ is said to be a prime cycle if and only if it cannot be expressed as a power of another walk. Let $\mathfrak{P}$ be the set of prime cycles.

**Definition A.3** (Edge zeta function [Watanabe, 2010])**.** Let $r_{i\to a}$ be a natural number associated with an edge $(i, a) \in E$ and $u_{(i\to a),(j\to b)}$ be an $r_{i\to a} \times r_{j\to b}$ matrix for $(i \to a) \rightharpoonup (j \to b)$. Then, the edge zeta function is defined as

$$\zeta(\boldsymbol{u}) := \prod_{\mathfrak{p}=(e_1 \rightharpoonup e_2 \cdots \rightharpoonup e_n \rightharpoonup e_1) \in \mathfrak{P}/\sim} \frac{1}{\det\left(I_{r_{e_1}} - u_{e_1,e_2} u_{e_2,e_3} \cdots u_{e_n,e_1}\right)}$$

where $I_r$ is the identity matrix of size $r$.

**Lemma A.4** (Sylvester's determinant theorem)**.** *For $n \times m$ matrix $A$ and $m \times n$ matrix $B$,*

$$\det(I_n + AB) = \det(I_m + BA).$$

*Proof.* One obtains the lemma from

$$\begin{bmatrix} I_n + AB & -A \\ 0 & I_m \end{bmatrix} \begin{bmatrix} I_n & 0 \\ B & I_m \end{bmatrix} = \begin{bmatrix} I_n & 0 \\ B & I_m \end{bmatrix} \begin{bmatrix} I_n & -A \\ 0 & I_m + BA \end{bmatrix}. \qquad \square$$

If a factor graph includes more than one cycle, the number of prime cycles is infinite. Hence, it is difficult to evaluate $\zeta(\boldsymbol{u})$ from the definition. The following lemma is generally useful for evaluating $\zeta(\boldsymbol{u})$.

**Lemma A.5** (Bass's formula)**.** *It holds $\zeta(\boldsymbol{u}) = \det(\mathcal{I}_{|E|} - \mathcal{M}(\boldsymbol{u}))^{-1}$ where $\mathcal{M}(\boldsymbol{u})_{e,e'} := u_{e,e'}$ if $e \rightharpoonup e'$ and $\mathcal{M}(\boldsymbol{u})_{e,e'} := 0$ otherwise, and where $\mathcal{I}_{|E|}$ is the identity matrix of the same size as $\mathcal{M}(\boldsymbol{u})$.*

*Proof.* From $\log\det(\cdot) = \mathrm{tr}(\log(\cdot))$, it holds

$$
\begin{aligned}
\log\zeta(\boldsymbol{u}) &= -\sum_{\mathfrak{p}=(e_1 \rightharpoonup e_2 \cdots \rightharpoonup e_n \rightharpoonup e_1)\in\mathfrak{P}/\sim} \log\left(\det\left(I_{r_{e_1}} - u_{e_1,e_2}u_{e_2,e_3}\cdots u_{e_n,e_1}\right)\right)\\
&= -\sum_{\mathfrak{p}=(e_1 \rightharpoonup e_2 \cdots \rightharpoonup e_n \rightharpoonup e_1)\in\mathfrak{P}/\sim} \mathrm{tr}\left(\log\left(I_{r_{e_1}} - u_{e_1,e_2}u_{e_2,e_3}\cdots u_{e_n,e_1}\right)\right)\\
&= \sum_{\mathfrak{p}=(e_1 \rightharpoonup e_2 \cdots \rightharpoonup e_n \rightharpoonup e_1)\in\mathfrak{P}/\sim} \sum_{k=1}^{\infty}\frac{1}{k}\mathrm{tr}\left(\left(u_{e_1,e_2}u_{e_2,e_3}\cdots u_{e_n,e_1}\right)^k\right)\\
&= \sum_{\mathfrak{p}=(e_1 \rightharpoonup e_2 \cdots \rightharpoonup e_n \rightharpoonup e_1)\in\mathfrak{P}} \frac{1}{n}\sum_{k=1}^{\infty}\frac{1}{k}\mathrm{tr}\left(\left(u_{e_1,e_2}u_{e_2,e_3}\cdots u_{e_n,e_1}\right)^k\right)\\
&= \sum_{\mathfrak{w}=(e_1 \rightharpoonup e_2 \cdots \rightharpoonup e_n \rightharpoonup e_1)\in\mathfrak{C}} \frac{1}{n}\mathrm{tr}\left(u_{e_1,e_2}u_{e_2,e_3}\cdots u_{e_n,e_1}\right)\\
&= \sum_{n=1}^{\infty}\frac{1}{n}\mathrm{tr}\left(\mathcal{M}(\boldsymbol{u})^n\right)\\
&= -\mathrm{tr}\left(\log\left(\mathcal{I}_{|E|} - \mathcal{M}(\boldsymbol{u})\right)\right) = -\log\left(\det\left(\mathcal{I}_{|E|} - \mathcal{M}(\boldsymbol{u})\right)\right).
\end{aligned}
$$

□

Furthermore, another expression of $\zeta(\boldsymbol{u})$ is known on some condition.

**Lemma A.6** (Ihara-Bass formula [Watanabe, 2010])**.** *Let $r_i$ be a natural number associated with a variable node $i \in V$. When $u_{(i\to a),(j\to b)}$ is an $r_i \times r_j$ matrix independent of $b$ and denoted by $u^a_{i\to j}$,*

$$
\zeta(\boldsymbol{u})^{-1} = \det(\mathcal{I}_N - \mathcal{D} + \mathcal{W})\prod_{a\in F}\det(\mathcal{U}^a)
$$

*where $\mathcal{D}$ is an $N \times N$ block diagonal matrix defined by $\mathcal{D}_{i,i} := d_i I_{r_i}$, where $\mathcal{U}^a$ is a $d_a \times d_a$ block matrix defined by $\mathcal{U}^a_{i,i} := I_{r_i}$ and $\mathcal{U}^a_{i,j} := u^a_{i\to j}$ for $i \neq j$, and where $\mathcal{W}$ is an $N \times N$ block matrix defined by $\mathcal{W}_{i,j} := \sum_{a:\{i,j\}\subseteq\partial a} w^a_{i\to j}$. Here, $w^a_{i\to j} := ((\mathcal{U}^a)^{-1})_{i,j}$.*

*Proof.* Let $\mathcal{T}$ be the $N \times |E|$ matrix defined by

$$
\begin{aligned}
\mathcal{T}_{i,(j\to a)} &:= \mathbb{I}\{i = j\} I_{r_j},\\
\mathcal{S}(\boldsymbol{u})_{(i\to a),(j\to b)} &:= \mathbb{I}\{i \neq j, a = b\} u^a_{i\to j}.
\end{aligned}
$$

Let $\mathcal{A} := \mathcal{T}^t\mathcal{T}$, which satisfies $\mathcal{A}_{(i\to a),(j\to b)} := \mathbb{I}\{i = j\} I_{r_i}$. Then,

$$\begin{aligned}\{\mathcal{S}(\boldsymbol{u})(\mathcal{A} - \mathcal{I}_{|E|})\}_{(i\to a, j\to b)} &= \sum_{k\to c} \mathbb{I}\{i \neq k, a = c\} u^a_{i\to k} \mathbb{I}\{k = j, c \neq b\} \\ &= u^a_{i\to j} \mathbb{I}\{j \in \partial a, i \neq j, a \neq b\}\end{aligned}$$

and hence, $\mathcal{M}(\boldsymbol{u}) = \mathcal{S}(\boldsymbol{u})(\mathcal{A} - \mathcal{I}_{|E|})$. One obtains

$$\begin{aligned}\det(\mathcal{I}_{|E|} - \mathcal{M}(\boldsymbol{u})) &= \det(\mathcal{I}_{|E|} - \mathcal{S}(\boldsymbol{u})(\mathcal{A} - \mathcal{I}_{|E|})) \\ &= \det(\mathcal{I}_{|E|} + \mathcal{S}(\boldsymbol{u})) \det(\mathcal{I}_{|E|} - \mathcal{S}(\boldsymbol{u})\mathcal{A}(\mathcal{I}_{|E|} + \mathcal{S}(\boldsymbol{u}))^{-1}) \\ &= \det(\mathcal{I}_{|E|} + \mathcal{S}(\boldsymbol{u})) \det(\mathcal{I}_{|V|} - \mathcal{T}(\mathcal{I}_{|E|} + \mathcal{S}(\boldsymbol{u}))^{-1}\mathcal{S}(\boldsymbol{u})\mathcal{T}^t) \\ &= \det(\mathcal{I}_{|E|} + \mathcal{S}(\boldsymbol{u})) \det(\mathcal{I}_{|V|} - \mathcal{T}(\mathcal{I}_{|E|} - (\mathcal{I}_{|E|} + \mathcal{S}(\boldsymbol{u}))^{-1})\mathcal{T}^t) \\ &= \det(\mathcal{I}_{|E|} + \mathcal{S}(\boldsymbol{u})) \det(\mathcal{I}_{|V|} - \mathcal{D} + \mathcal{T}(\mathcal{I}_{|E|} + \mathcal{S}(\boldsymbol{u}))^{-1}\mathcal{T}^t)\end{aligned}$$

where Lemma A.4 is used in the third equality. □

# A.2 Determinant of Hessian of the Bethe entropy

In this section, $(b_i)_{i\in V}$ and $(b_a)_{a\in F}$ are assumed to be members of arbitrary parametric families of distributions. The alphabet $\mathcal{X}$ is not necessarily finite. For $i \in V$, $b_i$ has a parameter $\boldsymbol{\eta}_i$. For $a \in F$, $b_a$ has a parameter $\boldsymbol{\eta}_a = (\boldsymbol{\eta}_{\langle a\rangle}, (\boldsymbol{\eta}_i)_{i\in\partial a})$. The condition (2.7) is assumed to be satisfied for any coordinate $((\boldsymbol{\eta}_i), (\boldsymbol{\eta}_{\langle a\rangle}))$. In the following, a parameter $\boldsymbol{\eta}$ is denoted by the normal font $\eta$ for simplicity. Let $\varphi_i := -\mathcal{H}(b_i)$ for $i \in V$ and $\varphi_a := -\mathcal{H}(b_a)$ for $a \in F$. The notation $B \succ 0$ means that a matrix $B$ is positive-definite.

**Lemma A.7.** *For* $((\eta_i), (\eta_{\langle a\rangle}))$ *satisfying*

$$\frac{\partial^2 \varphi_i}{\partial\eta_i \partial\eta_i} \succ 0, \quad \forall i \in V, \qquad\qquad \frac{\partial^2 \varphi_a}{\partial\eta_{\langle a\rangle} \partial\eta_{\langle a\rangle}} \succ 0, \quad \forall a \in F$$

*it holds that*

$$\det\left(\nabla^2\left(-\mathcal{H}_{\mathrm{Bethe}}((\eta_i), (\eta_{\langle a\rangle}))\right)\right) = \prod_{i\in V} \det\left(\frac{\partial^2\varphi_i}{\partial\eta_i\partial\eta_i}\right) \prod_{a\in F} \det\left(\frac{\partial^2\varphi_a}{\partial\eta_{\langle a\rangle}\partial\eta_{\langle a\rangle}}\right) \det\left(\mathcal{I}_N - \mathcal{D} + \mathcal{G}\right)$$

*where*

$$\mathcal{G}_{i,j} := \left(\frac{\partial^2\varphi_i}{\partial\eta_i\partial\eta_i}\right)^{-\frac{1}{2}} \left[\sum_{a\in\partial i\cap\partial j} \left(\frac{\partial^2\varphi_a}{\partial\eta_i\partial\eta_j} - \frac{\partial^2\varphi_a}{\partial\eta_i\partial\eta_{\langle a\rangle}} \left(\frac{\partial^2\varphi_a}{\partial\eta_{\langle a\rangle}\partial\eta_{\langle a\rangle}}\right)^{-1} \frac{\partial^2\varphi_a}{\partial\eta_{\langle a\rangle}\partial\eta_j}\right)\right] \left(\frac{\partial^2\varphi_j}{\partial\eta_j\partial\eta_j}\right)^{-\frac{1}{2}}.$$

*Proof.* It is easy to see that

$$\frac{\partial^2\left(-\mathcal{H}_{\text{Bethe}}\right)}{\partial\eta_i\partial\eta_j} = \sum_{a\in\partial i\cap\partial j}\frac{\partial^2\varphi_a}{\partial\eta_i\partial\eta_j} - \delta_{i,j}(d_i-1)\frac{\partial^2\varphi_i}{\partial\eta_i\partial\eta_i}$$
$$\frac{\partial^2\left(-\mathcal{H}_{\text{Bethe}}\right)}{\partial\eta_{\langle a\rangle}\partial\eta_{\langle b\rangle}} = \delta_{a,b}\frac{\partial^2\varphi_a}{\partial\eta_{\langle a\rangle}\partial\eta_{\langle b\rangle}}, \qquad \frac{\partial^2\left(-\mathcal{H}_{\text{Bethe}}\right)}{\partial\eta_i\partial\eta_{\langle a\rangle}} = \frac{\partial^2\varphi_a}{\partial\eta_i\partial\eta_{\langle a\rangle}}.$$

Let $\mathcal{V}$ be a block diagonal matrix defined by

$$\mathcal{V}_{i,i} := \frac{\partial^2\varphi_i}{\partial\eta_i\partial\eta_i}, \qquad \mathcal{V}_{a,a} := \frac{\partial^2\varphi_a}{\partial\eta_{\langle a\rangle}\partial\eta_{\langle a\rangle}}$$

and $\mathcal{C} := \nabla^2\left(-\mathcal{H}_{\text{Bethe}}((\eta_i),(\eta_{\langle a\rangle}))\right) - \mathcal{V}$. Then, one obtains

$$\nabla^2\left(-\mathcal{H}_{\text{Bethe}}((\eta_i),(\eta_{\langle a\rangle}))\right) = \mathcal{V}^{\frac{1}{2}}(\mathcal{I}_{N+|F|} + \mathcal{V}^{-\frac{1}{2}}\mathcal{C}\mathcal{V}^{-\frac{1}{2}})\mathcal{V}^{\frac{1}{2}}.$$

For $\mathcal{F} := \mathcal{V}^{-\frac{1}{2}}\mathcal{C}\mathcal{V}^{-\frac{1}{2}}$, it holds that

$$\mathcal{F}_{i,j} = \mathcal{V}_{i,i}^{-\frac{1}{2}}\sum_{a\in\partial i\cap\partial j}\frac{\partial^2\varphi_a}{\partial\eta_i\partial\eta_j}\mathcal{V}_{j,j}^{-\frac{1}{2}} - \delta_{i,j}d_i\mathcal{I}_{r_i}, \qquad \mathcal{F}_{a,b} = 0$$
$$\mathcal{F}_{i,a} = \mathcal{V}_{i,i}^{-\frac{1}{2}}\frac{\partial^2\varphi_a}{\partial\eta_i\partial\eta_{\langle a\rangle}}\mathcal{V}_{a,a}^{-\frac{1}{2}}, \qquad \mathcal{F}_{a,i} = \mathcal{V}_{a,a}^{-\frac{1}{2}}\frac{\partial^2\varphi_a}{\partial\eta_{\langle a\rangle}\partial\eta_i}\mathcal{V}_{i,i}^{-\frac{1}{2}}.$$

From $\det\left(\nabla^2\left(-\mathcal{H}_{\text{Bethe}}((\eta_i),(\eta_{\langle a\rangle}))\right)\right) = \det(\mathcal{V})\det(\mathcal{I}_{N+|F|} + \mathcal{F})$ and

$$\det(\mathcal{V}) = \prod_{i\in V}\det\left(\frac{\partial^2\varphi_i}{\partial\eta_i\partial\eta_i}\right)\prod_{a\in F}\det\left(\frac{\partial^2\varphi_a}{\partial\eta_{\langle a\rangle}\partial\eta_{\langle a\rangle}}\right)$$

we only have to prove $\det(\mathcal{I}_{N+|F|} + \mathcal{F}) = \det(\mathcal{I}_N - \mathcal{D} + \mathcal{G})$. For $u\times u$, $u\times v$ and $v\times u$ matrices $A$, $B$ and $C$, respectively, it holds

$$\begin{bmatrix} A & B \\ C & I_v \end{bmatrix}\begin{bmatrix} I_u & 0 \\ -C & I_v \end{bmatrix} = \begin{bmatrix} A-BC & B \\ 0 & I_v \end{bmatrix}$$

and hence

$$\det\left(\begin{bmatrix} A & B \\ C & I_v \end{bmatrix}\right) = \det(A-BC). \tag{A.1}$$

Therefore,

$$\det(\mathcal{I}_{N+|F|} + \mathcal{F}) = \det(\mathcal{I}_N + \mathcal{F}_{\text{VV}} - \mathcal{F}_{\text{VF}}\mathcal{F}_{\text{VF}}^t) = \det(\mathcal{I}_N - \mathcal{D} + \mathcal{G}). \qquad \square$$

For an exponential family, the determinant of Hessian of the Bethe entropy is connected to the edge zeta function. The following lemma is useful for our purpose.

**Lemma A.8** (Schur complement)**.** *For $u \times u$, $u \times v$, $v \times u$ and $v \times v$ matrices $A$, $B$, $C$ and $D$, respectively,*

$$X = \begin{bmatrix} A & B \\ C & D \end{bmatrix}, \qquad X^{-1} = \begin{bmatrix} \hat{A} & \hat{B} \\ \hat{C} & \hat{D} \end{bmatrix}$$

*Then, it holds*

$$A^{-1} = \hat{A} - \hat{B}\hat{D}^{-1}\hat{C} \tag{A.2}$$

$$\det(A) = \det(X)\det(\hat{D}) \tag{A.3}$$

*Proof.* The first equality is obtained by eliminating $B$ from an equation system

$$A\hat{A} + B\hat{C} = I, \qquad A\hat{B} + B\hat{D} = 0.$$

From

$$\begin{bmatrix} \hat{A} & \hat{B} \\ \hat{C} & \hat{D} \end{bmatrix} \begin{bmatrix} I_u & 0 \\ -\hat{D}^{-1}\hat{C} & I_v \end{bmatrix} = \begin{bmatrix} \hat{A} - \hat{B}\hat{D}^{-1}\hat{C} & B \\ 0 & \hat{D} \end{bmatrix}$$

it holds $\det(X)^{-1} = \det(A)^{-1}\det(\hat{D})$. □

**Lemma A.9.** *Assume that*

$$\left[\left(\frac{\partial^2 \varphi_a}{\partial \eta_a \partial \eta_a}\right)^{-1}\right]_{i,i} = \left(\frac{\partial^2 \varphi_i}{\partial \eta_i \partial \eta_i}\right)^{-1}. \tag{A.4}$$

*Then,*

$$\zeta^{-1}(\boldsymbol{u}) = \det\left(\nabla^2\left(-\mathcal{H}_{\mathrm{Bethe}}((\eta_i)_{i\in V}, (\eta_{\langle a\rangle})_{a\in F})\right)\right) \prod_{i\in V} \det\left(\frac{\partial^2 \varphi_i}{\partial \eta_i \partial \eta_i}\right)^{d_i - 1} \prod_{a\in F} \det\left(\frac{\partial^2 \varphi_a}{\partial \eta_a \partial \eta_a}\right)^{-1}$$

*where $r_i$ is the number of parameters of $b_i$ for $i \in V$ and where*

$$u^a_{i\to j} = \left(\frac{\partial^2 \varphi_i}{\partial \eta_i \partial \eta_i}\right)^{\frac{1}{2}} \left[\left(\frac{\partial^2 \varphi_a}{\partial \eta_a \partial \eta_a}\right)^{-1}\right]_{i,j} \left(\frac{\partial^2 \varphi_j}{\partial \eta_j \partial \eta_j}\right)^{\frac{1}{2}}. \tag{A.5}$$

*Proof.* From Lemma A.6, it holds

$$\zeta(\boldsymbol{u})^{-1} = \det(\mathcal{I}_N - \mathcal{D} + \mathcal{W}) \prod_{a\in F} \det(\mathcal{U}^a).$$

On the choice of variables (A.5), it holds

$$\mathcal{U}^a_{i,j} = \left(\frac{\partial^2 \varphi_i}{\partial \eta_i \partial \eta_i}\right)^{\frac{1}{2}} \left[\left(\frac{\partial^2 \varphi_a}{\partial \eta_a \partial \eta_a}\right)^{-1}\right]_{i,j} \left(\frac{\partial^2 \varphi_j}{\partial \eta_j \partial \eta_j}\right)^{\frac{1}{2}}$$

from the condition (A.4) for any $i, j \in V$ and $a \in F$ satisfying $i, j \in \partial a$. Let $\mathcal{K}^a$ be a submatrix of $\left(\frac{\partial^2 \varphi_a}{\partial \eta_a \partial \eta_a}\right)^{-1}$ including only its $(i, j)$ matrix elements for $i, j \in V$. From (A.2),

$$\left(\mathcal{K}^{a-1}\right)_{i,j} = \frac{\partial^2 \varphi_a}{\partial \eta_i \partial \eta_j} - \frac{\partial^2 \varphi_a}{\partial \eta_i \partial \eta_{\langle a \rangle}} \left(\frac{\partial^2 \varphi_a}{\partial \eta_{\langle a \rangle} \partial \eta_{\langle a \rangle}}\right)^{-1} \frac{\partial^2 \varphi_a}{\partial \eta_{\langle a \rangle} \partial \eta_j}.$$

Then, it holds

$$\begin{aligned} {}^{a}_{i\to j} = ((\mathcal{U}^a)^{-1})_{i,j} &= \left(\frac{\partial^2 \varphi_i}{\partial \eta_i \partial \eta_i}\right)^{-\frac{1}{2}} \left((\mathcal{K}^a)^{-1}\right)_{i,j} \left(\frac{\partial^2 \varphi_j}{\partial \eta_j \partial \eta_j}\right)^{-\frac{1}{2}} \\ &= \left(\frac{\partial^2 \varphi_i}{\partial \eta_i \partial \eta_i}\right)^{-\frac{1}{2}} \left(\frac{\partial^2 \varphi_a}{\partial \eta_i \partial \eta_j} - \frac{\partial^2 \varphi_a}{\partial \eta_i \partial \eta_{\langle a \rangle}} \left(\frac{\partial^2 \varphi_a}{\partial \eta_{\langle a \rangle} \partial \eta_{\langle a \rangle}}\right)^{-1} \frac{\partial^2 \varphi_a}{\partial \eta_{\langle a \rangle} \partial \eta_j}\right) \left(\frac{\partial^2 \varphi_j}{\partial \eta_j \partial \eta_j}\right)^{-\frac{1}{2}}. \end{aligned}$$

Hence, $\mathcal{W} = \mathcal{G}$ where $\mathcal{G}$ is what appears in Lemma A.7. Now, one obtains

$$\begin{aligned} \zeta(\boldsymbol{u})^{-1} = &\det(\nabla^2 \left(-\mathcal{H}_{\text{Bethe}}((\eta_i)_{i\in V}, (\eta_{\langle a \rangle})_{a\in F})\right)) \\ &\cdot \prod_{i\in V} \det\left(\frac{\partial^2 \varphi_i}{\partial \eta_i \partial \eta_i}\right)^{-1} \prod_{a\in F} \det\left(\frac{\partial^2 \varphi_a}{\partial \eta_{\langle a \rangle} \partial \eta_{\langle a \rangle}}\right)^{-1} \prod_{a\in F} \det(\mathcal{U}^a) \end{aligned}$$

from Lemmas A.6 and A.7. Finally, the lemma is obtained from

$$\begin{aligned} \prod_{a\in F} \det(\mathcal{U}^a) &= \prod_{i\in V} \det\left(\frac{\partial^2 \varphi_i}{\partial \eta_i \partial \eta_i}\right)^{d_i} \prod_{a\in F} \det(\mathcal{K}^a) \\ &= \prod_{i\in V} \det\left(\frac{\partial^2 \varphi_i}{\partial \eta_i \partial \eta_i}\right)^{d_i} \prod_{a\in F} \left[\det\left(\frac{\partial^2 \varphi_a}{\partial \eta_a \partial \eta_a}\right)^{-1} \det\left(\frac{\partial^2 \varphi_a}{\partial \eta_{\langle a \rangle} \partial \eta_{\langle a \rangle}}\right)\right]. \end{aligned}$$

In the above equation, the last equality is obtained by (A.3). □

**Corollary A.10** (Watanabe-Fukumizu formula [Watanabe and Fukumizu, 2009; Watanabe, 2010])**.** *Let $(\eta_i)_{i\in V}$ and $(\eta_a)_{a\in F}$ be the expectation parameters for $(b_i)_{i\in V}$ and $(b_a)_{a\in F}$ corresponding to sufficient statistics $(t_i(x_i))_{i\in V}$ and $((t_i(x_i))_{i\in\partial a}, t_{\langle a \rangle}(\boldsymbol{x}_{\partial a}))_{a\in F}$, respectively. Then, it holds*

$$\begin{aligned} \zeta(\boldsymbol{u})^{-1} = &\det\left(\nabla^2 \left(-\mathcal{H}_{\text{Bethe}}((\eta_i)_{i\in V}, (\eta_{\langle a \rangle})_{a\in F})\right)\right) \\ &\cdot \prod_{i\in V} \det(\text{Var}_{b_i}[t_i(X_i)])^{1-d_i} \prod_{a\in F} \det(\text{Var}_{b_a}[t_a(X_{\partial a})]) \end{aligned}$$

*where* $u^a_{i\to j} = \text{Cor}_{b_a}[t_i(X_i), t_j(X_j)]$.

The Watanabe-Fukumizu formula has many applications. One example is the convexity of the Bethe free energy [Heskes, 2004]. In [Watanabe and Fukumizu, 2009;

Watanabe, 2010], it is proved by using the Watanabe-Fukumizu formula that the Bethe free energy and the Bethe entropy are globally convex if and only if the factor graph has at most one cycle. This strengthens the result in [Pakzad and Anantharam, 2002], which shows that if the factor graph has at most one cycle, the Bethe entropy is convex. Furthermore, the Watanabe-Fukumizu formula gives a clear sufficient condition of local convexity of the Bethe free energy, which is directly related to the local stability condition of the belief propagation [Watanabe, 2010], and gives another proof of the sufficient condition of uniqueness of the fixed point of belief propagation obtained in [Mooij and Kappen, 2007], in which the sufficient condition is proved by using the fixed point theorem.

# Bibliography

A. Al-Bashabsheh and Y. Mao. Normal factor graphs and holographic transformations. *IEEE Transactions on Information Theory*, 57(2):752 –763, Feb. 2011.

J. Almeida and D. Thouless. Stability of the Sherrington–Kirkpatrick solution of a spin glass model. *Journal of Physics A: Mathematical and General*, 11(5):983–990, May 1978.

S. Amari and H. Nagaoka. *Methods of Information Geometry*, volume 191 of *Translations of Mathematical monographs*. Oxford University Press, 2000.

G. An. A note on the cluster variation method. *Journal of Statistical Physics*, 52(3): 727–734, 1988.

H. A. Bethe. Statistical theory of superlattices. *Proceedings of the Royal Society of London. Series A, Mathematical and Physical Sciences*, 150(871):552–575, 1935.

P. Billingsley. Statistical methods in Markov chains. *The Annals of Mathematical Statistics*, 32:12–40, 1961.

A. Billoire. Some aspects of infinite-range models of spin glasses: Theory and numerical simulations. In *Rugged Free Energy Landscapes*, volume 736 of *Lecture Notes in Physics*, pages 11–46. Springer Berlin Heidelberg, 2008.

J. P. Boyd. The devil's invention: Asymptotic, superasymptotic and hyperasymptotic series. *Acta Applicandae Mathematica*, 56:1–98, 1999.

R. W. Butler. *Saddlepoint approximations with applications*, volume 22. Cambridge University Press, 2007.

V. Chandrasekaran, M. Chertkov, D. Gamarnik, D. Shah, and J. Shin. Counting independent sets using the Bethe approximation. *SIAM Journal on Discrete Mathematics*, 25(2):1012–1034, 2011.

V. Y. Chernyak and M. Chertkov. Loop calculus and belief propagation for $q$-ary alphabet: Loop tower. In *Proc. 2007 IEEE Int. Symposium on Inform. Theory, Nice, France*, pages 316–320, June 24–29, 2007.

M. Chertkov and V. Y. Chernyak. Loop calculus in statistical physics and information science. *Physical Review E*, 73(6):065102, 2006a.

M. Chertkov and V. Y. Chernyak. Loop series for discrete statistical models on graphs. *Journal of Statistical Mechanics: Theory and Experiment*, 2006(06):P06009, 2006b.

M. Chertkov and V. Y. Chernyak. Loop calculus helps to improve belief propagation and linear programming decodings of low-density-parity-check codes. In *Proceedings 44th Allerton Conference*, Sept. 27–29, 2006c.

F. Comets and J. Neveu. The Sherrington–Kirkpatrick model of spin glasses and stochastic calculus: The high temperature case. *Communications in Mathematical Physics*, 166(3):549–564, Jan. 1995.

S. Condamin. Study of the weight enumerator function for a Gallager code. *Project Report Cavendish Laboratory, University of Cambridge, Cambridge, UK*, 2002.

I. Csiszár. The method of types. *IEEE Transactions on Information Theory*, 44(6): 2505–2523, 1998.

I. Csiszár and J. Körner. *Information theory: coding theorems for discrete memoryless systems*. Cambridge University Press, 2nd edition, 2011.

A. Dembo and A. Montanari. Ising models on locally tree-like graphs. *The Annals of Applied Probability*, 20(2):565–592, 2010.

A. Dembo, A. Montanari, and N. Sun. Factor models on locally tree-like graphs. 2011. URL `http://arxiv.org/abs/1110.4821`.

D. L. Donoho, A. Maleki, and A. Montanari. Message passing algorithms for compressed sensing: I. motivation and construction. In *Information Theory Workshop (ITW), 2010 IEEE*, pages 1 –5, Jan. 6–8, 2010.

M. Dyer, A. Frieze, and M. Jerrum. On counting independent sets in sparse graphs. *SIAM Journal on Computing*, 31(5):1527–1541, 2002.

M. Dyer, L. A. Goldberg, C. Greenhill, and M. Jerrum. The relative complexity of approximate counting problems. *Algorithmica*, 38(3):471–500, 2004.

S. Edwards and R. Jones. The eigenvalue spectrum of a large symmetric random matrix. *Journal of Physics A: Mathematical and General*, 9(10):1595–1603, Oct. 1976.

M. E. Fisher. Statistical mechanics of dimers on a plane lattice. *Physical Review*, 124 (6):1664, 1961.

P. Flajolet and R. Sedgewick. *Analytic Combinatorics*. Cambridge University Press, 2009.

G. D. Forney, Jr. and P. O. Vontobel. Partition functions of normal factor graphs. In *Proc. 2011 IEEE Inf. Theory and App. Workshop, La Jolla, CA*, 2011.

S. Franz, M. Leone, F. Ricci-Tersenghi, and R. Zecchina. Exact solutions for diluted spin glasses and optimization problems. *Phys. Rev. Lett.*, 87:127209, Aug 2001.

A. Globerson and T. S. Jaakkola. Approximate inference using planar graph decomposition. In *Advances in Neural Information Processing Systems 19: Proceedings of the 2006 Conference*, volume 19, page 473. The MIT Press, 2007.

V. Gómez, J. M. Mooij, and H. J. Kappen. Truncating the loop series expansion for belief propagation. *Journal of Machine Learning Research*, 8:1987–2016, Sept. 2007.

F. Guerra and F. L. Toninelli. The high temperature region of the Viana–Bray diluted spin glass model. *Journal of statistical physics*, 115(1):531–555, 2004.

L. Gurvits. Unleashing the power of Schrijver's permanental inequality with the help of the Bethe approximation, 2011. URL `http://arxiv.org/abs/1106.2844v11`.

T. Heskes. Stable fixed points of loopy belief propagation are local minima of the Bethe free energy. In *Advances in Neural Information Processing Systems*, volume 14, pages 343–350, 2002.

T. Heskes. On the uniqueness of loopy belief propagation fixed points. *Neural Computation*, 16(11):2379–2413, Mar. 2004.

M. Jerrum and A. Sinclair. Approximating the permanent. *SIAM journal on computing*, 18(6):1149–1178, 1989.

M. Jerrum and A. Sinclair. Polynomial-time approximation algorithms for the Ising model. *SIAM Journal on computing*, 22(5):1087–1116, 1993.

M. Jerrum, A. Sinclair, and E. Vigoda. A polynomial-time approximation algorithm for the permanent of a matrix with nonnegative entries. *Journal of the ACM (JACM)*, 51 (4):671–697, 2004.

Y. Kabashima. A CDMA multiuser detection algorithm on the basis of belief propagation. *Journal of Physics A: Mathematical and General*, 36(43):11111, 2003.

M. Kac and J. C. Ward. A combinatorial solution of the two-dimensional Ising model. *Physical Review*, 88:1332–1337, Dec. 1952.

P. W. Kasteleyn. The statistics of dimers on a lattice: I. the number of dimer arrangements on a quadratic lattice. *Physica*, 27(12):1209–1225, 1961.

R. Kikuchi. A theory of cooperative phenomena. *Physical Review*, 81(6):988, 1951.

R. Köetter, W.-C. W. Li, P. O. Vontobel, and J. L. Walker. Pseudo-codewords of cycle codes via zeta functions. In *Information Theory Workshop, 2004. IEEE*, pages 7–12. IEEE, 2004.

W. Krauth and M. Mézard. Storage capacity of memory networks with binary couplings. *Journal de Physique*, 50(20):3057–3066, 1989.

S. Kudekar. *Statistical physics methods for sparse graph codes*. PhD thesis, École Polytechnique Fédérale de Lausanne, 2009.

S. Kudekar and N. Macris. Decay of correlations for sparse graph error correcting codes. *SIAM Journal on Discrete Mathematics*, 25(2):956–988, 2011.

N. Macris. Griffith–Kelly–Sherman correlation inequalities: A useful tool in the theory of error correcting codes. *IEEE Transactions on Information Theory*, 53(2):664–683, 2007.

F. J. MacWilliams and N. J. A. Sloane. *The Theory of Error-Correcting Codes*. North-Holland Amsterdam, 1977.

O. C. Martin, M. Mézard, and O. Rivoire. Frozen glass phase in the multi-index matching problem. *Phys. Rev. Lett.*, 93:217205, Nov 2004.

O. C. Martin, M. Mézard, and O. Rivoire. Random multi-index matching problems. *Journal of Statistical Mechanics: Theory and Experiment*, 2005(09):P09006, 2005.

C. Méasson, A. Montanari, T. J. Richardson, and R. L. Urbanke. The generalized area theorem and some of its consequences. *IEEE Transactions on Information Theory*, 55(11):4793–4821, 2009.

M. Mézard and A. Montanari. *Information physics and computation*. Oxford Graduate Texts, 2009.

M. Mézard, G. Parisi, and M. A. Virasoro. *Spin glass theory and beyond*, volume 9. World Scientific, 1987.

R. Monasson. Optimization problems and replica symmetry breaking in finite connectivity spin glasses. *Journal of Physics A: Mathematical and General*, 31(2):513–529, Jan. 1998.

R. Monasson and R. Zecchina. Statistical mechanics of the random K-satisfiability model. *Physical Review E*, 56(2):1357–1370, 1997. ISSN 1550-2376.

A. Montanari. The glassy phase of Gallager codes. *The European Physical Journal B-Condensed Matter and Complex Systems*, 23(1):121–136, 2001.

A. Montanari. Tight bounds for LDPC and LDGM codes under MAP decoding. *IEEE Transactions on Information Theory*, 51(9):3221–3246, 2005.

A. Montanari. The asymptotic error floor of LDPC ensembles under BP decoding. *44th Allerton Conference on Communications, Control and Computing, Monticello*, pages 1168–1172, Oct 2006.

A. Montanari and T. Rizzo. How to compute loop corrections to the Bethe approximation. *Journal of Statistical Mechanics: Theory and Experiment*, 2005(10):P10011, 2005.

A. Montanari, F. Ricci-Tersenghi, and G. Semerjian. Clusters of solutions and replica symmetry breaking in random k-satisfiability. *Journal of Statistical Mechanics: Theory and Experiment*, 2008(04):P04004, 2008.

J. M. Mooij. *Understanding and improving belief propagation*. PhD thesis, Radboud University Nijmegen, May 2008. URL http://webdoc.ubn.ru.nl/mono/m/mooij_j/undeanimb.pdf.

J. M. Mooij and H. J. Kappen. Sufficient conditions for convergence of the sum-product algorithm. *IEEE Transactions on Information Theory*, 53(12):4422–4437, Dec. 2007.

R. Mori. Connection between annealed free energy and belief propagation on random factor graph ensembles. In *Proc. 2011 IEEE Int. Symp. Inf. Theory, St. Petersburg, Russia*, pages 2010–2014, July 31–Aug. 5, 2011. URL http://arxiv.org/abs/1102.3132v2.

R. Mori and T. Tanaka. Statistical mechanical analysis of low-density parity-check codes on general Markov channel. In *Proc. IEICE Symp. Inf. Theory App., Iwate, Japan*, Nov. 29– Dec. 2 2011. URL `http://arxiv.org/abs/1110.1930`.

R. Mori and T. Tanaka. Central approximation in statistical physics and information theory. In *Proc. 2012 IEEE Int. Symposium on Inform. Theory, St. Petersburg, Russia*, pages 1652–1656, July 1–6, 2012a.

R. Mori and T. Tanaka. New generalizations of the Bethe approximation via asymptotic expansion. In *The 35th Symposium on Information Theory and its Applications (SITA2012), Beppu, Oita, Japan*, Dec. 11–14, 2012b. URL `http://arxiv.org/abs/1210.2592`.

R. Mori, T. Tanaka, K. Kasai, and K. Sakaniwa. Effects of single-cycle structure on iterative decoding of low-density parity-check codes. *IEEE Transactions on Information Theory*, 59(1):238–253, Jan. 2013.

T. Morita. Cluster variation method of cooperative phenomena and its generalization i. *Journal of the Physical Society of Japan*, 12(7):753–755, 1957.

P. Mottishaw and C. de Dominicis. On the stability of randomly frustrated systems with finite connectivity. *Journal of Physics A: Mathematical and General*, 20(6):L375, 1987.

T. Murayama, Y. Kabashima, D. Saad, and R. Vicente. Statistical physics of regular low-density parity-check error-correcting codes. *Physical Review E*, 62(2):1577–1591, 2000.

T. Nakajima and K. Hukushima. Thermodynamic construction of a one-step replica-symmetry-breaking solution in finite-connectivity spin glasses. *Phys. Rev. E*, 80: 011103, Jul 2009.

H. Nishimori. *Statistical Physics of Spin Glasses and Information Processing: An Introduction*, volume 111 of *International Series of Monographs on Physics*. Oxford University Press, USA, 2001.

L. Onsager. Crystal statistics. I. a two-dimensional model with an order-disorder transition. *Physical Review*, 65(3-4):117, 1944.

M. Opper and O. Winther. From naive mean field theory to the TAP equations. In M. Opper and D. Saad, editors, *Advanced mean field methods: theory and practice*, chapter 2, pages 7–20. MIT Press, 2001.

P. Pakzad and V. Anantharam. Belief propagation and statistical physics. In *Proceedings of the Conference on Information Sciences and Systems, Princeton University*, number 225, Mar. 20–22, 2002.

G. Parisi, F. Ritort, and F. Slanina. Critical finite-size corrections for the Sherrington–Kirkpatrick spin glass. *Journal of Physics A: Mathematical and General*, 26(2):247–259, Jan. 1993.

J. Pearl. *Probabilistic reasoning in intelligent systems: networks of plausible inference*. Morgan Kaufmann Publishers Inc., San Francisco, CA, USA, 1988.

R. E. Peierls. On Ising's model of ferromagnetism. *Proceedings of the Cambridge Philosophical Society*, 32:477, 1936.

A. Pelizzola. Cluster variation method in statistical physics and probabilistic graphical models. *Journal of Physics A: Mathematical and General*, 38(33):R309, 2005.

T. Plefka. Convergence condition of the tap equation for the infinite-ranged Ising spin glass model. *Journal of Physics A: Mathematical and General*, 15(6):1971, June 1982.

T. J. Richardson and R. L. Urbanke. *Modern Coding Theory*. Cambridge University Press, 2008.

N. Ruozzi. The Bethe partition function of log-supermodular graphical models. In *Advances in Neural Information Processing Systems, Lake Tahoe, NV, USA*, 2012.

A. Schlijper. Convergence of the cluster-variation method in the thermodynamic limit. *Phys. Rev. B*, 27:6841–6848, Jun 1983.

S. Sherman. Combinatorial aspects of the Ising model for ferromagnetism. I. a conjecture of Feynman on paths and graphs. *Journal of Mathematical Physics*, 1(3): 202–217, 1960.

A. Sinclair, P. Srivastava, and M. Thurley. Approximation algorithms for two-state anti-ferromagnetic spin systems on bounded degree graphs. In *Proceedings of the Twenty-Third Annual ACM-SIAM Symposium on Discrete Algorithms*, pages 941–953. SIAM, 2012.

A. Sly and N. Sun. The computational hardness of counting in two-spin models on d-regular graphs. In *Proceedings of FOCS*, 2012. URL `http://arxiv.org/abs/1203.2602`.

H. M. Stark and A. A. Terras. Zeta functions of finite graphs and coverings, part II. *Advances in Mathematics*, 154(1):132–195, 2000.

E. B. Sudderth, M. J. Wainwright, and A. S. Willsky. Loop series and Bethe variational bounds in attractive graphical models. In *Advances in neural information processing systems*, volume 20, pages 1425–1432, 2008.

T. Tanaka. A statistical-mechanics approach to large-system analysis of CDMA multiuser detectors. *IEEE Transactions on Information Theory*, 48(11):2888–2910, Nov. 2002.

S. C. Tatikonda and M. I. Jordan. Loopy belief propagation and gibbs measures. In *Proceedings of the Eighteenth conference on Uncertainty in artificial intelligence*, pages 493–500. Morgan Kaufmann Publishers Inc., 2002.

Y. W. Teh and M. Welling. The unified propagation and scaling algorithm. *Advances in neural information processing systems*, 14:953–960, 2002.

S. Toda. PP is as hard as the polynomial-time hierarchy. *SIAM Journal on Computing*, 20(5):865–877, 1991.

H. Touchette. The large deviation approach to statistical mechanics. *Physics Reports*, 478(1-3):1–69, 2009.

S. P. Vadhan. The complexity of counting in sparse, regular, and planar graphs. *SIAM Journal on Computing*, 31(2):398–427, 2001.

L. G. Valiant. The complexity of computing the permanent. *Theoretical Computer Science*, 8:189–201, 1979.

L. G. Valiant. Holographic algorithms. *SIAM Journal on Computing*, 37(5):1565–1594, 2008.

L. Viana and A. J. Bray. Phase diagrams for dilute spin glasses. *Journal of Physics C: Solid State Physics*, 18(15):3037, 1985.

R. Vicente, D. Saad, and Y. Kabashima. Low-density parity-check codes—A statistical physics perspective. *Advances in imaging and electron physics*, 125:231–353, 2003.

P. O. Vontobel. Connecting the Bethe entropy and the edge zeta function of a cycle code. In *Proc. 2010 IEEE Int. Symposium on Inform. Theory, Austin, TX*, pages 704–708, June, 13–18 2010a.

P. O. Vontobel. Counting in graph covers: A combinatorial characterization of the Bethe entropy function, 2010b. URL http://arxiv.org/abs/1012.0065v1.

P. O. Vontobel. A combinatorial characterization of the Bethe and the Kikuchi partition functions. In *Information Theory and Applications Workshop (ITA), 2011*, pages 1–10. IEEE, 2011a.

P. O. Vontobel. The Bethe permanent of a non-negative matrix, 2011b. URL http://arxiv.org/abs/1107.4196v3.

P. O. Vontobel and R. Köetter. Graph-cover decoding and finite-length analysis of message-passing iterative decoding of LDPC codes, 2005. URL http://arxiv.org/abs/cs/0512078.

M. J. Wainwright, T. S. Jaakkola, and A. S. Willsky. Tree-based reparameterization framework for analysis of sum-product and related algorithms. *IEEE Transactions on Information Theory*, 49(5):1120–1146, 2003.

M. J. Wainwright, T. S. Jaakkola, and A. S. Willsky. A new class of upper bounds on the log partition function. *IEEE Transactions on Information Theory*, 51(7):2313–2335, 2005.

Y. Watanabe. *Discrete geometric analysis of message passing algorithm on graphs*. PhD thesis, The Graduate University for Advanced Studies, Mar. 2010.

Y. Watanabe. A conjecture on independent sets and graph covers, 2011. URL http://arxiv.org/abs/1109.2445v3.

Y. Watanabe and M. Chertkov. Belief propagation and loop calculus for the permanent of a non-negative matrix. *Journal of Physics A: Mathematical and Theoretical*, 43 (24):242002, 2010.

Y. Watanabe and K. Fukumizu. Graph zeta function in the Bethe free energy and loopy belief propagation. In *Advances in Neural Information Processing Systems*, volume 22, pages 2017–2025, 2009.

D. Weitz. Counting independent sets up to the tree threshold. In *Proceedings of the thirty-eighth annual ACM symposium on Theory of computing*, pages 140–149. ACM, 2006.

P. Whittle. Some distribution and moment formulae for the Markov chain. *Journal of the Royal Statistical Society. Series B (Methodological)*, pages 235–242, 1955.

W. Wiegerinck and T. Heskes. Fractional belief propagation. In *Advances in Neural Information Processing Systems*, pages 455–462. The MIT Press, 2003.

K. Y. M. Wong and D. Sherrington. Intensively connected spin glasses: Towards a replica-symmetry-breaking solution of the ground state. *Journal of Physics A: Mathematical and General*, 21:L459, 1988.

J.-Q. Xiao and H. Zhou. Partition function loop series for a general graphical model: Free-energy corrections and message-passing equations. *Journal of Physics A: Mathematical and Theoretical*, 44(42):425001, 2011.

J. S. Yedidia, W. T. Freeman, and Y. Weiss. Constructing free-energy approximations and generalized belief propagation algorithms. *IEEE Transactions on Information Theory*, 51(7):2282–2312, 2005.

A. L. Yuille. CCCP algorithms to minimize the Bethe and Kikuchi free energies: Convergent alternatives to belief propagation. *Neural Computation*, 14(7):1691–1722, 2002.

H. Zhou, C. Wang, J.-Q. Xiao, and Z. Bi. Partition function expansion on region graphs and message-passing equations. *Journal of Statistical Mechanics: Theory and Experiment*, 2011(12):L12001, 2011.

# Curriculum Vitae

## Ryuhei Mori

36-1 Yoshida-Honmachi, Sakyo-ku,
Kyoto, 606-8501, Japan
Phone: +81 075 753 4822
Email: rmori@sys.i.kyoto-u.ac.jp

## Education

**Tokyo Institute of Technology, Tokyo, Japan** Mar. 2008.
**Bachelor of Engineering**
Thesis Title: Asymptotic finite-length and finite-iteration analysis of low-density parity-check codes.

**Graduate School of Informatics, Kyoto University, Kyoto, Japan** Mar. 2010.
**Master of Informatics**
Thesis Title: Properties and construction of polar codes.

**Graduate School of Informatics, Kyoto University, Kyoto, Japan** Mar. 2013.
**Ph.D. in Informatics**
Thesis Title: New understanding of the Bethe approximation and the replica method.

## Another Position

**Research Fellow of the Japan Society for the Promotion of Science**
Apr. 2010 – Mar. 2013.

## Research Interest

Information theory, coding theory, low-density parity-check codes, polar codes, statistical physics, Bethe approximation, replica method, constraint satisfaction problems.

## Honors

2012, Nov. Ericsson Best Student Award.

2010, Dec. Society of Information Theory and its Applications Encouragement Award.

## Invited Talks

2011, "Researches about polar codes," Symposium on Information Theory and its Applications, Iwate, Nov. 30.

2011, "On the growth rate of spatially coupled LDPC codes," Workshop on Spatially Coupled Codes and Related Topics, Tokyo Institute of Technology, Sep. 21.

2010, "On the polar codes," LDPC codes workshop, Tohoku Gakuin University, Sep. 21.

## Professional Activities

Student Member of IEEE. Reviewer for IEEE Trans. Inf. Theory, IEEE Trans. Commun., IEEE Commun. Lett., IEICE Trans. Fundamentals, IEEE International Symposium on Information Theory, IEEE Information Theory Workshop, and IEICE International Symposium on Information Theory and its Applications.

## Journal Publications

[J4] R. Mori and T. Tanaka, "Source and channel polarization over finite fields and Reed-Solomon matrix," submitted for publication in the IEEE Trans. Inf. Theory in Nov. 2012.

[J3] S. H. Hassani, R. Mori, T. Tanaka, and R. L. Urbanke, "Rate-dependent analysis of the asymptotic behavior of channel polarization," accepted for publication in the IEEE Trans. Inf. Theory in Aug. 2012, doi:10.1109/TIT.2012.2228295.

[J2] R. Mori, T. Tanaka, K. Kasai, and K. Sakaniwa, "Effects of single-cycle structure on iterative decoding of low-density parity-check codes," IEEE Trans. Inf. Theory, vol.59, no.1, pp.238–253, Jan. 2013.

[J1] R. Mori and T. Tanaka, "Performance of polar codes with the construction using density evolution," IEEE Commun. Lett., vol.13, no.7, pp.519–521, Jul. 2009.

## Conference Publications (Peer-Reviewed)

[C10] R. Mori and T. Tanaka, "Central approximation in statistical physics and information theory," 2012 IEEE International Symposium on Information Theory, Cambridge, MA, pp.1657–1661, Jul. 1–6, 2012.

[C9] S.H. Hassani, N. Macris, and R. Mori, "Near concavity of the growth rate for coupled LDPC chains," 2011 IEEE International Symposium on Information Theory, St. Petersburg, Russia, pp.356–360, Jul. 31–Aug. 5, 2011.

[C8] R. Mori, "Connection between annealed free energy and belief propagation on random factor graph ensembles," 2011 IEEE International Symposium on Information Theory, St. Petersburg, Russia, pp.2016–2020, Jul. 31–Aug. 5, 2011.

[C7] R. Mori and T. Tanaka, “Non-binary polar codes using Reed-Solomon codes and algebraic geometry codes,” 2010 IEEE Information Theory Workshop, Dublin, Ireland, Aug. 30–Sep. 3, 2010.

[C6] T. Tanaka and R. Mori, “Refined rate of channel polarization,” 2010 IEEE International Symposium on Information Theory, Austin, TX, pp.889–893, Jun. 13–18, 2010.

[C5] R. Mori and T. Tanaka, “Channel polarization on q-ary discrete memoryless channels by arbitrary kernels,” 2010 IEEE International Symposium on Information Theory, Austin, TX, pp.894–898, Jun. 13–18, 2010.

[C4] R. Mori and T. Tanaka, “Performance and construction of polar codes on symmetric binary-input memoryless channels,” 2009 IEEE International Symposium on Information Theory, Seoul, Korea, pp.1496–1500, Jun. 28–Jul. 3, 2009.

[C3] R. Mori, T. Tanaka, K. Kasai, and K. Sakaniwa, “Finite-length analysis of irregular expurgated LDPC codes under finite number of iterations,” 2009 IEEE International Symposium on Information Theory, Seoul, Korea, pp.2497–2501, Jun. 28–Jul. 3, 2009.

[C2] R. Mori, K. Kasai, T. Shibuya, and K. Sakaniwa, “Asymptotic gaps between BP decoding and local-MAP decoding for low-density parity-check codes,” 2008 5th International Symposium Turbo Codes and Related Topics, Lausanne, Switzerland, pp.162–167, Sep. 1–5, 2008.

[C1] R. Mori, K. Kasai, T. Shibuya, and K. Sakaniwa, “Asymptotic bit error probability of LDPC codes for the binary erasure channel with finite number of iterations,” 2008 IEEE International Symposium on Information Theory, Toronto, Canada, pp.449–453, Jul. 6–11, 2008.

## Domestic Conferences (Not Peer-Reviewed)

[D7] R. Mori and T. Tanaka, “New generalizations of the Bethe approximation via asymptotic expansion,” 2012 IEICE Symposium on Information Theory and its Application, Oita, Japan, Dec. 11–14, 2012.

[D6] R. Mori and T. Tanaka, “Statistical mechanical analysis of low-density parity-check codes on general Markov channel,” 2011 IEICE Symposium on Information Theory and its Application, Iwate, Japan, Nov. 29–Dec. 2, 2011.

[D5] R. Mori, “Growth rate of spatially coupled LDPC codes,” LDPC workshop, Tokyo Institute Technology, Japan, Sep. 29–30, 2011.

[D4] R. Mori, “Lower bound of growth rate of Reed-Muller codes and polar codes,” LDPC workshop, Tokyo Institute Technology, Japan, Sep. 29–30, 2011.

[D3] R. Mori and T. Tanaka, "Rate-dependent analysis of speed of polarization," 2009 Symposium on Information Theory and its Application, Yamaguchi, Japan, Dec. 1–4, 2009.

[D2] R. Mori and T. Tanaka, "Solution-space geometry of random k-satisfiability problem," 2008 Symposium on Information Theory and its Application, Tochigi, Japan, Oct. 7–10, 2008.

[D1] R. Mori, K. Kasai, T. Shibuya, and K. Sakaniwa, "Analysis of error floor of belief propagation decoding for detailedly represented irregular LDPC codes," 2007 Symposium on Information Theory and its Application, Mie, Japan, Oct. 27–30, 2007.